\documentclass[a4paper,12pt]{article}
\usepackage{graphicx}
\usepackage{latexsym}
\usepackage{amsfonts}
\usepackage{amsmath}
\usepackage{psfrag}
\usepackage[sc,small]{caption2}
\setcaptionwidth{14cm}

\numberwithin{equation}{section}

\begin{document}
\begin{flushright}
hep-th/0507285\\
NSF-KITP-2005-54\\
NA-DSF-32-2005 \\
ROM2F/2005/16
\end{flushright}
\vfill
\thispagestyle{empty}

\begin{center}
\textbf{\Large Bulk Dynamics in Confining Gauge Theories}\\[48pt]%
Marcus Berg$^1$, Michael Haack$^1$ and Wolfgang M\"uck$^2$\\[24pt]%
\textit{$^1$ Kavli Institute for Theoretical Physics, University of
  California}\\ 
\textit{Santa Barbara, California 93106-4030, USA} \\%
E-mails: \texttt{mberg,mhaack@kitp.ucsb.edu}\\[24pt]
\textit{$^2$ Dipartimento di Scienze Fisiche, Universit\`a degli Studi di
  Napoli ``Federico II''}\\%
\textit{via Cintia, 80126 Napoli, Italy}\\%
E-mail: \texttt{mueck@na.infn.it}
\end{center}
\vfill

\begin{abstract}
We consider gauge/string duality (in the supergravity approximation) for
confining gauge theories.
The system under scrutiny is a 5-dimensional consistent
truncation of type IIB supergravity obtained using the
Papadopoulos-Tseytlin ansatz with boundary momentum added. 
We develop a gauge-invariant and sigma-model-covariant
approach to the dynamics of 5-dimensional bulk
fluctuations.
For the Maldacena-Nunez subsystem, 
we study glueball mass spectra.
For the Klebanov-Strassler subsystem, 
we compute the linearized
equations of motion for the 7-scalar system, and show that a
3-scalar sector containing the scalar dual to the gluino bilinear
decouples in the UV. We solve the fluctuation equations exactly
in the "moderate UV" approximation and check this
approximation numerically. 
Our results demonstrate the feasibility of analyzing the generally
coupled equations for scalar bulk fluctuations,
and constitute a step on the way towards
computing correlators in confining gauge theories.
\end{abstract}
\vfill \vfill \vfill
\newpage


\newcommand{\ie}{i.e.,\ }
\newcommand{\eg}{e.g.,\ }

\newcommand{\const}{\operatorname{const.}} 

\newcommand{\rmd}{\,\mathrm{d}}

\newcommand{\Tr}{\operatorname{tr}}

\newcommand{\e}[1]{\operatorname{e}^{#1}}

\newcommand{\tg}{\tilde{g}}
\newcommand{\tR}{\tilde{R}}
\newcommand{\tG}[2]{\tilde{\Gamma}^{#1}_{\;#2}}
\newcommand{\tn}{\tilde{\nabla}}

\newcommand{\bp}{\bar{\phi}}
\newcommand{\bq}{\bar{q}}

\newcommand{\vp}{\varphi}

\newcommand{\G}[2]{\mathcal{G}^{#1}_{\;\;#2}}
\newcommand{\R}{\mathcal{R}}

\newcommand{\mfa}{\mathfrak{a}}
\newcommand{\mfb}{\mathfrak{b}}
\newcommand{\mfc}{\mathfrak{c}}
\newcommand{\mfd}{\mathfrak{d}}
\newcommand{\mfe}{\mathfrak{e}}

\newcommand{\ha}{\hat{\mfa}}
\newcommand{\ca}{\check{\mfa}}

\newcommand{\vev}[1]{\left\langle{#1}\right\rangle}

\newcommand{\Order}[1]{\mathcal{O}\left(#1\right)}
\newcommand{\Of}{\Order{f}}
\newcommand{\Ofn}[1]{\Order{f^{#1}}}

\newcommand{\htt}{{h^{TT}}}

\newcommand{\tE}{\tilde{E}}

\newcommand{\K}{\mathcal{K}}

\newcommand{\diag}{\operatorname{diag}}

\newcommand{\rmK}{\operatorname{K}}

\newcommand{\rmM}{\operatorname{M}}


\newcommand{\tF}{\tilde{F}}

\newcommand{\be}{\begin{equation}}
\newcommand{\ee}{\end{equation}}
\newcommand{\beqn}{\begin{eqnarray}}
\newcommand{\eeqn}{\end{eqnarray}}

\newcommand{\non}{\nonumber \\}
\newcommand{\hmm}[1]{{\bf [#1]}\marginpar[\hfill${\bf \Longrightarrow}$]%
                  {${\bf \Longleftarrow}$} }


\section{Introduction}
\label{intro}

Gauge/string duality offers an alternative approach to 
aspects of supersymmetric non-Abelian gauge
theories that are hard to describe with conventional techniques. For
example, at strong coupling many non-Abelian gauge theories 
exhibit \emph{confinement}, the familiar yet still somewhat mysterious
phenomenon that the only finite-energy states are singlets under the
color gauge group: at colliders, we never see quarks directly, only
colorless hadrons. The details of confinement, and of other 
nonperturbative phenomena
such as chiral 
symmetry breaking, are difficult to capture with conventional
gauge theory methods. In
the dual picture, the nonperturbative gauge theory regime is
typically described by weakly coupled closed strings
propagating on a space of higher dimensionality (the {\it bulk}), 
and their dynamics can be approximated by
classical supergravity. 

One of the most powerful applications
of gauge/string duality is the calculation of field
theory correlation functions from the dual bulk dynamics. This idea
was developed in \cite{Maldacena:1997re,Gubser:1998bc, Witten:1998qj} for
superconformal gauge theories, whose gravity duals are Anti-de~Sitter (AdS)
spaces. Since then, in a program known as \emph{holographic
  renormalization} (\cite{Bianchi:2001de, Bianchi:2001kw, Martelli:2002sp,
Berg:2002hy,Papadimitriou:2004ap, Papadimitriou:2004rz} and references
therein), it has been systematically generalized to
gauge theories that are conformal in the ultraviolet (UV), whose
gravity duals are asymptotically AdS spaces.
There are several reasons to push ahead with this line of research.
First, confining gauge theories have duals
that are not asymptotically AdS. Despite some progress, the holographic
calculation of correlators in confining gauge theories has not yet
been carried out in any controlled approximation (we shall specify later
what we mean by that). Second, interesting new supergravity
solutions have been found recently
\cite{Lin:2004nb,Lunin:2005jy,Gursoy:2005cn,Halmagyi:2005pn}.
They are regular and thus qualify as dual configurations of
ground states in gauge theories. It is then 
only natural to
investigate the possibility of calculating correlation functions for
their dual field theories. In this paper, we report on progress
towards this general goal and sharpen some of the remaining challenges. 

Some of the asymptotically AdS backgrounds studied in the literature
(such as the GPPZ flow \cite{Girardello:1999bd}) were originally
envisaged as toy-model duals of confining gauge theories. The
obstruction to being full-fledged duals is a naked curvature
singularity at finite distance from the boundary into the bulk. 
One would have liked to interpret this distance
as the (dynamically generated) scale of onset of
confinement in the dual field theory, but unbounded curvature
invalidates the use of the supergravity approximation to string
theory. Although string theory appears to cure these curvature
singularities by the enhancon mechanism \cite{Johnson:1999qt} or
the Myers effect \cite{Myers:1999ps}, correlators 
are always computed in the supergravity approximation. 
In practice, this involves imposing regularity conditions on the bulk
fluctuations at the curvature singularity.
It would be interesting to quantify the precise effect of the
string theory resolution on the explicit correlators computed in singular
supergravity backgrounds, but it would be simpler if 
one could compute correlators directly from regular duals.

An even simpler approach to computing correlators
is the {\it hard-wall approximation}, which has been used 
 in the effort to connect
gauge/string duality to real QCD. This ``AdS/QCD
correspondence'' studies problems like 
meson-hadron coupling universality
\cite{Hong:2004sa,Hong:2005np,Sakai:2005yt} 
and deep inelastic scattering \cite{Polchinski:2002jw}. 
In the hard-wall approximation, one replaces the regular solution by
AdS space cut off at a 
minimal radius $r_{\text{IR}}$, the idea being that some of the physics
should be insensitive to the details of the geometry in the deep
infrared (IR) region, while retaining conformal UV behaviour.
Then, the issue arises which boundary conditions to impose at the IR
boundary. 
If one were able to compute correlators directly in the regular
solution at least for some simple cases, a qualitative picture of
which hard-wall boundary conditions best mimic the behaviour in the
regular case could be pieced together.\footnote{Incidentally, in
  \cite{Kachru:2003sx}, cut-off AdS was 
  used to model the dynamics of D-brane inflationary cosmology on the
  Klebanov-Strassler background. It is not unreasonable to hope that
  our methods will also prove useful in that context, for the same
  reasons as for AdS/QCD.} 

\begin{figure}[t]
\label{fig:approx}
\begin{center}
\psfrag{g}[bc][bc][.8][0]{gauge theory}
\psfrag{IR}[bc][bc][.8][0]{IR}
\psfrag{regular}[bc][bc][1][0]{regular}
\psfrag{floor}[bc][bc][1][0]{hard-wall approximation}
\psfrag{singular}[bc][bc][1][0]{singular approximation}
\psfrag{=}[bc][bc][1][0]{$\approx$}
\includegraphics[width=0.8\textwidth]{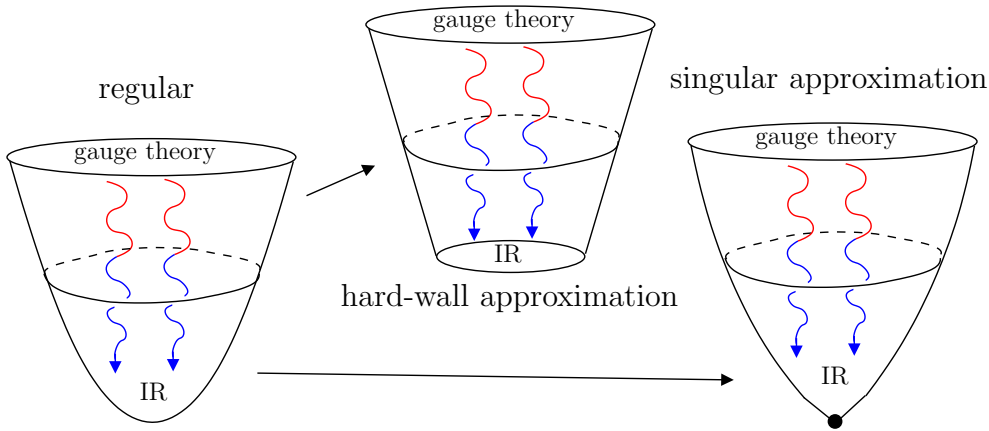}
\caption{Bulk toy models used in the literature.
In the hard-wall approximation, the bulk is exactly AdS, so couplings
in the gauge theory do not run (represented by straight sides in the
figure). In singular approximations, like the singular conifold, there
is logarithmic running, but also a curvature singularity (represented
by the black dot).} 
\end{center}
\end{figure}

Thus, we are interested in the question 
to what extent it is feasible to compute correlators directly from regular
supergravity duals of confining gauge theories. 
The first example of such a bulk configuration was 
the warped deformed conifold solution found by Klebanov and
Strassler (KS) \cite{Klebanov:2000hb}. The
$\mathcal{N}=1$ supersymmetric gauge theory dual to this solution
undergoes a cascade of Seiberg dualities, as recently explained in
greater detail in the lecture notes by Strassler \cite{Strassler:2005}.\footnote{See
also \cite{Herzog:2001xk} for a nice review of the KS solution.}
Importantly, the curvature remains small everywhere, so the
supergravity approximation can be used at all energies.
Another example is the wrapped D5-brane, also known as the
Maldacena-Nunez (MN) solution \cite{Maldacena:2000yy}. In the
infrared, it shares many properties with $\mathcal{N}=1$ SYM theory 
\cite{Apreda:2001qb, DiVecchia:2002ks, Bertolini:2002yr, Muck:2003zf},
but it becomes six-dimensional little string theory in the UV. 
We are mostly interested in the KS solution, since there the supergravity
approximation is under full control.  

Even before addressing the implementation of gauge/string duality in
such bulk configurations, it is worth noting that the dual field theory
interpretation of non-asymptotically-AdS supergravity configurations
poses a conceptual problem (which we do not resolve): in 
holographic renormalization, the asymptotically AdS bulk region
corresponds to the presence of a Wilsonian renormalization group (RG)
UV fixed point in 
the 4-dimensional gauge theory. Thus, in its absence, one might wonder
whether the dual gauge theory is well-defined in the Wilsonian
sense. Several viewpoints on this are possible. One may defer the
UV completion of the field theory to string theory, as in the
MN solution. Alternatively, one can attempt to \emph{define} the field
theory by its holographic dual, as advocated in
\cite{Aharony:2005zr}. Another hope may be to embed the KS solution into a
more complicated configuration with an asymptotically AdS region, so
that the dual field theory is UV-conformal, but there is an intermediate
energy range where couplings do run logarithmically as in the KS
solution. Here, we adopt a pragmatic approach, somewhat like
\cite{Aharony:2005zr}. We extrapolate from AdS/CFT that the bulk
dynamics encodes information about some dual field theory, which might
only be an effective theory, and try to see which of its features can be
extracted by existing holographic renormalization technology.

Optimistically, then, we would like to investigate whether techniques
similar to holographic renormalization can be used to calculate (effective)
field theory correlation functions from supergravity duals in
non-asymptotically AdS setups. First results in this direction were
obtained by Krasnitz \cite{Krasnitz:2000ir, Krasnitz:2002ct,
  Krasnitz:2003pj} for certain 2-point functions in the singular
conifold (Klebanov-Tseytlin) background \cite{Klebanov:2000nc}, and in 
the limit of very large energy. No counterterms were obtained, 
but it was argued that the particular correlators studied would only
have received minor corrections, had counterterms been included.
Counterterms 
were recently studied 
in a tour de force by Aharony, Buchel and Yarom 
\cite{Aharony:2005zr}, who obtained renormalized one-point functions
of the stress-energy tensor. 

To address the problem of computing correlators systematically, one
must face three interrelated issues: 
\begin{enumerate}
\item
define precisely the duality relations between supergravity
fields and field theory operators (the ``dictionary'' problem);
\item
renormalize the bulk prescription for correlation functions, that is,
compute the requisite covariant counterterms and show the absence of
divergences (the ``renormalization'' problem);
\item
solve for the
dynamics of supergravity fluctuations about the background of
interest, where the fluctuations must be allowed to vary along the
external spacetime coordinates (the ``fluctuation'' problem).
\end{enumerate}
In this paper, we will mostly address the last issue.\footnote{Although
it may seem that the first and second issues should be
resolved first, this point is moot: ultimately all three questions
have to be addressed, and as we shall see, the solution to one may
help with the others.}
We focus on gauge theories dual to the regular supergravity solutions
discussed above: the MN solution and especially the KS solution of
type IIB supergravity in ten dimensions. 

In holographic renormalization, 
the bulk dynamics is 5-dimensional (for a 4-dimensional gauge
theory). Thus, we need to find a sector of type IIB supergravity
which can be described by a 5-dimensional system, while allowing
for the background solutions we are interested in. Papadopoulos and
Tseytlin (PT) \cite{Papadopoulos:2000gj} found an effective 1-dimensional 
action (subject to a Hamiltonian constraint) that is general enough 
to describe both the MN and KS background solutions,
where the fields only depend on the radial coordinate. 
This suggests that one can
suitably generalize their ansatz to allow the parameterizing scalar
fields to depend also on the coordinates of the 4 dimensions of the 
gauge theory. In other words, we add boundary momentum
to the PT ansatz, which leads to a 5-dimensional effective theory. 
We will show that, after imposing an integrability constraint that is
automatically satisfied for the MN and KS systems, this generalization
constitutes a consistent truncation of type IIB supergravity and gives
rise to a 
non-linear sigma  model of scalars coupled to 5-dimensional gravity. Moreover, 
the resulting 5-dimensional system falls into a general class of
actions dubbed ``fake supergravity'' actions in
\cite{Freedman:2003ax}, since the scalar potential is determined by a
function resembling a superpotential. We will mostly stick to this
terminology (\ie ``fake supergravity''), even though the background solutions we consider have
been shown to preserve some supersymmety \cite{Gubser:2000vg,Grana:2000jj}, and
one might expect the full system to be embeddable in a supersymmetric
system (see Sec.~\ref{sugratrunc} for some further comments on this). 
These fake supergravity actions are formally similar to those
governing holographic RG flow backgrounds in standard AdS/CFT, which
suggests that they can be studied using appropriately generalized
AdS/CFT techniques.

Thus, we need to study the dynamics of fluctuations about the (MN and
KS) background solutions in the effective five-dimensional bulk
system. To this end, we generalize the
gauge-invariant formalism developed in \cite{Bianchi:2003ug}
to generic multi-scalar systems.
The gauge-invariant formalism
overcomes technical difficulties 
encountered in early work on correlation functions in
holographic RG flows \cite{DeWolfe:2000xi,
  Arutyunov:2000rq, Muck:2001cy}. These difficulties
arose from the fact that the fluctuations of ``active'' scalars (those
with a non-trivial radial background profile) couple to the
fluctuations of the five-dimensional metric already at the linear
level, making it inconsistent to set the metric fluctuations to zero
when studying the scalar fluctuations, or vice versa. Consistent
treatment of the coupled system typically involved, even in the
simplest cases, third-order differential equations containing spurious
gauge redundancies that needed to be painstakingly factored out by hand. 
Happily, fluctuations are manifestly disentangled 
at the linear level in the gauge-invariant
formalism, and their equations of motion are
second-order.\footnote{Gauge-invariant variables for linearized 
  scalar-gravity systems have been studied in cosmology since the
  early 1980s \cite{Bardeen:1980kt,Bardeen:1983qw}. Those variables
  are similar to the ones used in holographic renormalization in
  \cite{DeWolfe:1999cp,Bianchi:2000sm,Bianchi:2001de};
  typical cosmological backgrounds are themselves very similar 
  to the Poincar\'e-sliced AdS 
  domain walls used in the simplest RG flow geometries. The
  connection between the linearized cosmology variables and linearized
  holographic-renormalization variables was studied in
  \cite{Larsen:2004kf}. Also, holography of finite-temperature field theories
  using linearized gauge-invariant variables was initiated in
  \cite{Kovtun:2005ev}.
  Although the applications in this paper are
  worked out at the linear level, our gauge-invariant formalism is
  defined non-linearly.} 
The formalism was applied to the holographic
calculation of three-point functions and scattering amplitudes in
\cite{Muck:2004qg} (see also \cite{Bianchi:2003bd} for earlier work
on three-point functions), and the main ideas for the generalization that we
undertake here were presented in \cite{Muck:2004ih}.

These two ingredients---consistent truncation to a five-dimensional fake
supergravity system, and a general gauge-invariant formalism to describe its
fluctuations---put us in a position to tackle 
the ``fluctuation problem'' in the list above,
and now we proceed to summarize the new results of our work.
We first note that in most cases, one can only expect to solve the
fluctuation equations numerically, but there are notable exceptions
and simplifying limits where analytical solutions are possible. 
Still, with the hope that issues 1 and 2 in the above list will be
solved in the future, we wish to emphasize that numerical integration
of \emph{classical} gauge-invariant ordinary differential equations is
a vastly simpler problem than numerically computing the corresponding
correlators by lattice methods directly in the gauge theory, so the
formulation of equations well-suited for numerical analysis should be
important even in the absence of analytical solutions. Moreover, even
without solving issues 1 and 2, there are physical quantities 
that should not depend on the counterterms and which
can, therefore, be addressed directly using our methods. For example,
glueball masses in the gauge theory correspond to the
existence of normalizable bulk modes 
and do not depend on renormalization details. As an 
example, we calculate the mass spectra of states
for the ${\mathcal N}=1$ gauge theory dual of the MN solution,
up to a caveat discussed in Sec.~\ref{MN}.
We note that mass spectra obtained in the literature 
\cite{Ametller:2003dj,Caceres:2005yx} disagree with ours, which will
be discussed more thorougly in Secs.~\ref{MN} and \ref{outlook}.

For the KS
background, one can only hope to obtain numerical results, so we pose
the problem in terms of gauge-invariant variables and leave numerical
evaluation to future work. However, we \emph{can} analytically study
the scalar fluctuations of the KS system 
(\ie the 7 scalars present in the KS ansatz)
in the singular
Klebanov-Tseytlin (KT) background \cite{Klebanov:2000nc}, which is a
sensible approximation to the ultraviolet region of the KS background. 
In this case, we observe decoupling between the 4-scalar KT system and
the 3 additional scalars that are present in the KS system. 
We will refer to this group of 3 scalars as the \emph{gluino sector},
because it contains the scalar dual to the gluino bilinear $\Tr
\lambda \lambda$. 
The remaining equations are simple enough to
allow for analytical solutions in the 
``moderate UV''
regime considered by
Krasnitz \cite{Krasnitz:2000ir, Krasnitz:2002ct, Krasnitz:2003pj}, in
terms of combinations of Bessel functions and logarithms. 
For the ultraviolet physics of the KS gauge theory, we consider
the Krasnitz approximation to be controlled, since we will be able to check
it numerically by computing the same solutions in KS.
We leave a thorough check for future work and content ourselves 
with comparing our analytical results to numerical solutions of the
full equations in the KT background. The result is that the Krasnitz
approximation seems to work very well, so we expect our analytical
solutions to be useful as guidance in numerical work in the full KS
background. 

Now, let us outline the rest of the paper. 
In Sec.~\ref{adscft}, we start by reviewing briefly the essentials of
holographic renormalization in AdS/CFT,
in particular the dictionary and
renormalization problems. On the way, we will introduce generic fake
supergravity, which is the typical bulk system in holographic
RG flows. 
Then, in Sec.~\ref{sugratrunc}, we perform a consistent
truncation of type IIB supergravity to a 5-dimensional fake
supergravity system. Details of the calculation are given in
appendix A.

Sec.~\ref{gaugeinv} is dedicated to the generalization of the
gauge-invariant analysis of bulk fluctuations \cite{Bianchi:2003ug} to a
generic ``fake supergravity'' system, allowing for an arbitrary number
of scalars and an arbitrary (but invertible) sigma-model metric. 
The principle of reparametrization invariance is the beacon
that guides us to the main result of this section: a
system of (generally coupled) second order differential equations,
which describes the dynamics of the scalar fluctuations about
Poincar\'e-sliced domain walls in a manifestly gauge-invariant
fashion. The presentation of the
gauge-invariant method is intended to be pedagogical: the
principal line of argument is explained in the main text, while 
details are included in the appendices.

In Secs.~\ref{MN} and \ref{KS} we use our techniques to study the MN
and KS systems, respectively. For both, we shall first derive the most
general background solutions including all integration
constants. Although the regular bulk configurations correspond to a
unique choice of integration constants, we find it useful to keep the
constant governing the resolution of the singularity. It determines
the vacuum expectation value of the gluino bilinear, and by tuning it
one is able to consider regimes where analytic solutions to
the fluctuation equations are possible. In the MN system, we discover
a number of normalizable (sub-leading only) modes, which we link to
the mass values of glueball states. In the KS system, we perform the
calculation in the singular KT background and, in addition, apply
the Krasnitz approximation. The resulting solutions are very
similar to the ones Krasnitz found in simpler cases, 
but we refrain from trying to extract
correlators given that the ``dictionary'' and ``renormalization''
problems have yet to be solved.

Finally, Sec.~\ref{outlook} contains conclusions and a discussion of
possible further developments.

\section{Review: correlation functions from AdS/CFT}
\label{adscft}
In this section, we briefly review some essentials of
holographic renormalization, following the three-pronged list of
problems discussed in the introduction. We review how holographic
renormalization systematically 
resolves the "dictionary"
and "renormalization" problems for asymptotically AdS bulk
geometries. These two steps must ultimately be generalized to 
non-asymptotically AdS setups. We leave their general resolution to
future work (initial progress was made in \cite{Aharony:2005zr}), but 
we will comment on some of the specific challenges. 
A general approach to the
third issue in the list, the 
fluctuation problem, will be described in detail in
Sec.~\ref{gaugeinv}. 

Let us start by introducing a generic ``fake supergravity'' system, a
non-linear sigma model of scalar fields with a particular potential, 
coupled to gravity in $d+1$ dimensions (typically, $d=4$). 
Its action is given by\footnote{We follow the curvature conventions of
  MTW and Wald \cite{MTW,Wald:1984rg}, \ie the signature is mostly ``+'', and 
  $R^i{}_{jkl} = \partial_k \Gamma^i{}_{jl} +
  \Gamma^i{}_{km}\Gamma^m{}_{jl} - (k \leftrightarrow l)$. This has the 
  opposite sign of the convention used in \cite{Bianchi:2003ug}.}
\begin{equation}
\label{action5d}
  S = \int \rmd^{d+1}x \sqrt{g} \left[ - \frac14 R +\frac12
  G_{ab}(\phi) \partial_\mu \phi^a \partial^\mu \phi^b +V(\phi) \right]~,
\end{equation}
where the potential, $V(\phi)$, follows from a superpotential, $W(\phi)$, by  
\begin{equation}
\label{Vdef}
  V(\phi) = \frac12 G^{ab} W_a W_b -\frac{d}{d-1} W^2~.
\end{equation}
The matrix $G^{ab}(\phi)$ is the inverse of the sigma model metric
$G_{ab}(\phi)$. Our notation is as in \cite{Bianchi:2003ug}, \ie derivatives of $W$
with respect to fields are indicated as subscripts, as in $W_a=\partial
W/\partial\phi^a$. Moreover, the sigma model metric and its inverse
are used to lower and raise field indices.

Actions of the form \eqref{action5d} arise in a variety of cases, such
as the familiar truncation of $\mathcal{N}=8$, $d=5$ gauged 
supergravity, where several holographic RG flow background solutions
have been found. As we shall see in the next subsection, other
consistent truncations of type IIB supergravity can also give rise to
effective actions of the form \eqref{action5d}. This richness in
applications is our main motivation for considering 
the generic case in detail. 

We are interested in a particular class of solutions of the 
action \eqref{action5d} with $d$-dimensional
Poincar\'e invariance, called Poincar\'e-sliced domain
walls\footnote{As opposed to, for example, the
AdS-sliced domain walls studied in
\cite{LopesCardoso:2001rt,Clark:2004sb,Papadimitriou:2004rz}, 
where the $d$-dimensional boundary can be AdS instead of flat space.} 
or holographic RG flow backgrounds: 
\begin{equation}
\label{background1}
\begin{split}
  \rmd s^2 &= \rmd r^2 +\e{2A(r)} \eta_{ij} \rmd x^i \rmd x^j~,\\
  \phi^a &= \bp^a(r)~. 
\end{split}
\end{equation} 
That is, the radial domain wall in the metric 
is supported by a radial profile of one or several scalars (the
``active'' scalars). If the background fields are determined by the
following coupled first order equations (which is true in all the 
cases we consider):
\begin{equation}
\label{background2}
\begin{split}
  \partial_r A(r) &= -\frac2{d-1} W(\bp)~,\\
  \partial_r \bp^a(r) &= G^{ab} W_b~,
\end{split}
\end{equation}
the domain wall has been shown to be stable, cf.\ \cite{Skenderis:1999mm,Freedman:2003ax}.
These relations do not specify the background
uniquely (integration constants!), but they are sufficient for the general
analysis carried out in this section.
We also note that, although the various backgrounds we study in this
paper are "logarithmically warped" and
not usually given in the form \eqref{background1}, one can
always reach this form by a change of radial variable.

For the system \eqref{background2} to admit an asymptotically AdS
solution, it is necessary and sufficient that the superpotential $W$
possess a local 
extremum with a non-zero value, \ie $W_a(\phi_0)=0$ for all $a$. Then, 
$\phi_0$ is called a fixed point. Without loss of generality, we can
assume that the fixed-point value of $W$ is negative:\footnote{Note
  that an overall sign change of $W$ can be absorbed by changing the
  sign of the coordinate $r$.} 
$W(\phi_0)=-(d-1)/(2L)$, where $L$ is the characteristic AdS length
scale which is often set to $L=1$. 

Let us now briefly review how the issues discussed in
Sec.~\ref{intro} are addressed in AdS/CFT. We start with issue 1,
the dictionary between gauge theory operators and bulk fields. 
The action \eqref{action5d} is manifestly invariant under
field redefinitions---this is indeed the point of the
gauge-invariant formalism that we develop in Sec.~\ref{gaugeinv}---but
this invariance is given up when formulating the one-to-one
correspondence between bulk fields and primary conformal 
operators of the dual gauge theory. As is well known, conformal
invariance imposes that two-point functions of primary conformal
operators of different weights vanish:
\begin{equation}
\label{OOdiag}
  \vev{\mathcal{O}_{\Delta}(x_1) \, \mathcal{O}_{\Delta'}(x_2)} 
  = 0 \quad \text{ for } \Delta \neq \Delta'~.
\end{equation}
In AdS/CFT, this orthogonality property is achieved  for the holographically
calculated correlators by the following choice of field variables.
Let us consider Riemann normal coordinates (RNCs)
\cite{Petrov,Wald:1984rg} in field space around the fixed point $\phi_0$. This 
means that we choose field variables such that $\phi_0=0$, and that
the sigma model connections (defined later in \eqref{Gdef}) vanish at
$\phi_0$. This still leaves us the freedom to impose
$G_{ab}(\phi_0)=\delta_{ab}$ and, by means of a rotation, 
to diagonalize the symmetric matrix of second derivatives of $W$ at
$\phi_0$. With this choice of parametrization, $W$ has the
following expansion around the fixed point,
\begin{equation}
\label{fixedpointW}
  W = -\frac{(d-1)}{2L} - \frac12 \sum\limits_a \lambda_a (\phi^a)^2
  +\cdots~,
\end{equation}
where the ellipsis stands for terms that are at least cubic in $\phi$. 
Using the AdS/CFT dictionary, it is now a simple matter to establish that
the fields $\phi^a$ are dual to primary conformal operators of
dimensions\footnote{Usually the upper sign applies. The lower sign
  can be chosen if $|d/2-\lambda_a|<1$, and is accompanied by
  imposing irregular boundary conditions on the bulk
  fields 
  \cite{Breitenlohner:1982jf,Klebanov:1999tb, Berg:2002hy}.}  
\begin{equation}
\label{confdim}
  \Delta_a = \frac{d}2 \pm \left|\frac{d}2 -\lambda_a \right|~.
\end{equation}
For pure AdS, \eqref{fixedpointW} ensures that
the matrix of holographically calculated
two-point functions is diagonal, that is, equation \eqref{OOdiag}
follows. In general, \eqref{fixedpointW} is not enough to 
unambiguously identify a map between supergravity modes and field theory 
operators. For operators with the same dimension, 
one can usually distinguish them by other quantum numbers
like transformations under $R$-symmetry groups.
(When even that fails, one can try to use 
additional information from the correlators \cite{Berg:2002hy}.) It is fair to
say that the dictionary question is well understood in 
known asymptotically AdS examples.

The second issue, renormalization, is solved in general for 
bulk systems with asymptotically AdS bulk geometries by holographic
renormalization.  The reader is referred to the relevant papers
\cite{Bianchi:2001de, Bianchi:2001kw, 
Berg:2001ty,Martelli:2002sp, Berg:2002hy,Berg:2002ut,
Papadimitriou:2004ap,Papadimitriou:2004rz} and
lecture notes \cite{Skenderis:2002wp} for details. 
Holographic
renormalization systematically removes the divergences by first
formulating the bulk theory on a bulk space with cut-off boundary
located well in the asymptotic UV region. Covariant local counterterms
are added to the action so that removing the cutoff yields a finite
generating functional, and therefore finite correlation functions. 
The result of this procedure is most compactly described in terms of
the notions of \emph{sources} and \emph{responses}, which are
the coefficients in front of the leading and sub-leading series in the
asymptotic expansion of the bulk fields, respectively. That is,
a bulk scalar that is dual to an operator of dimension
$\Delta_a$ (with $+$ sign in \eqref{confdim})
displays asymptotic behavior of the schematic form
\begin{equation}
\label{scalarasymptotic}
  \phi^a(x,r) \approx \e{-(d-\Delta_a)r} \big[\hat{\phi}^a(x) +\cdots \big] + 
  \e{-\Delta_a r} \big[\check{\phi}^a(x) +\cdots \big]~,
\end{equation}
where $\hat{\phi}$ and $\check{\phi}$ denote the source and response
functions, respectively. Up to the addition of scheme-dependent local
terms, which arise from adding finite counterterms to the action, the response
function represents the \emph{exact one-point function} of the dual
operator, \ie the one-point function in the presence of sources. Thus,
in order to calculate higher correlation functions, one needs to solve
the dynamics of bulk fluctuations up to
the required order (\eg quadratic for 3-point functions), extract the response
function from their asymptotic behavior, and differentiate with respect
to the sources. 
It is important to note that although the local 
terms are scheme-dependent, in general they cannot
 just be dropped.
As was stressed in 
\cite{Bianchi:2001de, Bianchi:2001kw}, correlation functions 
computed in conflicting schemes
will in general fail to fulfill the requisite Ward identities.
The most efficient renormalization method to date is that presented in
\cite{Papadimitriou:2004rz}, 
that homes in on the minimal calculation needed for each correlator.

Here is an example of a correlation function calculated in this
fashion: take the GPPZ flow, which is $\mathcal{N}=4$ SYM deformed by
a $\Delta=3$ operator insertion. The result for the two-point
function of this operator for arbitrary boundary momentum
$p$ is \cite{Arutyunov:2000rq,Muck:2001cy,
  Bianchi:2001de}\footnote{The correlators given in these papers
  differ by a scheme-dependent local term.} 
\begin{equation}
\label{GPPZcorr}
  \vev{\mathcal{O}_{\phi}(p)\mathcal{O}_{\phi}(-p)} 
  = \frac{N^2}{2\pi^2} \frac{p^2}2 
  \left[
  \psi\left(\frac32+\frac12\sqrt{1-p^2}\right)
  +\psi\left(\frac32-\frac12\sqrt{1-p^2}\right) 
  -2\psi(1) \right]~,
\end{equation}
where $\psi(z) = \Gamma'(z)/\Gamma(z)$. 
Note that the only scale in this
expression is the asymptotic AdS length scale $L$, which has been set
to unity and is easily restored replacing $p \rightarrow pL$. 
The ultraviolet ($p^2 \to \infty$) asymptotics is that of the
limiting conformal theory, namely
$\vev{\mathcal{O}_{\phi}(p)\mathcal{O}_{\phi}(-p)} \to
p^{2\Delta-4}\log p$. 
The infrared regime (small $p^2$) encodes
the spectrum in a series of poles.
It would be very interesting to understand the connection, if any,
between AdS/CFT correlators of this type and high-energy 
correlators computed by integrability in QCD (see e.g.\
\cite{Belitsky:2004cz}), summing large numbers of certain classes of diagrams.
It is intriguing that those correlators also involve the $\psi$ function.

To end this section, let us outline how we imagine approaching 
the "dictionary" and "renormalization" problems
in the non-asymptotically AdS case. The absence of
a fixed point of $W$, as in the KS and MN solutions,
invalidates some of the strategies discussed above. First, the asymptotic
behavior must probably be studied on a case-by-case basis:
in general, there may not be a basis in which the scalars 
decouple asymptotically (we will encounter examples of
  this later in the MN and KS systems). This means 
that the bulk field/boundary
operator dictionary should be reformulated as finding suitable
source functions for the boundary operators.\footnote{Note that this
  is so even in generic AdS/CFT, where the bulk fields may not
  decouple in holographic RG flow backgrounds in some cases, making it ambiguous to
  speak of the dual bulk field of a specific gauge theory operator.} 
One possibility is to generalize the AdS/CFT definition of the
source function as follows: A 
system of $n$ coupled second-order differential equations for $n$
scalars has $2n$ independent asymptotic solutions, of which $n$
can be regarded as ``leading'' and $n$ as ``subleading''. The $n$
coefficients of the leading solutions can be defined to be sources of
dual gauge theory operators. It might be possible to exploit the
coupling between the bulk fields to describe operator mixing in the
dual gauge theory, but we leave this interesting question for the future.
For the KS case, we follow the strategy adopted so far in the
literature: to consider fields that are mass eigenstates in the
conformal limit. This does not, of course, resolve the dictionary issue
in the nonconformal case.

Second, in AdS/CFT, asymptotically AdS behavior implies that the divergent
terms of the bulk on-shell action can be ordered into a double
expansion in powers of the scalar fields and of the number of
boundary derivatives, with higher order terms being less
divergent. This means that the number of divergent terms is finite,
and that they can be cancelled by adding covariant local counterterms
at a cutoff boundary. At present, we have
no equivalent prescription
for general bulk systems, although the results of
\cite{Aharony:2005zr} are very promising.

A general approach to solving the "fluctuation problem" will be
described in detail in Sec.~\ref{gaugeinv}. But first, we need to derive
the system in which we will study fluctuations. 

\section{Adding boundary momentum to the PT ansatz} 
\label{sugratrunc}
The Papadopoulos-Tseytlin (PT) ansatz for type IIB supergravity
solutions with fluxes \cite{Papadopoulos:2000gj} reduces the problem
of finding 
these particular
 flux solutions to solving the equations of motion deriving
from an effective one-dimensional action subject to a zero-energy
constraint. This suggests that it should be possible to generalize the
PT ansatz in such a way that the scalars that parametrize the
10-dimensional solution depend not only on a ``radial'' variable, but
on all five ``external'' variables. 
This corresponds to allowing for non-zero momentum in the boundary
theory, as required for computing correlators as functions of
momentum. Such a generalization is indeed possible, and we shall
present the result in this section, with the technical details given
in appendix~\ref{construnc}.  
In order not to
unnecessarily overload the notation, we deviate slightly from the
convention used in the appendix by dropping tildes from the 5-dimensional
objects. In the main text, the meaning of the symbols
should be clear from the context, whereas a clearer distinction is
needed for the detailed calculations in the appendices.
The resulting five-dimensional action is of the form
\eqref{action5d}. It will be important for us that in many cases of
interest, including the KS and MN systems, a superpotential $W$
generating the potential $V$ via \eqref{Vdef} is known \cite{Papadopoulos:2000gj}.

The equations of motion of type IIB supergravity in the Einstein frame
are
\begin{align}
\label{einsteintend}
R_{MN} &= \frac12 \partial_M \Phi \partial_N \Phi + \frac12 \e{2\Phi}
\partial_M C \partial_N C + \frac{1}{96} g_s^2 \tF_{MPQRS}
\tF_N^{PQRS} \\
\notag 
&\quad +\frac{g_s}{4} (\e{-\Phi} H_{MPQ} H_N^{PQ} 
+ \e{\Phi} \tF_{MPQ} \tF_N^{PQ}) \\
\notag
&\quad - \frac{g_s}{48} g_{MN} (\e{-\Phi} H_{PQR} H^{PQR} 
+ \e{\Phi} \tF_{PQR} \tF^{PQR})~, \\
\label{dilatoneomtend}
\rmd \star \rmd\Phi &= \e{2 \Phi} \rmd C \wedge \star \rmd C 
- \frac{g_s}{2} \e{-\Phi} H_3 \wedge \star H_3 
+ \frac{g_s}{2} \e{\Phi} \tF_3 \wedge \star \tF_3~,\\
\label{ceomtend}
\rmd(\e{2 \Phi} \star \rmd C) &= -g_s \e\Phi H_3 \wedge \star
\tF_3~,\\
\label{f3eomtend}
\rmd(\e{\Phi} \star \tF_3) &= g_s F_5 \wedge H_3~,\\
\label{h3eomtend} 
\rmd \star (\e{-\Phi} H_3 - C \e{\Phi} \tF_3) &= -g_s F_5 \wedge F_3~,\\
\label{selfdual}
\star \tF_5 &= \tF_5~,
\end{align}
where we have used the notation
\begin{equation}
\notag
  F_3 = \rmd C_2~, \quad H_3 = \rmd B_2~, \quad F_5 = \rmd C_4~,
  \quad \tF_3 = F_3 - C H_3~, \quad \tF_5 = F_5 + B_2 \wedge F_3~. 
\end{equation}
From the last definition follows the Bianchi identity
\begin{equation}
\label{bif5}
\rmd\tF_5 = H_3 \wedge F_3~.
\end{equation}
In the following we set $g_s=1$ and $\alpha'=1$. 

Our ansatz for a consistent truncation follows PT closely, but allows the
scalar fields to depend on all five external coordinates. Thus, we
take
\begin{align}
\notag
\rmd s^2_{10} & = \e{2p-x} \rmd s_5^2 
     + (\e{x+g}+a^2 \e{x-g}) (e_1^2+ e_2^2) 
     + \e{x-g} [e_3^2+e_4^2-2a(e_1 e_3 + e_2 e_4)] +
     \e{-6p-x}e_5^2~,\\
\notag
\rmd s^2_5 & = g_{\mu\nu}\rmd y^\mu \rmd y^\nu~,\\
\notag
H_3 & =  h_2\, e_5 \wedge (e_4 \wedge e_2 + e_3 \wedge e_1) 
 +\rmd y^\mu \wedge \left[\partial_\mu h_1 (e_4\wedge e_3 + e_2\wedge
     e_1)+\right. \\
\notag &\quad \left. 
 + \partial_\mu h_2 (e_4 \wedge e_1 -e_3 \wedge e_2) 
 + \partial_\mu \chi (-e_4 \wedge e_3 + e_2 \wedge e_1)
   \right]~,\\
\notag
F_3 & = P \left\{ e_5 \wedge [e_4 \wedge e_3 + e_2 \wedge e_1 - b  
  (e_4 \wedge e_1 - e_3 \wedge e_2)] + \rmd y^\mu \wedge 
  [\partial_\mu b (e_4 \wedge e_2 + e_3 \wedge e_1)] \right\}~,\\
\notag
\Phi & = \Phi(y)~, \quad C=0~,\\
\label{PTansatz}
\quad \tF_5 &= \mathcal{F}_5 + \star \mathcal{F}_5~,\quad
    \mathcal{F}_5 = K\, e_1 \wedge e_2 \wedge e_3 \wedge e_4 \wedge e_5~,
\end{align}
where $p,x,g,a,b,h_1,h_2,K$ and $\chi$ are functions of the external 
coordinates $y^\mu$, and $P$ is a constant measuring the units of
3-form flux across the 3-cycle of $T^{1,1}$
in the UV. For readers familiar
with the KS background, it may be useful to note that
$\Phi=\chi=0$ in KS, and the other fields have backgrounds as given
later in section~\ref{KSbackground}.

We are using the KS convention for the
forms,\footnote{The relation to the PT and MN conventions can be found
  in footnote 7 of \cite{Papadopoulos:2000gj}.} 
\ie
\begin{equation}
\label{es}
\begin{aligned}
e_1 &= -\sin \theta_1\, \rmd \phi_1~, \quad 
  e_2 = \rmd \theta_1~, &
  e_3 &= \cos \psi \sin \theta_2 \, \rmd \phi_2 - \sin \psi \, \rmd
  \theta_2~,\\
e_4 &= \sin \psi \sin \theta_2 \, \rmd \phi_2 + \cos \psi \, \rmd \theta_2~,
  &
e_5 &= \rmd \psi + \cos \theta_1 \, \rmd \phi_1 + \cos \theta_2 \, \rmd
  \phi_2~.
\end{aligned}
\end{equation}
We note that the first term in the ansatz for $F_3$ is essentially 
$\omega_3 = g^5 \wedge \omega_2$
in KS notation, and as it will turn out that $b\rightarrow 0$ in the UV,
we see that the ansatz for $F_3$ indeed describes a flux piercing the 3-cycle
of $T^{1,1}$ in the UV.
Thus, we have parametrized the 10-d fields of type IIB
supergravity by a 5-d metric, $g_{\mu\nu}$, and a set
of ten scalars, $\Phi,p,x,g,a,b,h_1,h_2,K$ and $\chi$. As in
\cite{Papadopoulos:2000gj}, one finds 
(again, details are relegated to appendix \ref{construnc})
that some of the equations of
motion \eqref{einsteintend}--\eqref{bif5} impose constraints on this
system of fields, namely
\begin{equation} 
\label{kconstraint}
  K = Q + 2 P (h_1 + b h_2)~,
\end{equation}
for a constant $Q$
that sets the AdS scale when $P=0$, and 
\begin{equation}
\label{partialchi}
 \partial_\mu \chi = \frac{(\e{2g} + 2 a^2 + \e{-2g}a^4 - \e{-2g})
   \partial_\mu h_1 + 2 a (1 - \e{-2g}+ a^2 \e{-2g}) 
   \partial_\mu h_2 }{\e{2g}+(1-a^2)^2 \e{-2g}+2a^2}~.
\end{equation}
Although this latter constraint is a 5-d generalization of the
analogous constraint found by PT, 
unlike in their case
it does not only eliminate $\chi$
from the action, but also imposes restrictions on the
possible sets of
independent fields. These restrictions arise from the demand of 
integrability ($\partial_\nu \partial_\mu \chi = \partial_\mu \partial_\nu
\chi$) of the five
first-order partial differential
equations \eqref{partialchi}. Considering the four special cases given in
\cite{Papadopoulos:2000gj}, one finds that \eqref{partialchi} is
satisfied for the singular conifold
(the KT solution, a special case of the KS system), the deformed conifold (KS), and
the wrapped D5-brane (MN), but not in general for fluctuations about the resolved conifold
\cite{PandoZayas:2000sq}.
Further comments on this appear below.
 Thus, we
shall, in the following, consider only the KS and MN systems. 

Imposing the constraints \eqref{kconstraint} and \eqref{partialchi},
the remaining equations of motion can be derived from the 5-dimensional action
\begin{equation} 
\label{5dact}
  S_5 =  \int \rmd^5 y \sqrt{g} \left[ -\frac14 R
  + \frac12 G_{ab}(\phi) \partial_\mu \phi^a \partial^\mu \phi^b 
  + V(\phi) \right]~, 
\end{equation}
with sigma model metric
\begin{multline}
\label{actionPT}
 G_{ab}(\phi) \partial_\mu \phi^a \partial^\mu \phi^b =
 \partial_\mu x \partial^\mu x + \frac12 \partial_\mu g \partial^\mu g 
 + 6 \partial_\mu p \partial^\mu p 
 + \frac12 \e{-2g} \partial_\mu a \partial^\mu a 
 + \frac14 \partial_\mu \Phi \partial^\mu \Phi + \\
 + \frac12 P^2 \e{\Phi-2x} \partial_\mu b \partial^\mu b 
 + \frac{\e{-\Phi -2x}}{\e{2g} + 2a^2+\e{-2g}(1-a^2)^2} \left\{ 
 (1+2 \e{-2g} a^2) \partial_\mu h_1 \partial^\mu h_1
 +\phantom{\frac12}\right. \\
\left. 
 + \frac12 [\e{2g} +2a^2+\e{-2g}(1+a^2)^2] 
 \partial_\mu h_2 \partial^\mu h_2 + 2 a [\e{-2g}(a^2+1)+1]
 \partial_\mu h_1 \partial^\mu h_2 \right\}~,
\end{multline}
and potential
\begin{equation}
\label{VPT}
\begin{split}
 V(\phi) &= -\frac12 \e{2p-2x} [\e{g}+(1+a^2)\e{-g}] 
 + \frac18 \e{-4p-4x}[\e{2g}+(a^2-1)^2\e{-2g}+2a^2] +\\
&\quad
 + \frac14 a^2 \e{-2g+8p} + \frac18 P^2 \e{\Phi-2x+8p}[\e{2g} +
 \e{-2g} (a^2 - 2ab +1)^2 + 2 (a-b)^2] +\\
&\quad
  + \frac14 \e{-\Phi-2x+8p}h_2^2 +\frac18 \e{8p-4x} [Q+2P(h_1 + b
 h_2)]^2~. 
\end{split}
\end{equation}
As emphasized above, we must remember that integrability of \eqref{partialchi}
effectively restricts us to the KS and MN systems. With this restriction, 
the system \eqref{5dact} with kinetic terms \eqref{actionPT} and
potential \eqref{VPT} represents a consistent truncation of type IIB
supergravity. Moreover, the superpotential exists and is known in both
cases. We show the consistency of the truncation in appendix~\ref{construnc}.

It is an interesting and (as far as we know) open question how
the truncation of (\ref{actionPT}) and (\ref{VPT}) to the KS system can be made manifestly 
supersymmetric. As explained in \cite{Papadopoulos:2000gj} (and as we review in section 
\ref{KSbackground}), this truncation introduces one more constraint on the 
system of ten scalars, cf.\ \eqref{KS:agrel}. Together with \eqref{kconstraint} and \eqref{partialchi} this leaves seven independent scalars. To write down a manifestly supersymmetric effective action for them 
might require a generalization of the ansatz \eqref {PTansatz}.
To some readers
it may seem discouraging that the number of real
scalars in the KS system is odd, as
four-dimensional intuition would indicate
that the superpotential in a supersymmetric theory 
ought to be a holomorphic function in complex field variables. However, this intuition does not apply in
 odd dimensions. In ${\cal N} = 2$ theories in five 
dimensions, the vector multiplet only contains a real scalar, so it is conceivable that a potential of the form \eqref{Vdef} could
be appropriate for a supersymmetric theory, even if the derivatives
are with respect to real scalars. A similar situation arises in
${\cal N}=2$ supersymmetric theories in three dimensions (for example, those obtained from Calabi-Yau fourfold compactifications of M-theory).
There, the potential is given by an expression similar to
\eqref{Vdef} but involving two functions, one depending on the real
scalars of the vector multiplets and the other being a holomorphic
function depending on the remaining scalars
\cite{Haack:2001jz,Berg:2002es,deWit:2003ja,Hohm:2004rc}. 

It is also interesting to ask whether it is possible (at least in certain cases)
to rewrite the general form of the potential 
in a five-dimensional gauged ${\cal N}=2$ supergravity, 
given in \cite{Gunaydin:1984ak,Gunaydin:1999zx,Ceresole:2000jd,Bergshoeff:2004kh}, 
in a form that resembles \eqref{Vdef}. This question (and its generalization to 
${\cal N}=4$) was investigated in \cite{Celi:2004st,Zagermann:2004ac}.

We would also like to connect the discussion above to the work on the
Klebanov-Strassler Goldstone mode found in
\cite{Gubser:2004qj,Gubser:2004tf}
(this mode was predicted already in
\cite{Aharony:2000pp}).\footnote{See also \cite{Butti:2004pk}.} 
Since we argued that the analysis of
fluctuations about the resolved conifold does
not extend from the one-dimensional to the five-dimensional
truncation in any obvious way, one needs to generalize the ansatz to satisfy the
integrability constraint if one wants to study the dynamics of the
Goldstone mode multiplet. We have no reason to doubt that this is possible, but
we will not pursue it further here. 

\section{Real fluctuations in fake supergravity}
\label{gaugeinv}
\subsection{The sigma-model covariant field expansion}
\label{covnot}

It is our aim to study the dynamics of the fake
supergravity system \eqref{action5d}, \eqref{Vdef} on some
known backgrounds of the form \eqref{background1}, \eqref{background2}. 
In this section, we shall 
expand the fields around the background, exploiting the geometric
nature of the physical variables to 
formulate the fluctuation dynamics gauge-invariantly.
Our
arguments will closely follow the original development of the
gauge-invariant method for a single scalar in \cite{Bianchi:2003ug},
but important new ingredients will be needed in order to account for
the general sigma model. 

As is well known in gravity, reparametrization invariance of 
spacetime comes at the price of dragging along redundant metric
variables together with the physical degrees of freedom. 
One attempts to reduce redundancy by gauge fixing, but as
mentioned in the introduction, such an 
approach causes problems for fluctuations in holographic
RG flows, due to the coupling between metric and scalar
fluctuations. Thus, following \cite{Bianchi:2003ug}, we shall start
from a clean slate keeping all metric degrees of freedom and describe
in the next subsection how to isolate the physical ones. 

The geometry of the sigma-model target space is characterized by
the metric $G_{ab}(\phi)$, which we assume to be invertible,
the inverse being denoted $G^{ab}(\phi)$. 
One can define the sigma-model connection
\begin{equation}
\label{Gdef}
  \G{a}{bc} = \frac12 G^{ad} \left( \partial_c G_{db} 
  + \partial_b G_{dc} - \partial_d G_{bc} \right)~,
\end{equation}
and its curvature tensor
\begin{equation}
\label{Rdef}
  \R^a_{\;\;bcd} = \partial_c \G{a}{bd} - \partial_d \G{a}{bc} +
  \G{a}{ce} \G{e}{bd} - \G{a}{de}\G{e}{bc}~.
\end{equation}
We also define the covariant field derivative as usual, \eg
\begin{equation}
\label{covderdef}
  D_b A_a \equiv A_{a|b} \equiv \partial_b A_a -\G{c}{ab} A_c~.
\end{equation}
All indices after a bar "$|$" are intended as covariant field derivatives
according to \eqref{covderdef}. Moreover, field indices are lowered
and raised with $G_{ab}$ and $G^{ab}$, respectively.

\begin{figure}[th]
\begin{center}
\includegraphics[bb= 148 614 291 718, clip]{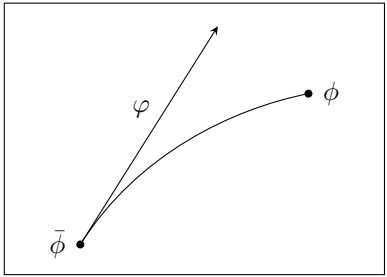}
\caption{\label{fig1}Illustration of the exponential map.}
\end{center}
\end{figure}

Armed with this notation, it is straightforward to expand
the scalar fields in a
sigma-model covariant fashion. The naive 
ansatz $\phi^a = \bp^a +\vp^a$, introducing $\vp^a$ simply as the
coordinate difference between the points $\phi$
and $\bp$ in field space, leads to non-covariant expressions at
quadratic and higher orders, because these $\vp^a$ do not form a vector
in (tangent) field space. In other words, the coordinate difference is
not a geometric object. However, it is well known that a covariant
expansion is provided by the \emph{exponential map} \cite{Wald:1984rg,Petrov},
\begin{equation}
\label{expmap}
  \phi^a = \exp_{\bp}(\vp)^a 
  \equiv \bp^a +\vp^a -\frac12 \G{a}{bc} \vp^b \vp^c +\cdots~,
\end{equation}
where the higher order terms have been omitted, and the connection
$\G{a}{bc}$ is evaluated at $\bp$. Geometrically, $\vp$
represents the tangent vector at $\bp$ of the geodesic curve
connecting the points $\bp$ and $\phi$, and its length is equal to the
geodesic distance between $\bp$ and $\phi$; see Fig.~\ref{fig1}.

It is also a standard result that the
components $\vp^a$ coincide with the Riemann normal coordinates (RNCs)
(with origin at $\bp$) of the point $\phi$ (see, \eg \cite{Petrov}). 
This fact can be used to simplify the task of writing equations in a
manifestly sigma-model covariant form. Namely, given a 
background point $\bp$, we can use RNCs to describe some
neighborhood of it and then employ the following properties at the
origin of the RNC system,
\begin{equation}
\label{RNCprop}
  \G{a}{bc} =0~,\qquad 
  \R^a_{\;\;bcd} = \partial_c \G{a}{bd} - \partial_d \G{a}{bc}~, 
\end{equation}
in order to express everything in terms of tensors.
Because the background fields depend on $r$, we
must be careful to use \eqref{RNCprop} only outside $r$-derivatives,
but the simplifications are still significant.

Finally, let us also define a ``background-covariant'' derivative
$D_r$, which acts on sigma-model tensors as, \eg
\begin{equation}
\label{Drdef}
  D_r \vp^a = \partial_r \vp^a + \G{a}{bc} W^b \vp^c~.
\end{equation}
If a tensor $A_a$ depends on $r$ only implicitly through its
background dependence, then we find the identity 
\begin{equation}
\label{DronW}
  D_r A_a(\bp) = W^b(\bp) D_b A_a(\bp)~.
\end{equation} 
The background-covariant derivative
$D_r$ will be important in our presentation of the field
equations in Sec.~\ref{fieldeq}.

\subsection{Gauge transformations and invariants}
\label{invars}
The form of the background
solution \eqref{background1} lends itself well
to the ADM (or time-slicing)
formalism for parametrizing the metric degrees of freedom
\cite{MTW,Wald:1984rg}. Instead of slicing in time, we shall write a
general bulk metric in the radially-sliced form 
\begin{equation}
\label{ADMmetric}
  \rmd s^2 = (n^2+ n_i n^i) \rmd r^2 + 2 n_i \rmd r \rmd x^i 
   + g_{ij} \rmd x^i \rmd x^j
\end{equation}
where $g_{ij}$ is the induced metric on the hypersurfaces
of constant $r$,
and $n$ and $n^i$ are the lapse function and shift vector,
respectively. It will be important to us that
the objects $n$, $n^i$ and $g_{ij}$ transform properly under
coordinate transformations of the radial-slice hypersurfaces.  
Details concerning the geometry of hypersurfaces are reviewed in
appendix~\ref{geom}. Again, we will not
put tildes on the bulk quantities in the main text, as the meaning
of the symbols should be clear from the context. In contrast, tildes
are used in the appendices in order to clearly distinguish bulk and
hypersurface quantities. 
 
We can now expand the radially-sliced metric around the background 
configuration:
\begin{equation}
\label{expandmetric}
\begin{split}
  g_{ij} &= \e{2A(r)} \left( \eta_{ij} + h_{ij} \right)~,\\
  n_i    &= \nu_i~,\\
  n      &= 1+ \nu~,
\end{split}
\end{equation}
where $h_{ij}$, $\nu_i$ and $\nu$ denote small
fluctuations. Henceforth, we shall adopt the notation that the indices
of metric fluctuations, as well as of derivatives $\partial_i$, 
are raised and lowered using the flat (Minkowski/Euclidean) metric,
$\eta_{ij}$.

Now let us turn to the question of isolating the physical degrees of
freedom from the set of fluctuations $\{h^i_j, \nu^i, \nu, \vp^a\}$
introduced so far. 
In the earlier AdS/CFT literature one usually
removed the redundancy following from diffeomorphism
invariance by partial gauge fixing, \ie by placing conditions on certain
components of the metric, such as $n\equiv 1$, $n^i \equiv 0$. 
And indeed, it is always possible to perform a change of coordinates
which transforms the metric into a form that satisfies the gauge
conditions.

Alas,
as discussed in the introduction, partial gauge fixing
can create problems in coupled systems. Instead, 
we will obtain the equations of
motion in gauge-invariant form. 
Let us start by considering
the effect of diffeomorphisms on the fluctuation fields. We consider 
a diffeomorphism of the form
\begin{equation}
\label{diffeo}
  x^\mu= \exp_{x'}[\xi(x')]^\mu = {x'}^\mu+ \xi^\mu(x') -\frac12
  \Gamma^{\mu}_{\nu\rho}(x') \xi^\nu(x')\xi^\rho(x') + \cdots~,
\end{equation}
where $\xi$ is infinitesimal. Notice that we found it convenient to apply
the diffeomorphism inversely, \ie we have expressed the old coordinates
$x^\mu$ in terms of the new coordinates ${x'}^\mu$. The use of the
exponential map  implies that also the transformation laws for the fields
can be written covariantly
(the functions
$\xi^\mu(x')$ are thought of 
as the components of a vector field).
For example, a scalar field transforms as 
\begin{equation}
\label{phitrafo}
 \delta \phi = \xi^\mu \partial_\mu \phi +\frac12 \xi^\mu \xi^\nu
 \nabla_\mu \partial_\nu \phi+\cdots~,
\end{equation}
whereas a covariant tensor of rank two transforms as 
\begin{equation}
\label{Etrafo}
\begin{split}
 \delta E_{\mu\nu} &= \xi^\lambda \nabla_\lambda E_{\mu\nu} 
 +(\nabla_\mu \xi^\lambda) (E_{\lambda\nu} 
 +\xi^\rho \nabla_\rho E_{\lambda\nu}) 
 +(\nabla_\nu \xi^\lambda) (E_{\mu\lambda} 
 +\xi^\rho \nabla_\rho E_{\mu\lambda})
 +\\  
 &\quad + (\nabla_\mu \xi^\lambda)(\nabla_\nu \xi^\rho)E_{\lambda\rho} 
 +\frac12 \xi^\rho\xi^\lambda (\nabla_\rho \nabla_\lambda E_{\mu\nu}
 -R^\sigma{}_{\lambda\mu\rho} E_{\sigma\nu} 
 -R^\sigma{}_{\lambda\nu\rho} E_{\mu\sigma}) +\\
 &\quad +\cdots~.
\end{split}
\end{equation}
For the metric tensor $g_{\mu\nu}$, \eqref{Etrafo} simplifies to 
\begin{equation}
\label{gtrafo}
  \delta g_{\mu\nu} = \nabla_\mu \xi_\nu + \nabla_\nu \xi_\mu 
   +(\nabla_\mu \xi^\lambda)(\nabla_\nu \xi_\lambda) 
   - R_{\mu\lambda\nu\rho} \xi^\lambda \xi^\rho +\cdots~.
\end{equation}
Eqs.\ \eqref{phitrafo} and \eqref{Etrafo} are most easily derived using
RNCs around $x'$ and using \eqref{RNCprop}. The second
order terms in $\xi$ have been included here in order to illustrate
the covariance of the transformation laws. For our purposes, the
linear terms will suffice.

Splitting the fake supergravity fields into background and
fluctuations, as defined in \eqref{expandmetric} and \eqref{expmap}, the
transformations \eqref{phitrafo} and \eqref{gtrafo} become gauge transformations for the fluctuations, to lowest
order:
\begin{equation}
\label{trafos}
\begin{split}
 \delta \varphi^a &= W^a \xi^r +\Of~,\\
 \delta \nu &= \partial_r \xi^r +\Of~,\\
 \delta \nu^i &= \partial^i \xi^r + \e{2A} \partial_r \xi^i+\Of~,\\
 \delta h^i_j &= \partial_j \xi^i +\partial^i (\eta_{jk} \xi^k) 
 -\frac{4}{d-1} W \delta^i_j \xi^r+\Of~.
\end{split}
\end{equation}
By $\Ofn{n}$ we mean terms of order $n$ in the fluctuations
$\{\varphi^a, h_{ij}, \nu_i, \nu\}$. 
Furthermore, let us decompose $h^i_j$ as follows,
\begin{equation}
\label{hsplit}
 h^i_j = \htt^i_j 
  + \partial^i \epsilon_j +\partial_j \epsilon^i
  + \frac{\partial^i \partial_j}{\Box} H + \frac1{d-1} \delta^i_j h~,
\end{equation}
where $\htt^i_j$ denotes the traceless transverse part, and $\epsilon^i$
is a transverse vector. It is straightforward to obtain from
\eqref{trafos}
\begin{equation}
\label{h_trafos}
\begin{split}
  \delta \htt^i_j &=\Of~,\\
  \delta \epsilon^i &= \Pi^i_j \xi^j+\Of~,\\ 
  \delta H &= 2 \partial_i \xi^i+\Of~,\\
  \delta h &= -4 W \xi^r+\Of~.
\end{split}
\end{equation}
The symbol $\Pi^i_j$ denotes the transverse projector,
\begin{equation}
\label{Pproj}
  \Pi^i_j = \delta^i_j - \frac1\Box \partial^i \partial_j~.
\end{equation}

The main idea of our approach is to construct gauge-invariant
combinations from the fields $\{\htt^i_j, \epsilon^i, h, H, \nu,
\nu^i, \vp^a\}$. Using the transformation laws \eqref{trafos} and
\eqref{h_trafos}, this is straightforward, and to lowest order, one
finds the gauge-invariant fields\footnote{The choice of
  gauge-invariant variables is not unique, of course, as any
  combination of them will be gauge-invariant as well.}
\begin{align}
\label{A}
 \mfa^a &= \varphi^a +W^a \frac{h}{4W}+\Ofn{2}~,\\
\label{B}
 \mfb &= \nu + \partial_r \left( \frac{h}{4W} \right)+\Ofn{2}~,\\
\label{C}
 \mfc &= \e{-2A} \partial_i \nu^i + \e{-2A} \Box \frac{h}{4W} 
 -\frac12 \partial_r H+\Ofn{2}~,\\ 
\label{D}
 \mfd^i &= \e{-2A} \Pi^i_j \nu^j - \partial_r \epsilon^i+\Ofn{2}~,\\
\label{E} 
  \mfe^i_j &=\htt^i_j+\Ofn{2}~.
\end{align}
The variables $\mfc$ and $\mfd^i$ both arise from $\delta
\nu^i$, which has been split into its longitudinal and transverse
parts. We chose the $\mathfrak{Fraktur}$ typeface for the gauge invariant
variables in order to avoid confusion with the field indices, and still
keep notational similarity with \cite{Bianchi:2003ug}. Notice that
$\mfc$ and $\mfd^i$ have been rescaled with respect to
\cite{Bianchi:2003ug} for later convenience.  

Although we have carried out the construction of gauge-invariant
variables only to lowest order, and this is all
we will need here, it is necessary for consistency that the preceding
analysis can be extended to higher orders, in principle. In this
context it becomes clear that the geometric nature of the field
expansions, as expressed by the exponential map, is a crucial
ingredient of the method. 

Finally, let us prepare the ground for the arguments of the next
subsection, where we shall analyze the implications of
gauge-invariance on the equations of motion. 
Let us introduce some more compact notation. Consider the set of
gauge-invariant fields, $I=\{\mfa^a,\mfb,\mfc,\mfd^i,\mfe^i_j\}$. From
the definitions \eqref{A}--\eqref{E} we see that there is a one-to-one
correspondence between $I$ and a sub-set of the fluctuation fields,
$Y=\{\varphi^a, \nu, \nu^i, \htt^i_j\}$. We also collect the remaining
fluctuation variables into a set, $X=\{h,H,\epsilon^i\}$. 
Henceforth, the symbols $I$, $X$ and $Y$ shall be used also to denote
members of the corresponding sets. 

One can better understand the correspondence between $I$ and $Y$ by 
noting that \eqref{A}--\eqref{E} can be re-written as  
\begin{equation}
\label{Yeq}
  Y = I + y(X) +\Ofn{2}~,
\end{equation}
where $y$ is a linear functional of the fields $X$. Going to quadratic
order in the fluctuations, one would find  
\begin{equation}
\label{Yeq2}
  Y = I + y(X) + \alpha(X,X) +\beta(X,I) +\Ofn{3}~,
\end{equation}
where $\alpha$ and $\beta$ are bi-linear in their arguments. Terms
of the form $\gamma(I,I)$ do not appear, as they can be absorbed into
$I$.

We interpret the gauge-invariant variables $I$ as the physical degrees of
freedom, whereas the $(d+1)$ variables $X$ represent the redundant metric
variables. This can be seen by observing that one can solve the
transformation laws \eqref{h_trafos} for the generators
$\xi^\mu$, which yields equations of the form
\begin{equation}
\label{xi_delX}
  \xi^\mu = z^\mu(\delta X) +\Ofn{2} = \delta z^\mu(X)
  +\Ofn{2}~,
\end{equation}
with $z^\mu$ being a linear functional.

\subsection{Einstein's equations and gauge invariance}
\label{Einstein_inv}
It is our aim to cast the equations of motion into an explicitly 
gauge-invariant form. This means that the final equations
should contain only the variables $I$ and make no reference to $X$ and
$Y$. Reparametrization invariance suggests that this should be
possible, and we shall establish the precise details in this
subsection. 

Let us consider Einstein's equations, symbolically
written as
\begin{equation}
\label{Einstein}
  E_{\mu\nu} = 0~,
\end{equation}
but it is clear that the arguments given below hold also for the
equations of motion for the scalar fields. To start, let us expand the
left hand side of \eqref{Einstein} around the background solution,
which yields, symbolically,
\begin{equation}
\label{E_expand}
  E_{\mu\nu} = E^{(1)1}_{\mu\nu}(X) + E^{(1)2}_{\mu\nu}(Y)
              + E^{(2)1}_{\mu\nu}(X,X) + E^{(2)2}_{\mu\nu}(X,Y) 
	      + E^{(2)3}_{\mu\nu}(Y,Y) +\Ofn{3}~.
\end{equation}
Here, $E^{(1)}$ and $E^{(2)}$ denote linear and bilinear terms,
respectively. The background equations are satisfied identically. 
Substituting $I$ for $Y$ using \eqref{Yeq2} yields
\begin{equation}
\label{E_2}
  E_{\mu\nu} = \tE^{(1)1}_{\mu\nu}(X) + E^{(1)2}_{\mu\nu}(I)
              + \tE^{(2)1}_{\mu\nu}(X,X) + \tE^{(2)2}_{\mu\nu}(X,I) 
	      + E^{(2)3}_{\mu\nu}(I,I) +\Ofn{3}~.
\end{equation}
Notice that the functionals $E^{(1)2}$ and $E^{(2)3}$ are 
unchanged ($Y$ is just replaced by $I$), whereas the others are
modified by the $X$-dependent terms of \eqref{Yeq2}, which we indicate
by adorning them with a tilde. 
For example, $\tE^{(2)2}$ receives contributions from $E^{(2)2}$,
$E^{(2)3}$ and $E^{(1)2}$. 

In order to simplify \eqref{E_2}, we consider its
transformation under the diffeomorphism \eqref{diffeo}. On the one
hand, from the general transformation law of tensors \eqref{Etrafo} 
we find, using also \eqref{xi_delX}, that it should transform as
\begin{equation}
\label{E_trafoX}
  \delta E_{\mu\nu} = [\partial_\mu \delta z^\lambda(X)] E_{\lambda\nu}  
                    +[\partial_\nu \delta z^\lambda(X)] E_{\mu\lambda} 
		    +\delta z^\lambda(X) \partial_\lambda E_{\mu\nu}
                    +\Ofn{3}~.
\end{equation}
On the other hand, the variation of \eqref{E_2} is
\begin{equation}
\label{deltaE_2}
  \delta E_{\mu\nu} = \tE^{(1)1}_{\mu\nu}(\delta X) 
              + 2 \tE^{(2)1}_{\mu\nu}(\delta X,X) 
              + \tE^{(2)2}_{\mu\nu}(\delta X,I) 
	      +\Ofn{3}~.
\end{equation}
Let us compare \eqref{E_trafoX} and \eqref{deltaE_2} order by order. 
The absence of first-order terms on the right hand side of
\eqref{E_trafoX} implies that
\begin{equation}
  \tE^{(1)1}_{\mu\nu}(X) = 0~.
\end{equation}
It can easily be checked that this is indeed the case.
Then, substituting $E_{\mu\nu}=E^{(1)2}_{\mu\nu}(I)+\Ofn{2}$ 
into the right hand side of \eqref{E_trafoX} yields
\begin{equation}
\label{E_trafo2}
  \delta E_{\mu\nu} = \delta \left\{ 
                    [\partial_\mu z^\lambda(X)] E^{(1)2}_{\lambda\nu}(I)
		   +[\partial_\nu z^\lambda(X)] E^{(1)2}_{\mu\lambda}(I) 
		   +z^\lambda(X) \partial_\lambda E^{(1)2}_{\mu\nu}(I)
                    \right\} +\Ofn{3}~.
\end{equation}
Comparing \eqref{E_trafo2} with the second order terms of
\eqref{deltaE_2}, we obtain
\begin{equation}
\begin{split}
 \tE^{(2)1}_{\mu\nu}(X,X) &=0~,\\
 \tE^{(2)2}_{\mu\nu}(X,I) &=   
                  [\partial_\mu z^\lambda(X)] E^{(1)2}_{\lambda\nu}(I)
		 +[\partial_\nu z^\lambda(X)] E^{(1)2}_{\mu\lambda}(I) 
                 +z^\lambda(X) \partial_\lambda E^{(1)2}_{\mu\nu}(I)~.
\end{split}
\end{equation}
Hence, we find that a simple expansion of Einstein's equations yields
gauge-dependent second-order terms, but they contain the
(gauge-independent) first order equation, and so can 
consistently be dropped. Happily, we arrive at the following equation,
which involves only $I$:
\begin{equation}
\label{eqn}
  E^{(1)2}_{\mu\nu}(I) + E^{(2)3}_{\mu\nu}(I,I) +\Ofn{3} =0~.
\end{equation}
The argument generalizes recursively to higher orders. One will find that the
gauge-dependent terms of any given order can be consistently dropped,
because they contain the equation of motion of lower orders. 

Eq.~\eqref{eqn} and its higher-order generalizations 
are obtained using the following recipe: 
\begin{center}
\fbox{\parbox{0.8\textwidth}{\emph{Expand the
  equations of motion to the desired order dropping the fields $X$ and
  replacing every field $Y$ by its gauge-invariant counterpart $I$.} }}
\end{center} 
This rule is summarized by the following substitutions,
\begin{equation}
\label{field_subs}
  \varphi^a \to \mfa^a~,\quad
          \nu       \to \mfb~,\quad
  \e{-2A} \nu^i     \to \mfd^i + \frac{\partial^i}{\Box} \mfc~,\quad
          h^i_j     \to \mfe^i_j~.
\end{equation}
Since $\mfe^i_j$ is traceless and transverse, the calculational
simplifications arising from \eqref{field_subs} are
considerable. For the reader's reference, the expressions that
result from \eqref{field_subs} for some geometric
objects are listed at the end of appendix~\ref{eomint}.

Let us conclude with the remark that, although the rules
\eqref{field_subs} can be interpreted as the gauge choice $X=0$, the
equations we found are truly gauge invariant.

\subsection{Equations of motion}
\label{fieldeq}
In this section, we shall put the above preliminaries into
practice.
The equations of motion that follow from the action \eqref{action5d}
are
\begin{equation}
\label{scalar_eom}
  \nabla^2 \phi^a +\G{a}{bc}\, g^{\mu\nu} 
  (\partial_\mu \phi^b)(\partial_\nu \phi^c) -V^a = 0
\end{equation}
for the scalar fields, and Einstein's equations
\begin{equation}
\label{Einstein_eom}
  E_{\mu\nu} = -R_{\mu\nu} 
  + 2 G_{ab} (\partial_\mu \phi^a)(\partial_\nu \phi^b) 
  + \frac{4}{d-1} g_{\mu\nu} V =0~.
\end{equation}
Notice that we use the opposite sign convention for
the curvature with respect to \cite{Bianchi:2003ug,
  Muck:2004ih}.

We are interested in the physical, gauge-invariant content of
\eqref{scalar_eom} and \eqref{Einstein_eom} to quadratic order in the 
fluctuations around an RG flow background of the form
\eqref{background1}, \eqref{background2}.
As we saw in the last section, the physical content is obtained by 
expanding the fields according to \eqref{expandmetric} and
\eqref{expmap} and then applying the substitution rules
\eqref{field_subs}. Since we defined the expansion
\eqref{expmap} geometrically, we will
obtain sigma-model covariant expressions. To carry out this
calculation in practice, it is easiest to use RNCs at a given point in
field space, so that one can use the relations \eqref{RNCprop} outside
$r$-derivatives.

In the following, we shall present the linearized
equations of motion, and indicate higher order terms as sources, the 
relevant quadratic terms of which are listed in appendix~\ref{sources}. 
For intermediate steps we refer the reader to appendix \ref{eomint}.
Let us start with the equation of motion for the scalar fields
\eqref{scalar_eom}, which gives rise to the following fluctuation
equation,
\begin{equation}
\label{eqa1}
\begin{split}
  \left[D_r^2 -\frac{2d}{d-1}W D_r +\e{-2A} \Box \right] \mfa^a
  -\left( V^a_{\;\;|c} - \R^a_{\;\;bcd} W^b W^d \right) \mfa^c -&\\
  - W^a \left( \mfc +\partial_r \mfb \right) - 2 V^a \mfb &=J^a~.
\end{split}
\end{equation}
Note the appearance of the field-space curvature tensor in the potential
term.

Second, the normal component of Einstein's equations\footnote{More
  precisely, it is the equation obtained by multiplying
  \eqref{Einstein_eom} by $N^\mu N^\nu - g^{ij} X^\mu_i
  X^\nu_j$.} 
gives rise to
\begin{equation}
\label{eqc}
  - 4 W \mfc + 4 W_a (D_r \mfa^a) - 4 V_a \mfa^a - 8V\mfb =J~.
\end{equation}

Third, the mixed components of \eqref{Einstein_eom} yield
\begin{equation}
\label{eqbd}
  -\frac12 \Box \mfd_i -2 W \partial_i \mfb -2 W_a \partial_i
   \mfa^a =J_i~.
\end{equation}

The appearance of the fields $\mfa^a$, $\mfb$, $\mfc$ and $\mfd^i$ on
the left hand sides of \eqref{eqa1}--\eqref{eqbd} seems to indicate
the coupling between the fluctuations of active scalars (non-zero
$W_a$) to those of the metric, which is familiar from the AdS/CFT
calculation of two-point functions in the literature.  
However, the gauge-invariant formalism  resolves this
issue, because \eqref{eqc} and \eqref{eqbd} can be
solved algebraically (in momentum space) for the metric fluctuations
$\mfb$, $\mfc$ and $\mfd^i$, so that the coupling of metric and scalar
fluctuations at linear order is
completely disentangled. One easily obtains
\begin{align}
\label{b_sol} 
  \mfb &= -\frac1{W} W_a \mfa^a -\frac1{2W} \frac{\partial_i}{\Box}
  J^i~,\\
\label{c_sol}
  \mfc &= \frac{W_a}{W} \left(\delta^a_b D_r -W^a_{\;\;|b}
  +\frac{W^aW_b}{W} \right) \mfa^b- \\  
\notag
  &\quad 
  -\frac1{4W} J + \frac12\left( \frac{W_a W^a}{W^2} -\frac{2d}{d-1} \right)
  \frac{\partial^i}{\Box} J_i~,\\ 
\label{d_sol}
  \mfd_i &= -\frac{2}{\Box} \Pi^j_i J_j~.
\end{align}

We proceed by substituting \eqref{b_sol} and \eqref{c_sol} into
\eqref{eqa1}, using also the identities
\begin{equation}
\label{Vident}
\begin{split}
  V^a &= W^{a|b} W_b -\frac{2d}{d-1} W W^a~,\\
  V^a_{\;\;|c} &= D_r W^a_{\;\;|c} +\R^a_{\;\;bcd} W^b W^d +
   W^{a|b} W_{b|c} -\frac{2d}{d-1} \left(W^aW_c +W W^a_{\;\;|c} \right)~,
\end{split}
\end{equation}
which follow from \eqref{Vdef} and \eqref{background2}, and we end up with
the second-order differential equation
\begin{equation}
\label{eqphi}
  \left[ \left( \delta^a_b D_r +W^a_{\;\;|b} -\frac{W^aW_b}{W}
         -\frac{2d}{d-1} W \delta^a_b \right) 
         \left( \delta^b_c D_r -W^b_{\;\;|c} +\frac{W^bW_c}{W} \right)
  +\delta^a_c \e{-2A} \Box \right] \mfa^c =\tilde{J}^a~,
\end{equation}
where the source term $\tilde{J}^a$ is related to the sources $J^a$,
$J$ and $J_i$ by
\begin{equation}
\label{tildeJ}
  \tilde{J}^a = J^a -\frac{W^a}{4W} J - \frac12 \left( \delta^a_b D_r
  +W^a{}_{|b} -\frac{W^aW_b}{W} -\frac{2d}{d-1} W \delta^a_b \right)
  \left(\frac{W^b}{W} \frac{\partial^i}{\Box} J_i \right)~.
\end{equation}
Eq.~\eqref{d_sol} implies that we can drop $\mfd^i$ in the source
terms (to quadratic order). 
Eq.~\eqref{eqphi} is the main result of the gauge-invariant
approach and governs the dynamics of scalar fluctuations around
generic Poincar\'e-sliced domain wall backgrounds. Being a system of
second order differential equations, one can use the standard Green's
function method to treat the interactions perturbatively. 

A feature that is evident from the linearized version of \eqref{eqphi}
is the existence of a background mode in the fluctuations. It is
independent of the boundary variables $x^i$, and is simply given by 
\begin{equation}
\label{backgroundmode}
  \mfa^a = \alpha \frac{W^a}{W}~,
\end{equation}
where $\alpha$ is an infinitesimal constant. 
In standard holographic renormalization, one can use the background
mode \eqref{backgroundmode} to establish the existence of finite
sources (CFT deformations) and vacuum expectation values in the dual
field theory. 
Asymptotically each component of the fluctuation vector is dual to a conformal primary
operator (as explained in Sec.~\ref{adscft});
a component of $W^a/W$ that behaves asymptotically as the leading term
of the general solution of \eqref{eqphi} is interpreted as a background source
deforming the CFT action by the corresponding dual operator,
while a background mode component that behaves asymptotically as the
sub-leading term of the general solution represents a vacuum expectation value of the dual
operator. We believe that a statement of this kind can be made also in
the general non-asymptotically AdS case, and we shall present an
example for the MN system in Sec.~\ref{MN}. 

Let us also consider the tangential components of \eqref{Einstein_eom}. 
Because of the Bianchi identity, their trace and divergence are
implied by \eqref{eqa1}, \eqref{eqc} and \eqref{eqbd}, which is easily
checked at linear order. Thus, we can use the traceless transverse
projector, 
\begin{equation}
\label{TTproj_def}
  \Pi^{ik}_{jl} = \frac12 \left(\Pi^{ik} \Pi_{jl}  + \Pi^i_l \Pi^k_j
  \right) - \frac1{d-1} \Pi^i_j \Pi^k_l~,
\end{equation}
in order to obtain the independent components. This yields
\begin{equation}
\label{eqe}
  \left( \partial_r^2 -\frac{2d}{d-1} W \partial_r +\e{-2A} \Box
  \right) \mfe^i_j = J^i_j~.
\end{equation}
As expected, the physical fluctuations of the metric satisfy the
equation of motion of a massless scalar field.

\section{The Maldacena-Nu\~nez system}
\label{MN}
\subsection{Review of the background solution}
\label{MN:review}
The MN system is obtained by imposing the following relations on the
general effective 5-d action obtained in Sec.~\ref{sugratrunc}:\footnote{We
  correct formula (5.25) of \cite{Papadopoulos:2000gj}.} 
\begin{equation}
\label{MN:special}
\begin{aligned} 
  Q&=0~, \qquad h_1=h_2=0~,\qquad & b&=a~,\\ 
  \Phi&= -6p-g- 2 \ln P~, & x&=\frac12 g -3p~.
\end{aligned}
\end{equation}
Together with \eqref{MN:special}, the constraints \eqref{kconstraint}
and \eqref{partialchi} imply also $K=0$ and $\chi=0$. (Notice that a
constant in $\chi$ is irrelevant.) 
It is straightforward to check from the equations of motion in
appendix~\ref{construnc} that this truncation is consistent, \ie the
equations of motion for $b$, $h_1$, $h_2$, $\Phi$ and $x$ are
satisfied or implied by those for $a$, $p$ and $g$. 
Notice that, having absorbed the constant $P^2$ into $\e{\Phi}$, it
has disappeared from the equations of motion. 
Hence, the effective 5-d action reduces to the form \eqref{action5d},
with three scalar fields ($g,a,p$), the sigma model metric
\begin{equation}
\label{MN:G}
  G_{ab} \partial_\mu \phi^a \partial^\mu \phi^b = 
  \partial_\mu g \partial^\mu g 
  + \e{-2g} \partial_\mu a \partial^\mu a 
  + 24 \partial_\mu p \partial^\mu p~,
\end{equation}
and the superpotential\footnote{We have adjusted the overall factor of the superpotential of
  \cite{Papadopoulos:2000gj} to our conventions.}
\begin{equation}
\label{MN:W}
  W = -\frac12 \e{4p} \left[ (a^2-1)^2 \e{-4g} +2(a^2+1) \e{-2g} +1
    \right]^{1/2}~.
\end{equation}

Let us briefly summarize the most general Poincar\'e-sliced domain wall
background solution \eqref{background1}
for this system. It is obtained by solving
\eqref{background2} and coincides with the family of
solutions found in \cite{Gubser:2001eg}. In the following, $g$, $a$ and
$p$ will denote the background fields, while the fluctuations are
described by the gauge-invariant variables $\mfa^a$. Introducing a
new radial coordinate, $\rho$, by
\begin{equation}
\label{MN:rhodef}
  \partial_\rho = 2 \e{-4p} \partial_r~,
\end{equation}
one can show from the equations for $g$ and $a$ that
\begin{equation}
\label{MN:auxsol} \left[(a^2-1)^2 +2(a^2+1) \e{2g} +\e{4g}
  \right]^{1/2} = 4\rho~.
\end{equation}
The integration constant arising here has no physical meaning and
has been used to fix the origin of $\rho$. Then, one easily obtains 
\begin{equation}
\label{MN:ahsol}
  a = \frac{2\rho}{\sinh(2\rho+c)}~,\quad 
  \e{2g} =4\rho\coth(2\rho+c) -(a^2+1)~,
\end{equation}
where $c$ is an integration constant with allowed values $0\leq c
\leq\infty$. We shall discuss the interpretation of $c$ in
the next subsection. The MN solution corresponds to $c=0$ and is the
only regular solution. All others suffer from a naked curvature
singularity.

It is also easy to show from \eqref{background2} that
\begin{equation}
\label{MN:pAsol}
  \e{-2A} \e{-8p} =C^2~,
\end{equation}
where the integration constant $C$ determines the 4-d reference
scale. We shall set $C^2=1/4$ for later convenience. 
The explicit solution for $p$ can be found by
plugging $\Phi$ from the literature into \eqref{MN:special}, but it
will not be needed here.

\subsection{The role of $c$}
\label{MN:roleofc}
The family of background solutions of the MN system suffers from 
naked singularities for all $c$ except for the case $c=0$, which is
regular. Hence, on the supergravity side the integration constant $c$
governs the resolution of the singularity. However, the
scalar $a(\rho)$ is the dual of the gluino bilinear $\lambda^2$
\cite{Apreda:2001qb}, so $c$, which enters $a(\rho)$ in
\eqref{MN:ahsol}, also determines the ``measured'' value of
the gluino condensate, $\vev{\lambda^2}$, which is of non-perturbative
field theory origin. In other words, $c$ identifies the ``amount'' of
non-perturbative physics that is captured by the supergravity
solution. 

In this subsection, we will attempt to flesh out this picture 
qualitatively,
applying Mathur's coarse graining argument \cite{Mathur:2004sv,
  Mathur:2005zp}, before we analyze the fluctuations in the next subsection.
Although only regular solutions
qualify as gravity duals of (pure) field theory quantum states,
the coarse graining 
argument indicates that certain singular solutions have a
meaning as an approximation to the duals of mixed
states. 
In this point of view,
singularities appear because the ``space-time foam''
 that is dual
to the mixture of pure states cannot be resolved by supergravity.
(We are using the terminology of \cite{Mathur:2004sv,
  Mathur:2005zp} here. See also \cite{Gubser:2000nd} 
  for some earlier discussion of the admissibility of singular solutions.)
In the case at hand, the possible pure
states are naturally identified as the $N$ equivalent vacua of
$\mathrm{SU}(N)$ $\mathcal{N}=1$ SYM theory, which are distinguished
by a phase angle in the gluino condensate.\footnote{In the 10-d MN solution, the location of the Dirac
  string for the magnetic 2-form $C_2$ is specified by an angular
  variable $\psi$ that can take $2N$ different values for the same field
  theory $\theta$-angle, but the solution is symmetric under a shift
  by $\pi$ of $\psi$, leaving $N$ different configurations.  
  Equivalently, one has $N$ different ways of placing probe $D5$-branes
  in the background, in order to obtain the same field theory action  
  \cite{Maldacena:2000yy,DiVecchia:2002ks,Muck:2003zf}.} 
Let us denote these $N$ vacua by $|n\rangle$, where
$n=0,1,2,\ldots,N-1$. The gluino condensate in these vacua takes the
values  
\begin{equation}
\label{MN:gluinocond}
  \langle n| \lambda^2 |n \rangle = \Lambda^3 \e{2\pi i n/N}~,
\end{equation}
where we have absorbed the $\theta$ angle of the gauge theory in the
phase of $\Lambda^3$.  

Now, let us form mixed states by defining the density matrix 
\begin{equation}
\label{MN:dmatrix}
  \varrho = \sum\limits_{n=0}^{N-1} p_n |n \rangle \langle n|~, \quad
  \text{with} \, \sum\limits_{n=0}^{N-1} p_n =1~. 
\end{equation}
Clearly, for equal weights, $p_n=1/N$, we would measure
$\vev{\lambda^2} = \Tr(\lambda^2 \varrho) =0$. 
For a generic mixed state, the measured value $\vev{\lambda^2}$
lies somewhere within the $N$-polygon spanned
by the $N$ pure-state values \eqref{MN:gluinocond}. 
Using standard thermodynamics arguments,
it is straightforward to determine the unique distribution $\{p_n\}$
maximizing the entropy for a given fixed $\vev{\lambda^2}$. Notice,
however, that the $N$ vacua are equivalent, and that, for large $N$, 
which is the regime described by the supergravity approximation, the
vacuum values of $\vev{\lambda^2}$ effectively span a circle of radius
$|\Lambda^3|$. Thus, up to $1/N$ corrections, the phase of some given 
$\vev{\lambda^2}$ is irrelevant, making the relevant parameter space
for the probability distribution $\{p_n\}$ effectively one-dimensional.
 
Thus, from the point of view advocated in \cite{Mathur:2004sv,
  Mathur:2005zp}, the integration constant $c$ can be interpreted
as a parameter that interpolates between the uniform distribution
($c=\infty$) and a pure state ($c=0$), with fixed phase of
$\vev{\lambda^2}$. 
It would be interesting to make this interpretation
precise by attempting to match the statistical entropy of a
mixed state with the area of the apparent horizon surrounding the dual
``space-time foam''. We leave such
investigations for the future. 

Instead, let us confirm the role of $c$ in determining the measured value
of the gluino condensate from the perspective of holographic
renormalization. Being a one-point function, the gluino condensate
should appear as a background \emph{response} function in a supergravity
field (cf.\ the discussion in Sec.~\ref{adscft}).
Thus, consider the background mode \eqref{backgroundmode} of
the fluctuation equation for an arbitrary value of $c$. 
As noted in Sec.~\ref{fieldeq}, the background mode, $W^a/W$,
is always a solution of \eqref{eqphi} independent of $x^i$.  
Let us determine its asymptotic behaviour (large $\rho$) and see
whether it is leading or sub-leading. For an arbitrary value
of $c$, we obtain 
\begin{equation}
\label{MN:asymp}
  \frac{W^a}{W} \sim \left( -\frac1{2\rho}, 8\e{-c}\rho\e{-2\rho} ,
  \frac16 \right)~.
\end{equation}
The first and third components are independent of $c$, \ie universal
for all background solutions, and they are leading compared to the
general solutions that we shall find in the 
next subsection. We have, at present, no specific interpretation of their
role, although the arguments outlined in Sec.~\ref{adscft} indicate
that they should correspond to finite field theory sources
(couplings). In contrast, the second
component is sub-leading and depends on $c$. Hence, we argue in
analogy with AdS/CFT (again, we refer to Sec.~\ref{adscft}) 
that its coefficient represents a
response function, so it determines the 
vacuum expectation value of the dual operator. In this case, 
the dual operator is the gluino bilinear.
Restoring dimensions, this yields
\begin{equation}
\label{MN:gluinoc}
  \vev{\lambda^2} = \Lambda^3 \e{-c}~, 
\end{equation}
which fits nicely with the preceding discussion involving mixed states.

\subsection{Fluctuations and mass spectra}
\label{MN:masspectra}
In the following, we shall consider the equation of motion for scalar
fluctuations about the singular background with $c=\infty$. 
Although we argued in the introduction that singular solutions as
supergravity duals should be taken with a grain of salt, doing so is
quite instructive and serves mainly two purposes: 
First, this solution elegantly describes the asymptotic
region (large $\rho$) of all background solutions, including the
regular MN solution, so that we can learn something about the
asymptotic behaviour of the field fluctuations, which will be
important for the ``dictionary'' and ``renormalization'' problems. 
Second, the matrix equation for fluctuations becomes diagonal and
analytically solvable. Thus, we can hope to get a qualitative glimpse
of the particle spectrum of the dual field theory. 

Consider the equation of motion for scalar fluctuations
\eqref{eqphi}. In terms of $\rho$ and going to 4-d momentum space, 
as well as neglecting the source terms on
the right hand side, \eqref{eqphi} becomes 
\begin{equation}
\label{MN:eqphi}
\left[ (\delta^a_b \partial_\rho + 2 M^a_b)
 (\delta^b_c \partial_\rho - 2 N^b_c) - k^2 \right] \mfa^c =0~,
\end{equation}
where we have fixed the 4-d scale by the choice $C^2=1/4$, which will
turn out convenient later. The matrices $M^a_b$ and $N^a_b$ are given by
\begin{equation}
\label{MN:AB}
\begin{split}
  N^a_b &=\e{-4p} \left(\partial_b W^a -\frac{W^a W_b}{W} \right)~,\\
  M^a_b &= N^a_b +2 \e{-4p} \left(\G{a}{bc}W^c -W \delta^a_b\right)~.
\end{split}
\end{equation}
Notice that the $p$-dependence in $M^a_b$ and $N^a_b$ cancels out. 
For the case $c=\infty$, the matrices $M^a_b$ and $N^a_b$ are
diagonal, 
\begin{equation}
\label{MN:ABasymp}
\begin{split}
  N^a_b &= \diag \left( -\frac1{2\rho}, \frac1{2\rho}-1,0 \right)~,\\
  M^a_b &= \frac1{4\rho-1} \diag \left( 4\rho -2 +\frac1{2\rho},
  1-\frac1{2\rho}, 4\rho \right)~.
\end{split}
\end{equation}

We are mostly interested in the field $\mfa^2$ (the middle component),
since its dual operator is the gluino bilinear ($+$ its hermitian
conjugate). From \eqref{MN:eqphi} and \eqref{MN:ABasymp}, its
equation of motion reads 
\begin{equation}
\label{MN:eqmot2}
  \left( \partial_\rho^2 +4\frac{2\rho-1}{4\rho-1} \partial_\rho
  +\frac{4}{4\rho-1} -k^2 \right) \mfa^2 =0~.
\end{equation}
Performing a change of variable by defining 
\begin{equation}
\label{MN:zdef}
  \rho-\frac14 = \alpha z~,
\end{equation}
with a constant $\alpha$ to be determined later, 
and using the following ansatz for the solution,
\begin{equation}
\label{MN:ansatz2}
  \mfa^2 = \e{az} z^b f(z)~,
\end{equation}
with constant $a$ and $b$, we find that the choices
\begin{equation}
\label{MN:constants2}
  a= -\alpha~, \quad b=\frac14~,\quad \alpha^2 (1+k^2) = \frac14
\end{equation}
lead to the equation
\begin{equation}
\label{MN:eqmot2z}
  \left( \partial_z^2  -\frac14 + \frac{3\alpha}{2z} +\frac{5}{16z}
  \right) f=0~.
\end{equation}
This can be recognized as Whittaker's equation, the solutions of which
are linear combinations of the two Whittaker functions
\begin{equation}
\label{MN:sol2M}
  f = \left\{ \rmM_{\frac32\alpha,\frac34}(z) ~,~
  \rmM_{\frac32\alpha,-\frac34}(z) \right\}~.
\end{equation}
Hence, using \eqref{MN:ansatz2} and the relation of Whittaker's
functions to confluent hypergeometric functions 
$\Phi$ and $\Psi$  \cite{Gradshteyn,
  Abramowitz}, we find 
\begin{equation}
\label{MN:sol2Phi}
  \mfa^2 \sim \e{-(\alpha+1/2)z} 
  \begin{cases}
    (\alpha z)^{3/2}\, 
    \Phi\left( \frac54-\frac32\alpha,\frac52;z\right)~,\\
    \Phi\left(-\frac14-\frac32\alpha,-\frac12;z\right)~.
  \end{cases}
\end{equation}
In standard AdS/CFT, one would impose a regularity condition in the
bulk interior in order to obtain a linear combination of the two
solutions, which uniquely fixes the relation between the response and
the source functions. Here, however, we were not able to find such a
condition, probably due to the curvature singularity of the
background. However, there is a useful feature that can guide us
in the choice of suitable solutions. From \eqref{MN:zdef}, we should
demand that the solution be invariant under a simultaneous change of
sign of $z$ and $\alpha$. Due to the identity \cite{Gradshteyn}
\begin{equation}
\label{MN:Phiident}
  \Phi(a,b;z) = \e{z} \Phi(b-a,b;-z)~,
\end{equation}
the particular solutions \eqref{MN:sol2Phi} are invariant under this
symmetry. This implies two things. First, we are free to choose the
solution for $\alpha$ in the 
last equation of \eqref{MN:constants2} such that
$\mathrm{Re}\,\alpha>0$, which implies also $\mathrm{Re}\, z
>0$. Notice that the square root in the definition 
of $\alpha$ demands a branch cut in $k^2$-space, which we place at
$k^2+1 <0$. This branch cut is an indication for a continuum in the
particle spectrum, for $m^2 = -k^2 >1$. (Notice that this is relative to a
reference scale, since we are working in dimensionless variables. With
the earlier choice $C^2=1/4$ we place the onset of the continuum
conveniently at the branch point $k^2=-1$.) 
Second, linear combinations of the solutions should also reflect this
symmetry implying that proportionality factors can
depend only on $\alpha^2$. In particular, the choice of the functions
$\Psi(a,b;z)$ instead of $\Phi(a,b;z)$ is not allowed, cf.\ \cite{Gradshteyn}.

It is instructive to consider the asymptotic behavior of the
solutions. Let $\alpha$ be generic and fixed, so that we can
consider large $z$. One finds that both solutions in
\eqref{MN:sol2Phi}, and any generic linear combination of them,
behave as  
\begin{equation}
\label{MN:asympt2}
  \mfa^2 \sim \e{(1/2-\alpha)z} z^{1/4-3\alpha/2}~,
\end{equation}
but there are notable exceptions. Indeed, the confluent hypergeometric
functions $\Phi(a,b;z)$ reduce to polynomials (Laguerre polynomials,
to be precise), if the first index, $a$, is zero or a negative
integer. In these cases, the generic leading terms \eqref{MN:asympt2}
are absent. Generalizing the AdS/CFT argument \cite{Witten:1998zw}, 
we interpret the corresponding values of $-k^2$ as discrete particle
masses in the spectrum of the dual field theory.

Hence, the two solutions \eqref{MN:sol2Phi} give rise to two
different discrete spectra 
\begin{equation}
\label{MN:spectrum1}
  m_n^2 = 1-\frac9{(4n+3)^2}~,\quad n=0,1,2,\ldots~,
\end{equation}
and
\begin{equation}
\label{MN:spectrum2}
  m_n^2 = 1-\frac9{(4n+5)^2}~,\quad n=0,1,2,\ldots~.
\end{equation}
Notice that there is a massless state, for $n=0$ in
\eqref{MN:spectrum1}. Moreover, both spectra approach the branch
point, $-k^2=1$, for $n\to\infty$. 

Similarly, we consider the other components. The equation of motion
for $\mfa^3$ is
\begin{equation}
\label{MN:eqmot3}
  \left( \partial_\rho^2 +\frac{8\rho}{4\rho-1} \partial_\rho
   -k^2 \right) \mfa^3 =0~,
\end{equation}
for which we obtain the solutions
\begin{equation}
\label{MN:sol3Phi}
  \mfa^3 \sim \e{-(\alpha+1/2)z} 
  \begin{cases} 
    (\alpha z)^{1/2}\, 
    \Phi\left( \frac34+\frac12\alpha,\frac32;z\right)~,\\
    \Phi\left( \frac14+\frac12\alpha,\frac12;z\right)~.
  \end{cases}
\end{equation}
As before, $z$ and $\alpha$ are defined by \eqref{MN:zdef} and
\eqref{MN:constants2}, respectively.
Hence, we find again a continuum of states for
$-k^2>1$. However, although the solutions \eqref{MN:sol3Phi} are
similar to \eqref{MN:sol2Phi}, the sign in front of the $\alpha$-terms
in the first index of the confluent hypergeometric functions does not
allow them to reduce to polynomials. (Remember that
$\mathrm{Re}\,\alpha>0$.) Hence, there is no
discrete spectrum of states. 

We would like to note that the solution \eqref{MN:sol3Phi} is very
similar to (3.17) of \cite{Ametller:2003dj}. They considered fluctuations of the dilaton about the MN background and introduced a hard-wall cut-off, and found
an unbounded discrete spectrum of glueball masses. This procedure was subsequently criticized in \cite{Caceres:2005yx}. 
Due to the discussion in the previous paragraph, we
do not infer glueball masses from the component $\mfa^3$.

The treatment of component $\mfa^1$ is slightly more complicated. Its
equation of motion is 
\begin{equation}
\label{MN:eqmot1}
  \left( \partial_\rho^2 +\frac{8\rho}{4\rho-1} \partial_\rho
   -\frac2{\rho^2} +\frac8{4\rho-1} -k^2 \right) \mfa^1 =0~.
\end{equation}
The awkward double pole in $\rho$ can be removed by setting
$\mfa^1=\rho^{-1}f(\rho)$, which yields the equation 
\begin{equation}
\label{MN:eqmot1f}
  \left[ \partial_\rho^2 +
  \left( 2 +\frac{2}{4\rho-1} -\frac2{\rho} \right) \partial_\rho 
  -k^2 \right] f =0~.
\end{equation}
After changing variables to $z$ by using \eqref{MN:zdef} and making
the ansatz 
\begin{equation}
\label{MN:ansatz1}
  f(z) = \e{cz}  \tilde f(z)~,
\end{equation} 
we find that the choice
\begin{equation}
\label{MN:constants1}
  c = -\frac12 -\alpha~,
\end{equation} 
where $\alpha$ is defined as before, leads to the equation
\begin{equation}
\label{MN:eqmot1z}
  \left\{ 4\alpha z \left[z \partial_z^2 + \left(-\frac32-z \right)
  \partial_z + \frac34 +\frac32\alpha \right] + 
  \left[z \partial_z^2 + \left(\frac12-z \right)
  \partial_z - \frac14 -\frac12\alpha \right] \right\} \tilde f =0~.
\end{equation}
The two terms in square brackets represent differential equations for
confluent hypergeometric functions, which gives us a nice hint for
solving the equation. Indeed, we can explicitly find the solutions,
which, combined with \eqref{MN:ansatz1} and $\mfa^1=\rho^{-1} f$,
result in
\begin{equation}
\label{MN:sol1Phi}
  \mfa^1 \sim \frac{\e{-(\alpha+1/2)z}}{\alpha z+1/4} 
  \begin{cases}
  \Phi\left( -\frac34-\frac32\alpha,-\frac32;z\right) 
  -\frac{4\alpha^2-1}3 z^2\,
  \Phi\left( \frac54-\frac32\alpha,\frac52;z\right)~,\\
  (\alpha z)^{1/2} \left[ 
  \Phi\left( -\frac14-\frac32\alpha,-\frac12;z\right)
  + \frac{36\alpha^2-1}5 z^2\,
  \Phi\left( \frac74-\frac32\alpha,\frac72;z\right) \right]~.
  \end{cases}
\end{equation}
Notice that both solutions respect the symmetry of simultaneously
changing the signs of $\alpha$ and $z$. The sign of the $\alpha$-terms
in the first index of the confluent hypergeometric functions
indicates that, in addition to the continuum from the branch cut, we
have again a discrete spectrum of states for those values of $\alpha$,
where these functions reduce to polynomials. The corresponding spectra
are given again by \eqref{MN:spectrum1} and \eqref{MN:spectrum2}, but
in \eqref{MN:spectrum1} only values $n=1,2,3,\ldots$ are allowed,
which implies that the massless state is absent.

To conclude this section, let us discuss whether we can trust the mass
spectrum we have found. This question arises since the
calculation was performed in the singular background with $c=\infty$,
but the true supergravity dual of a field theory vacuum is the MN
solution, with $c=0$. Moreover, one typically expects the boundary
conditions in the interior to influence the dual IR physics, but we
have not directly imposed any conditions except symmetry of
simultaneously changing the signs of $\alpha$ and $z$. However, there are only three things that can happen to each particular mass value 
when the regular background with $c=0$ is considered. First, there
could exist a corresponding regular and sub-leading solution for which the mass
value changes as we go from $c=\infty$ to $c=0$. 
Second, there could exist a 
corresponding regular and sub-leading solution with the same mass.
Third, the
corresponding sub-leading solution may not be regular at $\rho=0$, 
in which case that particular mass value would not be in the spectrum. 
In the following, we will argue that the first of these scenarios is
excluded. Remember that the background with
$c=\infty$, which we have considered here, correctly 
describes the asymptotic region of
the regular background. 
Hence, the asymptotic behaviour of the
fluctuations we found is valid also for $c=0$, 
implying that the mass spectra
\eqref{MN:spectrum1} and \eqref{MN:spectrum2} are unchanged. 
One can verify this by a series expansion in
$\e{-c}$ of the equations of motion. Also, it is a
straightforward but important check that the component $\mfa^3$ 
decouples from the other two for any value of $c$ and, therefore, 
cannot spoil the sub-leading behaviour. (Remember that the solutions
for $\mfa^3$ did not give rise to mass spectra.) 

However, it might happen that imposing a regularity condition on the
fluctuations, which is required to calculate 2-point
functions, does not allow for the solution that corresponds to a given
mass value. This mechanism can be summarized as follows: For 
given $k^2$, 
$\mfa^1$ and $\mfa^2$ give four independent solutions,
two of which give rise to the mass spectrum
\eqref{MN:spectrum1}, the other two leading to \eqref{MN:spectrum2}. These
solutions evolve as we go from $c=\infty$ to $c=0$, but their
asymptotic behaviour does not change. For $c=0$, imposing regularity
conditions will select two linear combinations of these four
solutions. 
If such a linear combination involves only the two solutions
corresponding to the same mass spectrum, then this spectrum will survive. 
If, in contrast, the
linear combination involves solutions corresponding to different mass
spectra, no mass values will result from it. A particularly
interesting case is the massless state, which belongs to
the spectrum \eqref{MN:spectrum1}, but arises only from the component $\mfa^2$, not
from $\mfa^1$, in the analysis above. One does not expect a
massless glueball state to exist, and in fact, it is likely to be
excluded by this mechanism. It is less likely that only single masses,
as opposed to an entire spectum, will survive this mechanism. This is
in contrast to the result of \cite{Caceres:2005yx}, where only a
single glueball state was found. We will not answer these interesting questions in this paper, but we intend to come back to them.

\section{The Klebanov-Strassler system}
\label{KS}
In this section we review the \emph{warped deformed conifold},
or the Klebanov-Strassler solution \cite{Klebanov:2000hb}.
We will be particularly interested in the ``gluino sector'', 
the 3-scalar system of fluctuations that contains the field dual to
the gluino bilinear $\Tr \lambda \lambda$. 

\subsection{Review of the background}
\label{KSbackground}
The KS system is obtained from the general PT system by relating the
fields $a$ and $g$ by the relation
\begin{equation}
\label{KS:agrel}
  a = \tanh y~,\quad \e{-g}= \cosh y \qquad \text{(KS)}~,
\end{equation}
whereby a new field $y$ (not to be confused with the 5-d
coordinates $y^\mu$ used in Sec.~\ref{sugratrunc}) is introduced. 
This relation renders the constraint \eqref{partialchi} integrable and implies
$\chi=0$. Moreover, one can check that the equations of motion for $a$
and $g$, \eqref{eoma} and \eqref{eomg}, become equivalent. 

There exists an even more restricted truncation, which gives
rise to the singular 10-d conifold background of KT and certain
fluctuations thereof. It contains four scalars and is obtained by
imposing 
\begin{equation}
\label{KS:KTrels}
  a=b=g=h_2=0 \qquad \text{(KT)}~,
\end{equation}
which also implies $\chi=0$. We shall not consider the KT system
separately, but discuss the KS system in a way similar to the
treatment of the MN system in Sec.~\ref{MN}. That is, we will consider
a class of background solutions characterized by a parameter $c$,
that formally interpolates between the KT and KS backgrounds. As in the MN
case, the background solutions are typically 
singular, except for the KS endpoint of the family. 
\begin{table}[ht]
\caption{Comparison of symbol and field conventions used by Apreda
  \cite{Apreda:2003gc} (see also \cite{Bigazzi:2000uu}), 
  Papadopoulos and Tseytlin
  \cite{Papadopoulos:2000gj}, and Klebanov and Strassler (KS) 
  \cite{Klebanov:2000hb}. The entries N/A mean that
  these fields do not appear explicitly in the KS paper. 
  Apreda's fields diagonalize the mass matrix in the
  AdS background for $P=0$, $Q=2/\sqrt{27}$. The last two columns
  contain, respectively, the mass squared of the bulk fields and the
  conformal dimensions of the dual operators in the AdS background
  for $P=0$. \label{KS:conventions}}
\renewcommand\arraystretch{1.4}
\[
\begin{array}{|c|c|c|c|c|c|}
\hline
{\rm Apreda} & {\rm PT} & {\rm KS} & m^2 & \Delta 
\\ \hline\hline
q & \frac15(x-2p)+\frac{3}{20}\ln(3)+\frac{1}{10}\ln(2) 
& {\rm N/A} & 32 & 8 \\ \hline
f_{\rm Apreda} & \frac15(x+3p)+\frac{1}{10}\ln(2/3) & {\rm N/A}  
& 12 & 6 \\ \hline
y & \sinh^{-1}(a \e{-g}) & {\rm N/A} & -3 & 3 \\ \hline
\Phi & \Phi & \Phi & 0 & 4 \\ \hline
s & -2h_1 & M (k+f_{\rm KS}) & 0 & 4 \\ \hline
N_1 & -h_2-P_{\rm PT} (b+1) & \frac{M}{2}(k-f_{\rm KS})- M F 
& 21 & 7 \\ \hline
N_2 & -h_2+ P_{\rm PT}(b+1) & \frac{M}{2} (k-f_{\rm KS})+MF 
& -3 & 3 \\ \hline\hline
P_{\rm Apreda} & -P_{\rm PT} \equiv -P & M/2 & - & - \\ \hline
\end{array} 
\]
\end{table}

For the remaining fields of the KS system, 
there exist a variety of conventions in the
literature, some of which we list for reference in
Tab.~\ref{KS:conventions}.\footnote{Note that there 
are typos in the first three equations of (5.24) in
\cite{Papadopoulos:2000gj}, which relate 
the variables used in that paper to those used in \cite{Klebanov:2000hb}. 
The correct relations can be read off from Tab.~\ref{KS:conventions}.}
For the purpose of rederiving the
background solutions, we shall start with the PT variables
$(x,p,y,\Phi,b,h_1,h_2)$, where $y$ was introduced in
\eqref{KS:agrel}. The sigma-model metric \eqref{actionPT}
for the KS system reduces to
\begin{multline}
\label{KS:Gab}
  G_{ab} \partial_\mu \phi^a \partial^\mu \phi^b = 
  \partial_\mu x \partial^\mu x 
  + 6 \partial_\mu p \partial^\mu p
  + \frac12 \partial_\mu y \partial^\mu y
  + \frac14 \partial_\mu \Phi \partial^\mu \Phi 
  + \frac{P^2}2 \e{\Phi-2x} \partial_\mu b \partial^\mu b +\\
  +\frac14 \e{-\Phi-2x} \left[ 
     \e{-2y} \partial_\mu (h_1-h_2) \partial^\mu (h_1-h_2)      
    +\e{2y} \partial_\mu (h_1+h_2) \partial^\mu (h_1+h_2) \right]~,
\end{multline}
and the superpotential reads
\cite{Papadopoulos:2000gj} 
\begin{equation}
\label{KS:W}
  W = -\frac12 \left( \e{-2p-2x} +\e{4p} \cosh y \right) 
  +\frac14 \e{4p-2x} \left( Q + 2Pb h_2 + 2Ph_1 \right)~.
\end{equation}
Setting $P=0$, there exists an AdS fixed point. The choice
$Q=2/\sqrt{27}$ leads to the corresponding AdS background with unit length
scale, and Apreda's fields vanish in this background. 

For the KS system,
the background equations \eqref{background1}, \eqref{background2}
become%
\begin{equation}
\label{KS:backgrdeq}
\begin{aligned}
  \partial_r (x+3p) &= \frac32 \e{-2p-2x} -\e{4p} \cosh y~,\\ 
  \partial_r (x-6p) &= 2 \e{4p} \cosh y - \frac32 \e{4p-2x}  
    \left[ Q + 2Pbh_2 +2P h_1 \right]~,\\
  \partial_r y &= -\e{4p} \sinh y~,\\
  \partial_r \Phi &= 0~,\\
  \partial_r b &= \frac1{P\e{\Phi}} \e{4p} h_2~,\\
  \partial_r h_1 &= P\e{\Phi} \e{4p} \left[\cosh(2y) -b \sinh(2y)
    \right]~,\\
  \partial_r h_2 &= P\e{\Phi} \e{4p} \left[b \cosh(2y) -\sinh(2y)
    \right]~.
\end{aligned}
\end{equation}
We shall, in the following, rederive the background solutions of this
system by following the calculations of KS \cite{Klebanov:2000hb}, but
adding the relevant integration constants. 
From \eqref{KS:backgrdeq} we can immediately read off $\Phi=\Phi_0=\const$, and
after introducing the KS radial coordinate $\tau$ by 
\begin{equation}
\label{KS:taudef}  
  \partial_\tau = \e{-4p} \partial_r~,
\end{equation}
we easily find 
\begin{equation}
\label{KS:ysol}
  \e{y} = \tanh \frac{\tau+c}2~.
\end{equation}
For generality, we shall keep the integration constant $c$. In
particular, $c$ takes the values $\infty$ and $0$ for the KT and KS
solutions, respectively. Similar to the parameter $c$ in the MN
solution discussed in Sec.~\ref{MN}, it determines whether the supergravity 
solution is regular ($c=0$) or not ($c \neq 0$). We note that \eqref{KS:ysol}
restricts the range of $\tau$ to $\tau > -c$.  

From the equations for $b$, $h_1$ and $h_2$ one can derive the
differential equation
\begin{equation}
\label{KS:beq}
  \partial_\tau^2 b = b \cosh(2y) -\sinh(2y)~,
\end{equation}
whose general solution is 
\begin{equation}
\label{KS:bsol}
  b = b_1 \cosh(\tau+c) -\frac{(b_1+1)\tau +b_2}{\sinh(\tau+c)}~.
\end{equation}
We must set $b_1=0$ in order to avoid the exponential blow-up for
large $\tau$, and $b_2$ can be absorbed into a redefinition of
$\tau$ and $c$. Hence, we have 
\begin{equation}
\label{KS:bsol2}
  b = -\frac{\tau}{\sinh(\tau+c)}~,
\end{equation}
from which follows immediately
\begin{equation}
\label{KS:h2sol}
  h_2 = P\e{\Phi_0} \frac{\tau\coth(\tau+c)-1}{\sinh(\tau+c)}~.
\end{equation}
Then, we obtain also
\begin{equation}
\label{KS:h1sol}
  h_1 = P\e{\Phi_0}\coth(\tau+c) 
    [ \tau \coth(\tau+c) -1] +\tilde{h}~,
\end{equation}
where $\tilde{h}$ is an integration constant. 

The functions $b$, $h_1$ and $h_2$ determine the function $K$, which
measures the 5-form flux in the 10-d configuration
\eqref{PTansatz}.\footnote{Note that this is not the $K$ of KS.}  
From \eqref{kconstraint}, \eqref{KS:bsol2}, \eqref{KS:h2sol} and
\eqref{KS:h1sol} we find 
\begin{equation}
\label{KS:K}
  K = K_0  +P^2 \e{\Phi_0} \frac{\tau\coth(\tau+c)-1}{\sinh^2(\tau+c)} 
   [\sinh(2\tau+2c) -2\tau]~,
\end{equation}
where we have abbreviated 
\begin{equation}
\label{KS:K0}
  K_0 = Q +2 P \tilde{h}~.
\end{equation}

Now, let us calculate the backgound fields $x$ and $p$. It is
convenient to use Apreda's fields $f$ and $q$, the definitions of
which are given in Tab.~\ref{KS:conventions}.
Then, from the equation for $(x+3p)$ we find 
\begin{equation}
\label{KS:feq}
  5 \partial_\tau f = \e{-10f} -\cosh y~,
\end{equation}
with the general solution
\begin{equation}
\label{KS:fsol}
  \e{10f} = \coth(\tau+c) -\frac{\tau+f_0}{\sinh^2(\tau+c)}~,
\end{equation}
where $f_0$ is again an integration constant. The remaining background
equation gives rise to 
\begin{equation}
\label{KS:qeq}
  \left( \partial_\tau -\frac43 \coth y \right) \e{6q-8f/3} =
  - 2 \cdot3^{1/2} K \e{-20f/3}~,
\end{equation}
where $K$ is given by \eqref{KS:K}. Isolating the homogeneous solution
by the ansatz\footnote{The constant factor 
$2^{4/3}e^{-4c/3}$
has been inserted to normalize the 
forefactor to unity in the $c\to \infty$ limit.}  
\begin{equation}
\label{KS:hdef}
  \e{6q-8f/3} = 2^{4/3} \e{-4c/3} \sinh^{4/3}(\tau+c)\, h(\tau)~,
\end{equation}
we obtain from \eqref{KS:qeq} that $h(\tau)$ satisfies 
\begin{equation}
\label{KS:heq}
\begin{split}
  \partial_\tau h &= -2^{1/3} 3^{1/2} \e{4c/3} [\sinh(2\tau+2c)
  -2\tau-2f_0]^{-2/3} \times \\
  & \quad\times \left\{ K_0 +P^2 \e{\Phi_0}
    \frac{\tau\coth(\tau+c)-1}{\sinh^2(\tau+c)} [\sinh(2\tau+2c)
    -2\tau] \right\}~.
\end{split}
\end{equation}

It is instructive to consider the limit $c \to \infty$, which
describes the large-$\tau$ behaviour of all background solutions.
In this case, we obtain explicitly 
\begin{equation}
\label{KS:hsolcinf}
  h = \frac12 3^{3/2} \e{-4\tau/3} \left[ K_0 +2P^2 \e{\Phi_0}
  \left(\tau-\frac14\right) \right]+h_0~,
\end{equation}
The choice $h_0=0$, needed in order to avoid the exponential growth in
\eqref{KS:hdef}, removes the asymptotically flat region from the 10-d
solution.

Finally, one can show that the equation for the warp factor $A$ in
\eqref{background1} yields 
\begin{equation}
\label{KS:Asol}
  \e{-2A} = C^2 \e{-2x/3} (2\e{-c})^{-2/3} \sinh^{-2/3} (\tau+c)~,
\end{equation}
where the integration constant $C$ sets the 4-d scale and will be
fixed later.\footnote{Note that in this formula 
most of the complicated $\tau$-dependence of $\e{-2A}$
is hidden in the factor $\e{-2x/3}$.}

The regular KS solution is given by fixing
the integration constants as follows:
\begin{equation}
\label{KS:constants}
  c=f_0=K_0=0~,
\end{equation}
and imposing vanishing $h$ for large $\tau$, which yields
\begin{equation}
\label{KS:hsol}
  h= 2^{1/3} 3^{1/2} P^2 \e{\Phi_0} \int\limits_{\tau}^\infty
    \rmd \vartheta \, \frac{\vartheta\coth \vartheta -1}{\sinh^2
    \vartheta} \left[ \sinh(2\vartheta)-2\vartheta  
    \right]^{1/3}~.
\end{equation}
Note that our definition of $h$ differs from the one in
\cite{Klebanov:2000hb} by a constant involving a factor $\epsilon^{-8/3}$. (Although \cite{Klebanov:2000hb} fix $\epsilon$ to a numerical
value early on, it is clear from (65) in \cite{Herzog:2001xk}
that their $h \sim \epsilon^{-8/3}$.)
Our constant $C^2$ of \eqref{KS:Asol}, which
appears in front of the external 4-dimensional metric
in \eqref{background1}, corresponds to
$\epsilon^{-4/3}$ of \cite{Klebanov:2000hb} up to numerical factors.

\subsection{Fluctuation equations}
\label{KSeoms}
We are now in a position to write down the equations of
motion for fluctuations \eqref{eqphi} about the background solutions
found in the previous subsection. Let us begin by expressing 
the equation of motion
in terms of the KS radial coordinate $\tau$. After multiplying
\eqref{eqphi} by $\e{-8p}$ and using \eqref{KS:taudef} we obtain
\begin{equation}
\label{KS:eqmot}
  \left[ (\partial_\tau + M )(\partial_\tau -N) +\e{-8p-2A} \Box
  \right] \mfa =0~,
\end{equation}
where the matrices $M$ and $N$ are given by
\begin{equation}
\label{KS:MNdef}
\begin{split}
  N^a_b &= \e{-4p} \left( \partial_b W^a -\frac{W^aW_b}W \right)~,\\
  M^a_b &= N^a_b +2\e{-4p} \left( \G{a}{bc}W^c +\e{-2p-2x} \delta^a_b
  \right)~.
\end{split}
\end{equation}
When we substitute the KS background in \eqref{KS:MNdef},
the matrices become quite complicated and are relegated to Appendix \ref{sec:KSmatrices}.
We view it as an important step to have obtained them explicitly,
and we intend to come back to a more detailed study of
them at a later date.
 
In the following, we shall consider fluctuations about the KT
background, which is given by the choice of integration constants
\begin{equation}
\label{KS:KTbackground}
  c= \infty~,\quad K_0=f_0=h_0=\Phi_0=0~.
\end{equation}
The motivation for this choice is essentially the same as for the MN
system: this background describes correctly the asymptotic region of
the KS solution, and the equations of motion have a simpler form,
which can be treated analytically (with a further approximation
described in Sec.~\ref{KSnearIR}). 

For the background specified by the integration constants
\eqref{KS:KTbackground}, the matrices $M$ and $N$ have quite a simple form.
Using Apreda's variables for the fluctuation fields, $\mfa=
\delta(q,f,\Phi,s,y,N_1,N_2)$, 
and $P\equiv P_{\rm PT} = -P_{\rm Apreda}$,
we find
\begin{align}
\label{KS:Minf}
  M &= \begin{pmatrix}
       \frac{4(8\tau-3)}{3(4\tau+1)} & 0 & 0 &
       \frac{32(\tau-1)}{45P(16\tau^2-1)} & 0 & 0 & 0 \\ 
       0 &-\frac{2}{3} & 0 & -\frac{4}{15P(4\tau-1)} & 0 & 0 & 0 \\
       0 & 0 & \frac{4}{3} & -\frac{8}{3P(4\tau-1)} & 0 & 0 & 0 \\
       \frac{20P(4\tau-1)}{4\tau+1} & 0 & 0 &
       \frac{8(2\tau+1)(4\tau-3)}{3(16\tau^2-1)} & 0 & 0 & 0 \\
       0 & 0 & 0 & 0 & \frac{1}{3} & \frac{8}{3P(4\tau-1)} &
       \frac{8}{3P(4\tau-1)} \\
       0 & 0 & 0 & 0 & 0 & \frac{28\tau-19}{3(4\tau-1)} & 0 \\
       0 & 0 & 0 & 0 & 0 & 0 & \frac{4\tau-13}{3(4\tau-1)} 
       \end{pmatrix}~,\\
\label{KS:Ninf}
  N &= \begin{pmatrix}
      \frac{16(\tau-1)}{3(4\tau+1)} & 0 & 0 & \frac{4}{9P(4\tau+1)} & 
      0 & 0 & 0 \\
      0 & -2 & 0 & 0 & 0 & 0 & 0 \\  
      0 & 0 & 0 & 0 & 0 & 0 & 0 \\  
      \frac{32P(\tau-1)}{4\tau+1} & -8P & -2P & \frac{8}{3(4\tau+1)} & 
      0 & 0 & 0 \\
      0 & 0 & 0 & 0 & -1 & 0 & 0 \\
      0 & 0 & 0 & 0 & 2P & 1 & 0 \\
      0 & 0 & 0 & 0 & 2P & 0 & -1
       \end{pmatrix}~.
\end{align}
The block-diagonal form of these matrices is a nice
feature of the KT background. Remember that \eqref{KS:KTrels} defines
a consistent truncation of the KS to the KT system, so that the lower
left $3\times 4$ off-diagonal blocks of $M$ and $N$ are expected to be
zero, but vanishing of the upper right block 
is a bonus feature.
It is a particularly welcome bonus, since the gluino sector
$\delta(y,N_1,N_2)$ is where we would expect much of the interesting
physics to be encoded. 

We also see from \eqref{KS:Minf} and \eqref{KS:Ninf} that the UV limit
$\tau\rightarrow \infty$ and the conformal limit $P\rightarrow 0$ do
not commute. One might have considered performing an expansion in $P$
to study a ``near-conformal'' regime, but the
order of limits would pose a problem. This is not surprising, because among other things we have imposed $K_0=0$
on the solution, which is not possible for $P=0$, as can be seen from
\eqref{KS:K}. It is of course possible to study the conformal
(Klebanov-Witten \cite{Klebanov:1998hh}) system directly, but this would
require changing field variables.  

For the KT background, it is useful to change the radial variable by
introducing\footnote{Our $\sigma$ corresponds to $r$ of KT up to a
  multiplicative factor, whereas our $r$ corresponds to their $u$.}
\begin{equation}
\label{KS:sdef}  
  \tau= 3\ln \sigma +\frac14~.
\end{equation}
Using \eqref{KS:Asol}, \eqref{KS:hdef} and \eqref{KS:hsolcinf},
we find that the term in \eqref{KS:eqmot} with the 
4-dimensional box operator is proportional to
\begin{equation}
\label{KS:momentumterm}
  \e{-2A-8p} \sim \frac{C^2 P^2}{\sigma^2} \ln \sigma~,
\end{equation}
where we have suppressed a numerical factor. Hence, from \eqref{KS:eqmot}
and with a suitable choice of the constant $C$ follows, in momentum space,
\begin{equation}
\label{KS:eqmot2}
  \left[ (\sigma\partial_\sigma + 3 M )(\sigma \partial_\sigma -3N) -
  \frac{k^2 P^2}{\sigma^2} \ln \sigma \right] \mfa =0~.
\end{equation}
We see that fixing $C$ indeed sets the 4-dimensional energy scale,
as claimed in the previous section.
 
We further introduce\footnote{Our $v$ corresponds to Krasnitz's
  $y$.} 
\begin{equation}
\label{KS:ydef}  
  v = \frac{kP}{\sigma}~,
\end{equation}
in terms of which \eqref{KS:eqmot2} becomes
\begin{equation}
\label{KS:eqmot3}
  \left[ v^3 \partial_v v^{-3} \partial_v
  - Y v^{-1} \partial_v 
  - Z v^{-2} - \ln \frac{kP}{v} \right]\mfa =0~,
\end{equation}
where the matrices $Y$ and $Z$ are given by
\begin{equation}
\label{KS:YZdef}
  Y = 3 (M-N) - 4~,\quad Z = 9MN+3\sigma \partial_\sigma N~.
\end{equation}
In \cite{Krasnitz:2002ct}, fluctuations of the 4-scalar KT system were
studied in a particular gauge, leading to equations more complicated
than, but presumably equivalent to \eqref{KS:eqmot3}.

\subsection{``Moderate UV'' approximation}
\label{KSnearIR}

Despite its apparent simplicity,
equation  \eqref{KS:eqmot3} has no analytic solution.
A method to extract the 
response functions at leading order in the high-energy limit 
was developed by Krasnitz \cite{Krasnitz:2000ir, Krasnitz:2002ct}. We
proceed to briefly review this method, but first we pause for a short
comment on our motivation to use the method in the first place. 

We are, of course, ultimately interested in all energy ranges and the
confining phase, not just the high-energy limit. Nevertheless, we
have seen that the matrices  in appendix \ref{sec:KSmatrices} are
prohibitively complicated for analytical work, so we view the
approximation in this subsection as a simple way to get a handle on
the full problem in one particular regime (high energy),
which should provide good cross-checks for a numerical treatment. 
In addition, since renormalization is a UV problem, KT counterterms
should be sufficient to renormalize KS correlators, so the UV regime
seems a good place to start.

\begin{figure}[th]
\begin{center}
\psfrag{fmUV}[bc][bc][1][0]{$\phi_{\mathrm{mUV}}(v)$}
\psfrag{fxUV}[bc][bc][1][0]{$\phi_{\mathrm{xUV}}(v)$}
\psfrag{v2logkv}[bc][bc][.8][0]{$v^2 \log (k/v)$}
\psfrag{ll1}[bc][bc][.8][0]{$\ll 1$}
\psfrag{1/kllv}[bc][bc][.8][0]{$1/k \ll v$}
\psfrag{vll1/slogk}[bc][bc][.8][0]{$v \ll 1/\sqrt{\log k}$}
\psfrag{abslogv}[bc][bc][.8][0]{$|\log v|$}
\psfrag{llabslogk}[bc][bc][.8][0]{$\ll |\log k|$}
\psfrag{IR}[bc][bc][1][0]{IR}
\psfrag{arrow}[bc][bc][1.2][0]{$\longrightarrow$}
\psfrag{veq0}[bc][bc][1][0]{$v=0$}
\includegraphics[width=0.8\textwidth]{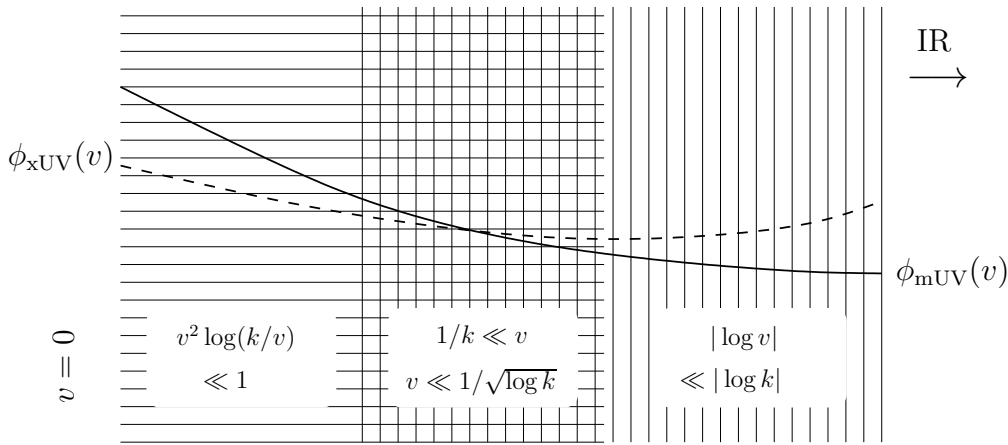}
\caption{Krasnitz matching for a generic field $\phi$.
The solution denoted $\phi_{\mathrm{mUV}}$ is regular in the IR,
and analogous to our solutions below. The solutions
are matched to approximately agree in the cross-hatched overlap
region.}  
\label{KS:krasnitzfig}
\end{center}
\end{figure}

Here is the brief review.
In \cite{Krasnitz:2000ir, Krasnitz:2002ct},
the KT solution was divided into two overlapping regions, which we
will call ``moderate UV'' (or ``mUV'') region and ``extreme UV'' (or ``xUV'')
region. For the purposes of this discussion, let us set $P=1$; it 
can be restored by $k \rightarrow kP$.
In the mUV region, $|\log v| \ll |\log k|$, so 
we can approximate the troublesome $\log (k / v)$ 
in \eqref{KS:eqmot3} by a constant $\log k$. This clearly does not
work for $v$ too small, hence ``moderate'' UV, but when it does work, 
exact solutions of the approximated equation can be found
\cite{Krasnitz:2000ir, Krasnitz:2002ct}. In the xUV region,
\cite{Krasnitz:2000ir, Krasnitz:2002ct} treats
$\log (k/v)$ as a perturbation, and expands iteratively in it. 
Then, there is an intermediate overlap region (see
Fig.~\ref{KS:krasnitzfig}) where both solutions should be valid
simultaneously. For large $k$, the solutions naively appear to differ
appreciably in the intermediate regime, unless there is some relation
between large-$k$ terms in the two solutions: this allows us to match the
leading-order terms in $k$. Analytic correlators can in principle be
extracted from this matching at leading order in $k$, but we reiterate
that the dictionary and renormalization problems should be completely
solved before any gauge theory correlators can be quoted with certainty.
(Thus we will not perform the xUV analysis here, but we mentioned it for
completeness).

Here, we will show that the 7-scalar system is analytically
solvable in the mUV approximation, generalizing 
the analysis of \cite{Krasnitz:2000ir, Krasnitz:2002ct}
to the present case.  
We will then check our solutions numerically.

The mUV regime is obtained
in two steps. 
First, we consider the UV region, \ie large
$\tau$, which implies large $\sigma$. To leading order, 
the matrices $Y$ and $Z$ become
\begin{equation}
\label{KS:Yapprox}
  Y = \begin{pmatrix}
       0 & 0 & 0 & 0 & 0 & 0 & 0 \\  
       0 & 0 & 0 & 0 & 0 & 0 & 0 \\  
       0 & 0 & 0 & 0 & 0 & 0 & 0 \\  
       36P & 24P & 6P & 0 & 0 & 0 & 0 \\
       0 & 0 & 0 & 0 & 0 & 0 & 0 \\
       0 & 0 & 0 & 0 & -6P & 0 & 0 \\
       0 & 0 & 0 & 0 & -6P & 0 & 0  
       \end{pmatrix}~,
\end{equation}
and
\begin{equation}
\label{KS:Zapprox}
  Z = \begin{pmatrix}
       32 & 0 & 0 & 0 & 0 & 0 & 0 \\
       0 & 12 & 0 & 0 & 0 & 0 & 0 \\
       0 & 0 & 0 & 0 & 0 & 0 & 0 \\
       336 P & -96P & -24P & 0 & 0 & 0 & 0 \\
       0 & 0 & 0 & 0 & -3 & 0 & 0 \\
       0 & 0 & 0 & 0 & 42P & 21 & 0 \\
       0 & 0 & 0 & 0 & 6P & 0 & -3
       \end{pmatrix}~.
\end{equation}
As a check, we see that $Z$ reproduces the masses in table
\ref{KS:conventions} in the conformal limit $P\rightarrow 0$, although
we noted earlier that one would need rescaled field variables to study
this limit. (The mass does not depend on the field normalization.)

Second, as discussed earlier,
the mUV region is isolated by considering
large external momenta\footnote{We recall that we use dimensionless variables.}
$|\log k| \gg |\log v|$. (As in the discussion above, note that this limits
$v$ from below as well as from above.). This 
means that we can neglect $\ln v$ from
\eqref{KS:eqmot3}. When this is done, $k$ is easily removed from
\eqref{KS:eqmot3} by defining 
\begin{equation}
\label{KS:zdef}
  z = \sqrt{\ln (kP)}\, v~,
\end{equation}
so that we obtain the equation
\begin{equation}
\label{KS:eqmotz}
  \left[ z^3 \partial_z z^{-3} \partial_z -Y z^{-1} \partial_z -
  Z z^{-2} -1 \right] \mfa =0~.
\end{equation}
The variable $z$ blows up in the conformal limit $P\rightarrow 0$ 
(cf.\ the order-of-limits discussion in the previous subsection). If needed, one can always go back to \eqref{KS:eqmot2} and set $P=0$ there.

With $Y$ and $Z$ given by the constant matrices \eqref{KS:Yapprox} and
\eqref{KS:Zapprox}, the equation
\eqref{KS:eqmotz} admits analytical solutions. We are, as
usual in AdS/CFT, interested in the solutions that are regular for
large $z$.  

For the four scalars of the KT system, we obtain
\renewcommand{\arraystretch}{0.8}
\begin{align}
\notag
  \mfa_1 &= z^2 \rmK_6(z) 
    {\scriptstyle \begin{pmatrix} 1 \\ 0 \\ 0 \\ 6P \\ \mathbf{0}_3\end{pmatrix}}
    -6P\left[ 4 z^2 \rmK_4(z) +6z \rmK_1(z)
    +6\rmK_0(z) -3 z^2 \rmK_2(z) \ln z \right] 
    {\scriptstyle \begin{pmatrix}  0 \\ 0 \\ 0 \\ 1 \\ \mathbf{0}_3 \end{pmatrix}}~,
  \\[1mm]
\notag
  \mfa_2 &= z^2 \rmK_4(z)
    {\scriptstyle  \begin{pmatrix} 0 \\ 1 \\ 0 \\ 0 \\ \mathbf{0}_3 \end{pmatrix}}
    -12P\left[ z^2 \rmK_4(z)+ 2z \rmK_1(z) +2 \rmK_0(z) - z^2
     \rmK_2(z) \ln z \right]
    {\scriptstyle \begin{pmatrix} 0 \\ 0 \\ 0 \\ 1 \\ \mathbf{0}_3 \end{pmatrix}}~,
  \\[1mm]
\notag
  \mfa_3 &= z^2 \rmK_2(z)
    {\scriptstyle \begin{pmatrix} 0 \\ 0 \\ 1 \\ 0 \\ \mathbf{0}_3 \end{pmatrix}}
    -3P\left[ 2z \rmK_1(z) + 2\rmK_0(z) -z^2 \rmK_2(z) \ln z \right] 
    {\scriptstyle \begin{pmatrix} 0 \\ 0 \\ 0 \\ 1 \\ \mathbf{0}_3 \end{pmatrix}}~,
  \\[1mm]
\label{KS:flucsols4}
  \mfa_4 &= z^2 \rmK_2(z)
    {\scriptstyle \begin{pmatrix} 0 \\ 0 \\ 0 \\ 1 \\ \mathbf{0}_3 \end{pmatrix}}~,
\end{align}
where the $K_n$ are Bessel functions of order $n$. For the gluino sector, we find
\begin{align}
\notag
  \mfa_5 &= z^2 \rmK_1(z) 
    {\scriptstyle \begin{pmatrix} \mathbf{0}_4 \\ 1 \\ 0 \\ 0 \end{pmatrix}}
    -3P\left[ z^2 \rmK_5(z) \ln z +21 z \rmK_4(z) +\frac76 z^2
      \rmK_1(z) + \right. \\
\notag   &\quad 
    \left. +\frac{80}{z} \rmK_2(z) +\frac{240}{z^2} \rmK_1(z)
    +\frac{384}{z^3} \rmK_0(z)\right] 
    {\scriptstyle \begin{pmatrix} \mathbf{0}_4 \\ 0 \\ 1 \\ 0 \end{pmatrix}} -\\
\notag &\quad
    -3P\left[z^2 \rmK_1(z)\ln z + z \rmK_0(z) \right]
    {\scriptstyle \begin{pmatrix} \mathbf{0}_4 \\ 0 \\ 0 \\ 1 \end{pmatrix}}~,
  \\
\notag
  \mfa_6 &= z^2 \rmK_5(z)
    {\scriptstyle \begin{pmatrix} \mathbf{0}_4 \\ 0 \\ 1 \\ 0 \end{pmatrix}}~, 
  \\[1mm]
\label{KS:flucsols3}
  \mfa_7 &= z^2 \rmK_1(z)
    {\scriptstyle \begin{pmatrix} \mathbf{0}_4 \\ 0 \\ 0 \\ 1 \end{pmatrix}}~.
\end{align}
\renewcommand{\arraystretch}{1}
A few comments on these solutions are in order. We note that from the
approximate matrices $Y$ and $Z$, given in \eqref{KS:Yapprox} and
\eqref{KS:Zapprox}, 
it could have been gleaned already that the component $y$ is a source
for $N_1$ and $N_2$, but not the other way around.
Hence, it is not surprising that the solutions $\mfa_6$ and $\mfa_7$,
where only $N_1$ or $N_2$ are non-zero, are significantly simpler
than $\mfa_5$, where also the $y$-component is turned on. Our next
observation is that $y$ sources the other two gluino-sector fields by
terms linear in $P$. Indeed, the matrices  $Y$ and $Z$ in \eqref{KS:Yapprox},
\eqref{KS:Zapprox} make manifest the fact that the Apreda basis
diagonalizes the mass matrix in the conformal ($P=0$) limit. It makes
it equally manifest that the gauge/gravity dictionary
problem is significantly more pressing in the $P \neq 0$ case than in
the conformal limit. 

These analytical solutions are remarkably simple. 
In the face of dark \emph{Shelob} horror like
the full KS matrices shown in appendix \ref{sec:KSmatrices},
these solutions may prove to be our saving
\emph{E\"arendil} light,\footnote{see \texttt{wikipedia.org}} 
provided we can convince ourselves that they actually do solve the
exact KT equation \eqref{KS:eqmot3} in a suitably approximate sense.
The Krasnitz approximation is valid for very large $k$,
so we give a representative check for moderately large $k$,
when the approximation should just begin to work. 

\begin{figure}[th]
\begin{center}
\psfrag{y}[bc][bc][1][0]{$y$}
\includegraphics[width=0.5\textwidth]{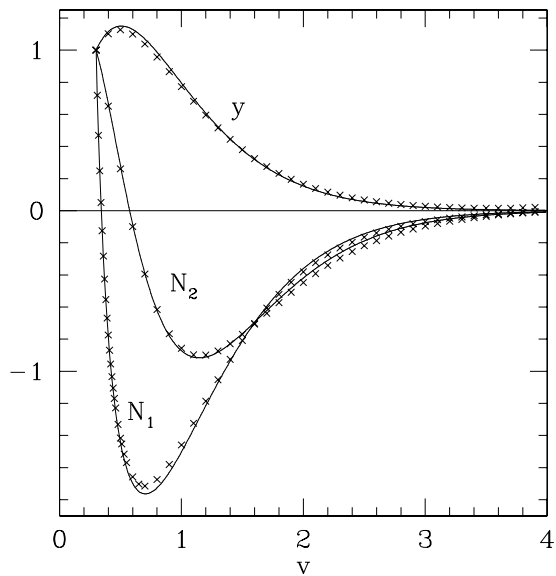}
\caption{\label{fig:compnumerics} 
  Moderate-UV analysis: comparison of the 
  analytical solutions \eqref{KS:flucsols3} of equation
  \eqref{KS:eqmotz} with the
  corresponding numerical solution of \eqref{KS:eqmot3} found by shooting  
  (marked by crosses) for $k=10^3$, $P=1$. The ``response functions''
  agree to an accuracy of 8\%.} 
\end{center}
\end{figure}

The numerical solutions were found by shooting for approximately
regular solutions of \eqref{KS:eqmot3}, that is, minimizing field
values at the grid endpoint by tuning the derivative at a UV
cutoff.\footnote{Further details will be given in future work.}
Superimposing a linear combination of
the solutions $\mfa_5$, $\mfa_6$, $\mfa_7$ of
the approximate equation \eqref{KS:eqmotz}, 
and normalizing them to unity at the cutoff,
we find good qualititative
agreement in Fig.~\ref{fig:compnumerics}. The derivatives at the cutoff essentially give the response
functions for the given value of $k$ (since we normalized the fields
at the cutoff to unity). The numerical responses agree with the
analytical solutions of the approximate equation 
within fairly good accuracy, and the accuracy will improve with energy.
Above and beyond any matching procedure \`a la Krasnitz, we
take the good agreement in Fig.~\ref{fig:compnumerics} as evidence
that the solutions \eqref{KS:flucsols4},
\eqref{KS:flucsols3} may give us the crutch we need as we embark on a
numerical study of the full KS system.

\section{Outlook}
\label{outlook}

In this paper we have investigated aspects of the bulk dynamics of
supergravity fluctuations about the duals of confining gauge theories,
in particular the KS and MN backgrounds. In our
to-do list in the introduction, we called this the ``fluctuation
problem''. This sets the stage for addressing the problem of
calculating correlators in confining gauge theories from
non-asymptotically AdS supergravity backgrounds.
To be able to perform our analysis we derived a consistent truncation 
of type IIB supergravity to a set of scalars coupled to 5-d gravity,
which is general enough to deal with fluctuations about the KS and MN
backgrounds. Importantly, we also developed a gauge-invariant and
sigma-model-covariant formulation of the dynamics of the field
fluctuations in generic, ``fake-supergravity''-type systems, which
should find many applications amongst the various configurations
studied in the literature. Moreover, we point out that the
gauge-invariant formalism naturally includes higher order
interactions. Hence, once the ``dictionary'' and ``renormalization'' 
problems 
for holographic renormalization of confining gauge theories 
(as introduced in the introduction)
are understood, the calculation of three-point functions and scattering
amplitudes (along the lines of \cite{Muck:2004qg}) should become straightforward.

Concerning our particular results, there are many open issues that
could and will be addressed in the near future. 
For the MN system, the most interesting question is to check the
validity of the mass spectra \eqref{MN:spectrum1} and
\eqref{MN:spectrum2} by numerically solving the fluctuation equation 
in the regular background. As discussed in detail at the end of
Sec.~\ref{MN:masspectra}, it is only the existence of the discrete
masses, not their particular values, which is in doubt. This question
is of particular interest also in view of the contrasting results of
\cite{Ametller:2003dj,Caceres:2005yx}. It is an interesting point,
though, that the existence of an upper bound on the masses, as is the
branch point in our case, was also found in 
\cite{Caceres:2005yx}. In any case, all MN results should be
considered in light of the fact that the supergravity approximation is
not under control in the UV region of the MN solution.

For the KS system, it should be straightforward to generalize the  
numerical analysis of Sec.~\ref{KSnearIR} to the full KS background. 
This will not only lead to a better understanding of the range of
validity of our approximate analytical solutions of
Sec.~\ref{KSnearIR}, but also pave the way for the extraction of some
dual physics, once progress has been made on the dictionary and
renormalization problems. Moreover, a more detailed analysis of the
fluctuation equations in the ``extreme UV'' region, for instance the asymptotic expansion 
used by Krasnitz \cite{Krasnitz:2000ir, Krasnitz:2002ct} which we have not performed here,
might shed further light on these problems.

We would also like to comment on the question of the glueball spectrum
in the KS theory. This has already been studied in
\cite{Caceres:2000qe,Amador:2004pz}, where it was argued 
that the glueball spectrum is an IR quantity. As the 3-cycle is an
$S^3$ in the IR, the fields were expanded in harmonics of $S^3$ in
these papers, which was argued to lead to a decoupling of,
for example, the
dilaton. From \eqref{eomdil}, \eqref{i1} and \eqref{i2} it is obvious
that this  
decoupling can never be exact. The complicated dependence on the
internal coordinates present in the PT ansatz simply drops out of the
10-d equation of motion for the dilaton, leaving the 
5-d equation given in \eqref{eomdil}. It might still be that the
expansions performed in \cite{Caceres:2000qe,Amador:2004pz} are
approximately correct, but it would be important to check to what
extent this is really a controllable approximation of the IR
physics of the KS gauge theory. We are optimistic that our formalism
presents a useful starting point to attempt such a check by solving
the linearized gauge-invariant equations for the scalars numerically.

Another interesting open issue was brought up in
Sec.~\ref{sugratrunc}. In the PT ansatz, there is an additional scalar
field, which does not appear in the KS system:
This is the superpartner of the Goldstone mode predicted in
\cite{Aharony:2000pp} and studied in
\cite{Gubser:2004qj,Gubser:2004tf}. Even though it seems to be an
ideal candidate for addressing the problem of calculating 2-point
functions in the KS background (at linear level it decouples from the other scalars
as long as it depends only on the radial variable
\cite{Gubser:2004qj}), it turns out that the dynamics of this mode
requires a generalization of the PT ansatz, once we allow the field to depend 
on all five external coordinates, since then it does not
satisfy the integrability constraint \eqref{partialchi} in general. It would be very 
interesting to find a generalization of the PT ansatz that would lead
to a 5-dimensional consistent truncation of the type IIB equations of
motion and include this additional mode.
Attempts along those lines might also lead to a form of the
5-dimensional effective theory which is manifestly supersymmetric. 

In all, we have found an efficient approach to (at
least an important subset of) the dynamics of fluctuations about the
supergravity duals of confining gauge theories, and demonstrated its
applicability in a number of examples. This is an important step towards a
full understanding of the ``fluctuation'' problem for holographic
renormalization. We are hopeful that our results make also the
``dictionary'' and ``renormalization'' problems more
accessible. We look forward to 
the day when exciting new physics of
confining gauge theories can be reliably extracted from gauge/string duality.

\section*{Acknowledgements}

It is a pleasure to thank O.~Aharony, D.~Berenstein, M.~Bianchi, A.~Buchel, G.~Dall'Agata, C.~Herzog, 
I.~Klebanov, A.~Lerda, S.~Mathur, P.~Ouyang, H.~Samtleben, A.~Tseytlin and 
M.~Zagermann for illuminating insights and helpful advice through discussions 
or email conversations. We thank M.~Bianchi for collaboration in an early stage of the work and 
C.~Herzog for comments on a draft of the paper. 
M.B.\ was supported by the 
Wenner-Gren Foundations, 
and M.H.\ by the German Science Foundation (DFG). 
Moreover, the research of M.B.\ and M.H.\ was supported in part by 
the National Science Foundation under
Grant No. PHY99-07949. 
W.M.\ acknowledges financial support from the MIUR-COFIN project
2003-023852, from the European Commission's $6^{\rm th}$ framework
programme, project MRTN-CT-2004-005104, and from INFN. This work was started while 
M.B.\ and M.H.\ were at Universit\`a di Roma 2 ('Tor Vergata') and they are 
thankful for the hospitality of the string theory group. 
There, 
M.~B.\ was supported by 
a Marie Curie Fellowship, contract number HPMF-CT-2001-01311.
Also, the work of M.~B.\ and M.~H.\ was supported in part by INFN,
in part by the MIUR-COFIN contract 2003-023852, by the EU contracts MRTN-CT-2004-503369 and MRTN-CT-2004-512194, by the INTAS contract 03-51-6346, and by the NATO grant PST.CLG.978785.
We would also 
like to thank CERN, and M.B.\ and M.H.\ the Perimeter Institute, for hospitality 
during part of the work. 


\begin{appendix}
\section{Consistent truncation of type IIB supergravity}
\label{construnc}
In this appendix, we present details of the consistent truncation of
type IIB supergravity to the effective 5-d system \eqref{5dact}. In
particular, we shall show how the constraints \eqref{kconstraint} and
\eqref{partialchi} and the 5-d equations of motion derivable from (\ref{5dact}) arise from the 10-d
supergravity equations. When comparing the results of this appendix with (\ref{5dact})-(\ref{VPT}),
one has to bear in mind that we omitted the tildes in those formulas for esthetic 
reasons.

To start, let us identify our conventions and present some useful formulas.
We use the metric and curvature conventions of MTW, Polchinski and Wald,
\ie the signature is mostly plus and 
\begin{equation}
\begin{split}
  R_{MNP}\, ^Q &= \partial_N \Gamma^Q_{MP} - \partial_M \Gamma^Q_{NP}
  + \Gamma^S_{MP} \Gamma^Q_{SN} - \Gamma^S_{NP} \Gamma^Q_{SM}\ , \\
  R_{MN} &= R_{MPN}\, ^P\ ,
\end{split}
\end{equation}
where the Christoffel symbols are defined as
\begin{equation}
\Gamma^P_{MN} = \frac12 g^{PQ} (\partial_M g_{NQ} + \partial_N g_{MQ} - 
\partial_Q g_{MN})\ . 
\end{equation}
With these conventions one has the following transformation rules
of the Ricci tensor and Ricci scalar of a $D$-dimensional manifold
under a conformal transformation
$\hat g_{MN} = \Omega^2 g_{MN}$, cf.\ appendix~D of \cite{Wald:1984rg}
\begin{align} 
\label{conftrans}
  \hat R_{MN} & =  R_{MN} - (D-2) \nabla_M \partial_N \ln \Omega
  - g_{MN} g^{PQ} \nabla_P \partial_Q \ln \Omega +\\
\notag
  &\quad + (D-2) (\partial_M \ln \Omega)(\partial_N \ln \Omega) 
  - (D-2) g_{MN} g^{PQ} (\partial_P \ln \Omega)(\partial_Q \ln
  \Omega)~, \\
\notag
  \hat R &= \Omega^{-2} \left[ R - 2 (D-1) g^{PQ} \nabla_P 
  \partial_Q \ln \Omega - (D-2)(D-1) g^{PQ} (\partial_P \ln \Omega)
  (\partial_Q \ln \Omega) \right]~. 
\end{align}
Moreover, our conventions for the Hodge star of a $p$-form 
$\omega_p$ are
\begin{equation}
  \star \omega_p = \frac{1}{p! (D-p)!} \omega_{M_1 \ldots M_p} 
  \epsilon^{M_1 \ldots M_p}\,\! _{N_1 \ldots N_{D-p}} 
  \rmd x^{N_1} \wedge \ldots \wedge \rmd x^{N_{D-p}}\ .
\end{equation}
Finally, we adopt the convention to adorn with a tilde objects 
derived from the metric $\rmd s_5^2= \tg_{\mu\nu} \rmd y^\mu \rmd
y^\nu$ (again, note the difference in notation compared to \eqref{PTansatz}; in the main text we 
suppressed the tildes for readability).  
For example, $\tilde \nabla$ denotes the covariant derivative
with respect to $\tg_{\mu\nu}$, and $\tilde F^{\mu ij} = \tg^{\mu\nu}
g^{ik} g^{jl} F_{\nu kl}$. Note the relation of $\tg_{\mu\nu}$ to the
external components of the metric, $g^{(\mathrm{ext})}_{\mu\nu} =
\e{2p-x} \tg_{\mu\nu}$, as follows from
\eqref{PTansatz}. $g^{(\mathrm{int})}$ denotes the remaining internal
part, although we usually omit the superscript 
${}^{\rm (int)}$ if it is clear from the indices $i,j,\ldots$
that we mean an internal component. Also note that the usage of the index $i$ to label the 
internal coordinates in the 10-dimensional metric (\ref{PTansatz}), i.e.\
$i \in \{ \psi, \theta_1, \theta_2, \phi_1, \phi_2\}$, differs from the usage in Sec.~\ref{gaugeinv} and Appendices \ref{geom}, \ref{eomint} and \ref{sources}. 

Let us turn to the analysis of the 10-d equations of motion.
The equation of motion for the RR scalar $C$, \eqref{ceomtend}, is
satisfied, because $H_{M_1 \ldots M_3} F^{M_1 \ldots M_3}=0$.

The equation of motion for $\tF_5$, \eqref{bif5}, 
leads to 
\begin{equation}
 \label{k}
 \partial_\mu K = 2 P \partial_\mu (h_1 + b h_2)~,
\end{equation}
from which the constraint \eqref{kconstraint} follows. 

The second constraint, \eqref{partialchi}, arises from
\eqref{h3eomtend}, in particular, from the mixed components
\begin{equation} 
 \label{eomh3}
 \partial_M \left(\e{-\Phi} H^{M \mu i} \sqrt{-g}\right) = 
 \partial_k \left(\e{-\Phi} H^{k \mu i} \sqrt{-g}\right) = 0~.
\end{equation}
Eq.~\eqref{eomh3} follows from \eqref{h3eomtend}, because
$\epsilon^{M_1 \ldots M_{10}} F_{M_1 \ldots M_5}  
F_{M_6 \ldots M_8}$ has no mixed components, and $C\equiv0$. 
Furthermore, in the first equality of \eqref{eomh3}
we have used that the components of $H$ have at most one 
external index. One can show that all components of 
\eqref{eomh3} are fulfilled, once \eqref{partialchi} is 
imposed.\footnote{For this and most of the following 
calculations we used symbolic
computation software, in particular {\tt Maple}
and the {\tt GRTensor} package \cite{GRTensor}.} 

The equation of motion for the dilaton can be checked as follows. If
we denote  
\begin{align} 
\label{i1}
I_1 &:= 2\e{8p}h_2^2 + 2 \partial_\mu h_2 \tilde\partial^\mu h_2 \\
\notag
  &\quad + \frac{4 (1+2 \e{-2g} a^2) \partial_\mu h_1 \tilde\partial^\mu h_1
  + 8 \e{-2g} a^2 \partial_\mu h_2 \tilde\partial^\mu h_2  
  + 8 a [\e{-2g}(a^2+1)+1] \partial_\mu h_1 \tilde\partial^\mu h_2}{\e{2g}
 + 2 a^2 + \e{-2g}(1-a^2)^2} \\
\notag
 &= \frac16 \e{2p+x} H_{MNP}H^{MNP}~,\\
\label{i2}
I_2 &:= P^2 \left\{ 2 \partial_\mu b \tilde\partial^\mu b + \e{8p} [\e{2g} +
  \e{-2g} (a^2 - 2ab +1)^2 + 2 (a-b)^2] \right\}\\
\notag
  &= \frac16 \e{2p+x} F_{MNP}F^{MNP}~,
\end{align}
then the dilaton-dependent terms in the 5-d action \eqref{actionPT}
are given by
\begin{equation} 
\label{5dactdil}
S_5^{\rm dil} =  \int \rmd^5 y \sqrt{\tg} \left( 
  \frac18 \partial_\mu \Phi \tilde\partial^\mu \Phi + \frac18 \e{-\Phi-2x} I_1
 + \frac18 \e{\Phi-2x} I_2 \right)~.
\end{equation}
The equation of motion that follows from \eqref{dilatoneomtend} and
the constraint \eqref{partialchi} is
\begin{equation} 
\label{eomdil}
  \e{x-2p} \tilde \nabla^2 \Phi = -\frac12 \e{-\Phi-2p-x} I_1 
  + \frac12 \e{\Phi-2p-x} I_2~.
\end{equation}
Obviously, \eqref{eomdil} is precisely the 
equation of motion that one would derive from the 5-d action
\eqref{5dactdil}.

Let us next consider the equation of motion for $F_3$,
\eqref{f3eomtend}, which reads
\begin{equation} 
\label{nablaf}
  \nabla_M (\e{\Phi} F^{MNP}) = - \frac{1}{3!\sqrt{g^{\rm (ext)}}} 
  F_{y_1 \ldots y_5} H_{ijk} \epsilon^{ijkNP}~.  
\end{equation}
The right hand side is only non-vanishing if $N$ and $P$ are 
internal indices. One can verify 
that the same holds
also for the left hand side. To see this, note that 
from the ansatz of $F_3$ and the block
structure of the metric, one only has to check
\begin{equation} 
\label{only}
  \partial_k (F^{k\mu i} \sin \theta_1 \sin \theta_2) = 0\ ,
\end{equation}
because $F_3$ can have at most one external index. 
The validity of \eqref{only} can  easily
be checked with the help of a
computer.  
Thus, the non-trivial part of the equation of motion for 
$F_3$ boils down to 
\begin{equation} \label{eomF}
\nabla_M (e^{\Phi} F^{Mlm}) = - \frac{1}{3!} K e^{3p-\frac32 x} 
H_{ijk} \epsilon^{ijklm}\ ,
\end{equation}
where we have made use of $F_{y_1 \ldots y_5} = K \e{3p-\frac32 x} 
\sqrt{g^{\rm (ext)}}$. It turns out that the angle dependences
of the left and right hand sides in \eqref{eomF} coincide for each 
value of $l$ and $m$. Moreover, the components 
{\it only} 
differ in their angle dependence. More precisely, on
the one hand we have
\begin{align} 
\label{flhs}
  \nabla_M (\e{\Phi} F^{Mlm}) &= \frac{\e{\Phi}}{\sin \theta_1 \sin
  \theta_2} \partial_k (\sin \theta_1 \sin \theta_2 F^{klm})
  + \e{x-2p} \tilde \nabla_\mu (\e{\Phi} \tilde F^{\mu lm}) \\
\notag
  &= P \Big\{ \e{\Phi +6p-x-2g} \left[\e{2g} (b-a) -a(a^2-2ab+1) \right] 
  \\
\notag
  &\quad 
  - \e{x-2p} \tilde \nabla_\mu (\e{\Phi -2x} \tilde \partial^\mu b) \Big\} 
  f^{lm}(\psi,\theta_1,\theta_2,\phi_1,\phi_2)~,
\end{align}
where $f^{lm}$ is some simple rational expression involving trigonometric
functions of the angles, whose precise form depends on $l$ and $m$. On
the other hand,
\begin{equation} 
\label{frhs}
  - \frac{1}{3!} K \e{3p-\frac32 x} 
  H_{ijk} \epsilon^{ijklm} = - K \e{6p-3x} h_2 
  f^{lm}(\psi,\theta_1,\theta_2,\phi_1,\phi_2)~.
\end{equation}
Taking \eqref{flhs} and \eqref{frhs} together leads to the equation of
motion for $b$ 
\begin{equation}
\label{eomb}
P^2 \tilde \nabla_\mu (\e{\Phi -2x} \tilde \partial^\mu b) =
P K \e{8p-4x} h_2 + P^2 \e{\Phi +8p-2x} 
[b - a - a \e{-2g} ( a^2 + 1-2ba)]~,
\end{equation}
which is exactly what one would derive from 
\eqref{5dact}.

Now, we come to the equation of motion for $H_3$, \eqref{h3eomtend},
which is equivalent to
\begin{equation} 
\label{nablaH}
  \nabla_M (\e{\Phi} H^{MNP}) = \frac{1}{3!\sqrt{g^{\rm (ext)}}} 
  F_{y_1 \ldots y_5} F_{ijk} \epsilon^{ijkNP}~.  
\end{equation}
Again, the right hand side is only non-vanishing for internal 
components $N$ and $P$. As we already said above, the same is true 
for the left hand side after imposing the constraint \eqref{partialchi}, cf.\ \eqref{eomh3}.
Thus, the non-trivial part of \eqref{nablaH} becomes 
\begin{align} 
\label{eomH}
  \nabla_M (\e{-\Phi} H^{Mlm}) &= \frac{\e{-\Phi}}{\sin \theta_1 \sin
  \theta_2} \partial_k (\sin \theta_1 \sin \theta_2 H^{klm})
  + \e{x-2p} \tilde \nabla_\mu (\e{-\Phi} \tilde H^{\mu lm}) \\
\notag 
  &= \frac{1}{3!} K \e{3p-\frac32 x} F_{ijk} \epsilon^{ijklm}~.
\end{align}
The expressions for $\tilde \nabla_\mu (\e{-\Phi} \tilde H^{\mu lm})$ are 
much more involved than the case of $F_3$ discussed above. 
Also the general structure of the equations is more complicated.
In particular, we have 
\begin{equation} 
\label{hrhs}
  \frac{1}{3!} K \e{3p-\frac32 x} 
  F_{ijk} \epsilon^{ijklm} = P K \e{6p-3x} 
  \left[ b f_1^{lm}(\psi,\theta_1,\theta_2,\phi_1,\phi_2)  
  + f_2^{lm}(\theta_1,\theta_2,\phi_1,\phi_2)\right]~,
\end{equation}
where $f_1$ and $f_2$ differ in such a way that $f_1$ always
(\ie for all values of $l$ and $m$) 
contains a factor $\cos(\psi)$ or $\sin(\psi)$, whereas 
$f_2$ is independent of $\psi$. Furthermore,
\begin{equation} 
\label{laplinth}
  \frac{\e{-\Phi}}{\sin \theta_1 \sin \theta_2}
  \partial_k (\sin \theta_1 \sin \theta_2 H^{klm}) = 
  -\e{-\Phi+6p-x} h_2\, f_1^{lm}(\psi,\theta_1,\theta_2,\phi_1,\phi_2)~,
\end{equation}
and
\begin{multline} 
\label{laplh}
  \e{x-2p} \tilde \nabla_\mu (\e{-\Phi} \tilde H^{\mu lm}) = 
  \e{x-2p} \tilde \nabla_\mu \left\{ \e{-\Phi-2x} \times
  \phantom{\frac12} \right. \\
  \times \left[ \frac{2a(1+\e{-2g}(1+a^2))}{\e{2g}+2a^2+\e{-2g}(1-a^2)^2}
  \tilde\partial^\mu h_1 
  + \frac{\e{2g}+2a^2+\e{-2g}(1+a^2)^2}{\e{2g}+2a^2+\e{-2g}(1-a^2)^2}
  \tilde\partial^\mu h_2 \right]  f_1^{lm} + \\ 
  \left. + 2 \e{-\Phi-2x} \left[ \frac{1+2a^2
  \e{-2g}}{\e{2g}+2a^2+\e{-2g}(1-a^2)^2} \tilde\partial^\mu h_1 
  + \frac{a (1 + \e{-2g} (1+a^2))}{\e{2g}+2a^2+\e{-2g}(1-a^2)^2} 
  \tilde\partial^\mu h_2 \right] f_2^{lm} \right\}~.
\end{multline}
It is not difficult to verify that the coefficients of 
$f_1^{lm}$ in \eqref{hrhs}, \eqref{laplinth} and \eqref{laplh}, when 
inserted into \eqref{eomH}, add up to give the equation of motion for 
$h_2$, as derived from \eqref{5dact}. That is,
\begin{multline} 
\label{eomh2}
  \tilde \nabla_\mu \left\{ \e{-\Phi-2x}
  \left[ \frac{2a(1+\e{-2g}(1+a^2))}{\e{2g}+2a^2+\e{-2g}(1-a^2)^2}
  \tilde\partial^\mu h_1 
  + \frac{\e{2g}+2a^2+\e{-2g}(1+a^2)^2}{\e{2g}+2a^2+\e{-2g}(1-a^2)^2}
  \tilde\partial^\mu h_2 \right] \right\} \\
  = \e{-\Phi+8p-2x} h_2 + P K \e{8p-4x} b~,
\end{multline}
whereas the coefficients of 
$f_2^{lm}$ give the equation of motion for 
$h_1$, as derived from \eqref{5dact}, \ie
\begin{multline}
\label{eomh1}
  \tilde \nabla_\mu \left\{ 2 \e{-\Phi-2x} \left[ 
  \frac{1+2a^2 \e{-2g}}{\e{2g}+2a^2+\e{-2g}(1-a^2)^2} 
  \tilde\partial^\mu h_1
  + \frac{a (1 + \e{-2g} (1+a^2))}{\e{2g}+2a^2+\e{-2g}(1-a^2)^2} 
  \tilde\partial^\mu h_2 \right] \right\} \\
   = P K \e{8p-4x}~.
\end{multline}

Finally, we consider Einstein's equation, \eqref{einsteintend}.
The mixed components are trivially satisfied, because both sides of
\eqref{einsteintend} are identically zero. For the relevant internal
components we notice that
\begin{equation}
 R_{ij} = R_{ikj}\, ^k + R_{i\mu j}\, ^\mu~.
\end{equation}
Using the fact that the only non-vanishing Christoffel symbols 
besides the pure components $\Gamma^i_{jk}$ and $\Gamma^\mu_{\nu \rho}$ 
are
\begin{equation}
 \Gamma^\mu_{ij} = -\frac12 g^{\mu \nu} \partial_\nu g_{ij}\ , \quad 
 \Gamma^i_{j\mu} = \frac12 g^{il} \partial_\mu g_{jl} = \Gamma^i_{\mu j}\ , 
\end{equation}
we derive
\begin{equation}
\label{rij}
  R_{ij} = R_{ij}^{(\rm int)} 
  -\frac14 (\partial^\mu g_{ij})(\partial_\mu \ln g^{(\rm int)}) 
  + \frac12 (\partial_\mu g_{ik})(\partial^\mu g_{jl}) g^{kl}
  -\frac12 \nabla_\mu \partial^\mu g_{ij}\ ,
\end{equation}
where $g^{(\rm int)}$ denotes the internal block of the metric, and
$R_{ij}^{(\rm int)}$ is the Ricci tensor that follows from it. Notice
the absence of tildes in \eqref{rij}. 

Hence, the internal components of Einstein's equation are
\begin{align} 
\label{rhseinpt}
0 &= R_{ij}^{(\rm int)} 
  -\frac14 (\partial^\mu g_{ij})(\partial_\mu \ln g^{(\rm int)}) 
  + \frac12 (\partial_\mu g_{ik})(\partial^\mu g_{jl}) g^{kl}
  -\frac12 \nabla_\mu \partial^\mu g_{ij} \\
\notag &\quad 
  -\frac{1}{96} F_{i m_1 \ldots m_4} F_j{}^{m_1 \ldots m_4} 
  - \frac14 \left(\e{-\Phi} H_{iPQ} H_j{}^{PQ} 
             + \e{\Phi} F_{iPQ} F_j{}^{PQ}\right) \\
\notag &\quad
  + \frac{1}{8} g_{ij} \left(\e{-\Phi -2p-x} I_1 + \e{\Phi -2p-x}
  I_2\right) \equiv S_{ij} ~, 
\end{align}
where $I_1$ and $I_2$ were defined in \eqref{i1} and
\eqref{i2}.
The expressions are quite complicated, but using 
{\tt Maple} we checked
that all components of \eqref{rhseinpt} are satisfied once 
$S_{\psi \psi}$, $S_{\theta_1 \theta_1}$, 
$S_{\theta_2 \theta_2}$ and $S_{\phi_1 \phi_2}$, for instance, are 
zero. Moreover, taking these four components and solving for 
the second derivatives $\tilde \nabla^2 p$, 
$\tilde \nabla^2 x$, $\tilde \nabla^2 a$ and
$\tilde \nabla^2 g$, leads to the 
same expressions as derived from the action 
\eqref{5dact}, \ie
\begin{multline} 
\label{eomp}
\tilde \nabla_\mu \tilde \partial^\mu p = 
  -\frac16 \left\{ \e{2p-2x-g} [\e{2g}+(1+a^2)] + \frac12 
  \e{-4p-4x}[\e{2g}+(a^2-1)^2 \e{-2g}+2a^2] - \right. \\
  - 2a^2 \e{-2g+8p} -2 \e{-\Phi-2x+8p}h_2^2 - \e{8p-4x} K^2 - \\
  \left. \phantom{\frac12} - P^2 \e{\Phi-2x+8p}
  [\e{2g} + \e{-2g} (a^2 - 2ab +1)^2 + 2 (a-b)^2] \right\}~,
\end{multline}
\begin{multline}
\label{eomx}
\tilde \nabla_\mu \tilde \partial^\mu x  =   
  \e{2p-2x-g} [\e{2g}+(1+a^2)] - \frac12
  \e{-4p-4x}[\e{2g}+(a^2-1)^2 \e{-2g}+2a^2] -\\
  - \frac14 P^2 \e{\Phi-2x} \left\{ \e{8p}[\e{2g} +
  \e{-2g} (a^2 - 2ab +1)^2 + 2 (a-b)^2] + 2 \partial_\mu b \tilde 
  \partial^\mu b \right\} - \\
  - \frac12 \e{8p-4x} K^2 -\frac14 \e{-\Phi-2x} \left\{ 2 \e{8p} h_2^2 + 
  2 \partial_\mu h_2 \tilde \partial^\mu h_2 + \phantom{\frac13} \right. \\
  + \left. 
  \frac{4 (1+2 \e{-2g} a^2) \partial_\mu h_1 \tilde \partial^\mu h_1
  + 8 \e{-2g} a^2 \partial_\mu h_2 \tilde \partial^\mu h_2  
  + 8 a [\e{-2g}(a^2+1)+1] \partial_\mu h_1 \tilde \partial^\mu
  h_2}{\e{2g} + 2a^2+ \e{-2g}(1-a^2)^2} \right\}~,
\end{multline}
\begin{multline}
\label{eoma}
\tilde \nabla_\mu (\e{-2g} \tilde \partial^\mu a) =
  - \e{-\Phi-2x} [\e{2g} + 2a^2+ \e{-2g}(1-a^2)^2]^{-2} \times \\
  \times \left\{ 4 a \e{-2g} (a^2-1) [1+(a^2+1)\e{-2g}] 
  \partial_\mu h_1 \tilde \partial^\mu h_1
  +4 a [\e{-4g} (a^4-1)-1] \partial_\mu h_2 \tilde \partial^\mu h_2 +
  \right. \\
  \left. 
  + 2 [\e{-4g}(a^6+5a^4-5a^2-1) -\e{2g} + \e{-2g} (a^4-1) -a^2 -1]
  \partial_\mu h_1 \tilde \partial^\mu h_2 \right\} - \\
  -2a \e{2p-2x-g}+a \e{-4p-4x}[(a^2-1)\e{-2g}+1] +a \e{-2g+8p} +\\
  +P^2(a-b) \e{\Phi-2x+8p}[ \e{-2g}(a^2-2ab+1)+1]~,
\end{multline}
\begin{multline}
\label{eomg}
\tilde \nabla_\mu \tilde \partial^\mu g = 
  - \e{-2g} \partial_\mu a \tilde \partial^\mu a 
  - 2 \e{-\Phi-2x} [\e{2g} +2a^2 +\e{-2g}(1-a^2)^2]^{-2} \times \\
  \times \left\{ [\e{2g} +4a^2 + \e{-2g} (3a^4 +2a^2-1)] 
  \partial_\mu h_1 \tilde \partial^\mu h_1 
  + 4a^2(1+a^2 \e{-2g}) \partial_\mu h_2 \tilde \partial^\mu h_2  
  +\right.\\
  \left. + 2a[\e{2g} +2(a^2+1) +\e{-2g} (a^4+4a^2-1)] 
  \partial_\mu h_1 \tilde \partial^\mu h_2 \right\} -\\
  - \e{2p-2x-g} [\e{2g}-(1+a^2)] 
  + \frac12 \e{-4p-4x}[\e{2g}-(a^2-1)^2 \e{-2g}] -\\
  - a^2 \e{-2g+8p} +\frac12 P^2 \e{\Phi-2x+8p} [\e{2g}-\e{-2g}(a^2-2ab+1)^2]~. 
\end{multline}
Thus, the equations of motion for $(p,x,a,g)$ arising from
\eqref{5dact} guarantee that all internal components  
of Einstein's equation are satisfied. 

For the external components of  Einstein's equation 
we note that 
\begin{equation}
\label{rmunu}
  R_{\mu \nu} = R_{\mu k\nu}{}^k + R_{\mu \rho \nu}{}^\rho 
  = - \frac{3}{2} \nabla_\mu \partial_\nu (x-2p) - \frac14 g^{ml} g^{ik}
  (\partial_\mu g_{il})(\partial_\nu g_{mk}) 
  + R_{\mu \nu}^{(\rm ext)}\ , 
\end{equation}
where $R_{\mu \nu}^{(\rm ext)}$ stands for the purely external part of the
Ricci-tensor, \ie the one that is calculated solely with 
$\Gamma^\mu_{\nu\rho}$. Using \eqref{conftrans} and 
\begin{align}
  \frac14 g^{ml} g^{ik}
  (\partial_\mu g_{il})(\partial_\nu g_{mk}) &= \e{-2g} \partial_\mu a 
  \partial_\nu a + \frac32 \partial_\mu p 
  \partial_\nu x + \frac32 \partial_\mu x 
  \partial_\nu p +\\
\notag &\quad
  + \partial_\mu g \partial_\nu g + 9 \partial_\mu p \partial_\nu p 
  + \frac54 \partial_\mu x \partial_\nu x~,
\end{align}
one arrives at 
\begin{align} 
\notag
  R_{\mu \nu} & = \tilde R_{\mu \nu} - 2 \partial_\mu x \partial_\nu x
  - 12 \partial_\mu p \partial_\nu p 
  - \e{-2g} \partial_\mu a \partial_\nu a 
  - \partial_\mu g \partial_\nu g 
  - \tg_{\mu \nu} \tilde \nabla_\rho \tilde \partial^\rho p 
  + \frac12 \tg_{\mu \nu} \tilde \nabla_\rho \tilde \partial^\rho x \\
\label{Rmunu}
  &= \frac12 \partial_\mu \Phi \partial_\nu \Phi 
  + \frac{1}{96} F_{\mu \rho \sigma \alpha \beta} F_\nu \, ^{\rho
  \sigma \alpha \beta} 
  + \frac14 \e{-\Phi} H_{\mu mn} H_\nu \, ^{mn} 
  + \frac14 \e{\Phi} F_{\mu mn} F_\nu \, ^{mn} - \\
\notag
  &\quad 
- \frac{1}{48} g_{\mu \nu} (\e{-\Phi} H_{MNP} H^{MNP} 
  + \e{\Phi} F_{MNP} F^{MNP})\ .
\end{align}
Finally, using
\begin{align}
\frac{1}{96} F_{\mu \rho \sigma \alpha \beta}
  F_\nu \, ^{\rho \sigma \alpha \beta} 
  &= - \frac14 K^2 \e{8p-4x} \tg_{\mu \nu}~, \\
  H_{\mu mn} H_\nu  \, ^{mn} &= 
  \frac{8 \e{-2x}}{\e{2g} +2a^2+\e{-2g}(1-a^2)^2} \left\{
   (1+2 \e{-2g} a^2) \partial_\mu h_1 \partial_\nu h_1
  +\phantom{\frac12} \right.\\
\notag &\quad 
  + \frac12 [\e{2g} +2a^2+\e{-2g}(1+a^2)^2] 
  \partial_\mu h_2 \partial_\nu h_2 +\\
\notag & \left. \phantom{\frac12} 
  + a [\e{-2g}(a^2+1)+1] (\partial_\mu h_1 \partial_\nu h_2 + 
  \partial_\mu h_2 \partial_\nu h_1) \right\}~, \\
F_{\mu mn} F_\nu  \, ^{mn} &= 
  4P^2 \e{-2x} \partial_\mu b \partial_\nu b~,  
\end{align}
the relations \eqref{i1} and \eqref{i2}, as well as \eqref{eomp} and
\eqref{eomx} in order to dispose of the second derivatives of $x$ and
$p$ in \eqref{Rmunu}, one verifies that 
\begin{equation}
\tilde R_{\mu \nu}
  = 2 G_{ab} \partial_\mu \phi^a \partial_\nu \phi^b + 
  \frac43 \tilde g_{\mu \nu} V\ , 
\end{equation}
with $G_{ab}$ and $V$ given in \eqref{actionPT} and (\ref{VPT}).

\section{Geometric relations for hypersurfaces}
\label{geom}
The time-slicing (or ADM) formalism \cite{Wald:1984rg,MTW}, which we 
employ in our analysis of Einstein's equations, makes essential use of
the geometry of hypersurfaces \cite{Eisenhart}. Therefore, we shall 
begin with a review of the basic relations governing their
geometry.

A hypersurface in a space-time with coordinates $X^\mu$
($\mu=0,\ldots, d$) and metric $\tg_{\mu\nu}$ is defined by a set of
$d+1$ functions, $X^\mu(x^i)$ ($i=1,\ldots, d$), where the $x^i$ are a
set of coordinates on the hypersurface (note the difference to appendix \ref{construnc}, where 
$i$ labeled the internal coordinates of the 10-dimensional space-time). 
The tangent vectors, $X^\mu_i \equiv
\partial_i X^\mu$, and the normal vector, $N^\mu$, of the hypersurface
can be chosen such that they satisfy the following orthogonality relations,
\begin{equation}
\label{geom:ortho}
\begin{split}
  \tilde g_{\mu\nu}\, X^\mu_i X^\nu_j &= g_{ij}~,\\
  X^\mu_i N_\mu &= 0~,\\
  N^\mu N_\mu &=1~,
\end{split}
\end{equation}
where $g_{ij}$ represents the (induced) metric on the hypersurface. 
Henceforth, a tilde will be used to label quantities
characterizing the ($d+1$)-dimensional space-time manifold, whereas
those of the hypersurface remain unadorned. 

The equations of Gauss and Weingarten define the second fundamental
form, $\K_{ij}$, of the hypersurface,
\begin{align}
\label{geom:gauss1}
  \partial_i X^\mu_j + \tG{\mu}{\lambda\nu} X^\lambda_i X^\nu_j
  - \Gamma^k_{\;\;ij} X^\mu_k &= \K_{ij} N^\mu~,\\
\label{geom:weingarten}
  \partial_i N^\mu + \tG{\mu}{\lambda\nu} X^\lambda_i N^\nu 
  &= -\K^j_i X^\mu_j~.
\end{align}
The second fundamental form describes the extrinsic curvature of the
hypersurface, and is related to the intrinsic curvature by another
equation of Gauss,
\begin{align}
\label{geom:gauss2}
  \tR_{\mu\nu\lambda\rho} X^\mu_i X^\nu_j X^\lambda_k X^\rho_l 
  &= R_{ijkl} + \K_{il} \K_{jk} - \K_{ik} \K_{jl}~.\\
\intertext{Furthermore, it satisfies the equation of Codazzi,}
\label{geom:codazzi}
  \tR_{\mu\nu\lambda\rho} X^\mu_i X^\nu_j N^\lambda X^\rho_k 
  &= \nabla_i \K_{jk} - \nabla_j \K_{ik}~. 
\end{align}
The symbol $\nabla$ denotes covariant derivatives with respect to the
induced metric $g_{ij}$.  

The above formulas simplify if (as in the familiar time-slicing
formalism), we choose space-time coordinates such that 
\begin{equation}
\label{geom:Xdef}
  X^0 = {\rm const.}~, \quad X^i=x^i~.
\end{equation}
Then, the tangent vectors are given by $X^0_i=0$ and $X^j_i
=\delta_i^j$. One conveniently splits up the space-time
metric as (shown here for Euclidean signature)
\begin{align}
\label{geom:split}
  \tg_{\mu\nu} &= \begin{pmatrix} n_i n^i +n^2 & n_j \\
				n_i & g_{ij} \end{pmatrix}~,\\
\intertext{whose inverse is given by}
\label{geom:splitinv}
  \tg^{\mu\nu} &= \frac1{n^2} \begin{pmatrix} 1& -n^j \\
			-n^i & n^2 g^{ij} +n^i n^j \end{pmatrix}~.
\end{align}
The matrix $g^{ij}$ is the inverse of $g_{ij}$, and is used to raise
hypersurface indices. The quantities $n$ and
$n^i$ are the lapse function and shift vector, respectively. 

The normal vector $N^\mu$ satisfying the orthogonality relations 
\eqref{geom:ortho} is given by 
\begin{equation}
\label{geom:normal}
  N_\mu = (n,0)~, \qquad N^\mu= \frac1{n}(1,-n^i)~.
\end{equation}
Then, one can obtain the second fundamental form from the equation of
Gauss \eqref{geom:gauss1} as 
\begin{equation}
\label{sli:Kij}
  \K_{ij} = n \tG{0}{ij} = - \frac1{2n} \left(\partial_0 g_{ij} -
  \nabla_i n_j - \nabla_j n_i \right)~.
\end{equation}

We are interested in expressing all bulk quantities in terms of
hypersurface quantities. Using the equations of Gauss and Weingarten,
some Christoffel symbols can be expressed as follows,
\begin{align}
\label{geom:conn_kij}
  \tG{k}{ij} &= \Gamma^k_{\;\;ij} - \frac{n^k}n \K_{ij}~,\\
\label{geom:conn_0i0}
  \tG{0}{i0} &= \frac1n \partial_i n + \frac{n^j}n \K_{ij}~,\\
\label{geom:conn_ki0}
  \tG{k}{i0} &= \nabla_i n^k - \frac{n^k}n \partial_i n 
  - n\K_{ij} \left( g^{jk}+ \frac{n^jn^k}{n^2} \right)~.
\end{align}
The remaining components, $\tG{0}{00}$ and $\tG{k}{00}$,
are easily found from their definitions using \eqref{geom:split}
and \eqref{geom:splitinv},
\begin{align}
\label{geom:conn_000}
  \tG{0}{00} &= \frac1n \left( \partial_0 n +n^j \partial_j n +n^i n^j
  \K_{ij} \right)~,\\
\label{geom:conn:k00}
  \tG{k}{00} &= \partial_0 n^k + n^i \nabla_i n^k - n \nabla^k n -2n
  \K^k_i n^i -n^k \tG{0}{00}~.
\end{align}

\section{Intermediate steps}
\label{eomint}
In this appendix, we provide the equations of motion in terms of the
geometric variables characterizing the time-slice hypersurfaces
introduced in appendix~\ref{geom}. The equations of motion that follow
from the action \eqref{action5d} are\footnote{Note that, as opposed to the main text, we use 
a tilde here to denote 5d quantities in order to distinguish them 
from the hypersurface quantities.} 
\begin{equation}
\label{scalar_eom_app}
  \tilde{\nabla}^2 \phi^a +\G{a}{bc} \tilde{g}^{\mu\nu} 
  (\partial_\mu \phi^b)(\partial_\nu \phi^c) -V^a = 0
\end{equation}
for the scalar fields, and Einstein's equations
\begin{equation}
\label{Einstein_eom_app}
  E_{\mu\nu} = -\tilde{R}_{\mu\nu} 
  + 2 G_{ab} (\partial_\mu \phi^a)(\partial_\nu \phi^b) 
  + \frac{4}{d-1} \tilde{g}_{\mu\nu} V =0~.
\end{equation}

In terms of hypersurface quantities, \eqref{scalar_eom_app} takes the
form
\begin{multline}
\label{scalar_eom_int}
  \left\{ \partial_r^2 -2n^i \partial_i \partial_r +n^2 \nabla^2 +n^i
  n^j \nabla_i \partial_j - ( n \K^i_i +\partial_r \ln n - n^i
  \partial_i \ln n) \partial_r + \phantom{\K^j_j} \right. \\
  \left. + \left[ n \nabla^i n - \partial_r n^i + n^j \nabla_j n^i 
  + n^i ( n \K^j_j +\partial_r \ln n - n^j \partial_j \ln n)
  \partial_i \right] \right\} \phi^a +\\
  + \G{a}{bc} \left[ (\partial_r \phi^b)(\partial_r \phi^c) -2n^i
  (\partial_i \phi^b) (\partial_r \phi^c) +(n^2 g^{ij} +n^i
  n^j)(\partial_i \phi^b)(\partial_j \phi^c) \right] - n^2
  \frac{\partial V}{\partial \phi^a} =0~.
\end{multline}

Eq.\ \eqref{Einstein_eom_app} splits  
into components that are normal, mixed, and
tangential to the hypersurface, obtained by
projecting with $N^\mu N^\nu -g^{ij}X^\mu_i X^\nu_j$, $N^\mu X^\nu_i$
and $X^\mu_i X^\nu_j$, respectively. 
The normal components become
\begin{multline}
\label{Einstein_normal_int}
  (n\K^i_j)(n\K^j_i) -(n\K^i_i)^2 + n^2 R -4 n^2 V +\\
  +2 G_{ab} \left[ (\partial_r \phi^a)(\partial_r \phi^b) -2n^i
  (\partial_i \phi^a) (\partial_r \phi^b) +(n^i n^j - n^2 g^{ij}) 
  (\partial_i \phi^a)(\partial_j \phi^b) \right] =0~.
\end{multline} 
The mixed components are 
\begin{multline}
\label{Einstein_mixed_int}
  \partial_i (n\K^j_j) - \nabla_j  (n\K^j_i) 
  - n \K^j_j \partial_i \ln n + n \K^j_i \partial_j \ln n 
  - 2 G_{ab} \left(\partial_r \phi^a -n^j \partial_j \phi^a\right)
  (\partial_i \phi^b) =0~.
\end{multline} 
For the tangential components one obtains
\begin{multline}
\label{Einstein_tang_int}
  -\partial_r (n\K^i_j) +n^k \nabla_k (n\K^i_j) + n\K^i_j (n\K^k_k
  +\partial_r \ln n -n^k \partial_k \ln n) + n \nabla^i \partial_j n
  + \\
  +n\K^i_k \nabla_j n^k - n\K^k_j \nabla_k n^i - n^2 R^i_j + 2n^2
  G_{ab} (\nabla^i \phi^a) (\partial_j \phi^b) +\frac{4 n^2 V}{d-1} 
  \delta^i_j=0~.
\end{multline} 
The equations of motion given in Sec.~\ref{fieldeq} are obtained from
\eqref{scalar_eom_int}--\eqref{Einstein_tang_int} upon expanding the
fields and using the substitution rules \eqref{field_subs}. For this,
the following expressions for geometric hypersurface quantities, up to
quadratic order in the gauge-invariant fluctuations, are useful. 
The extrinsic curvature tensor is
\begin{align}
\label{nK_exp}
\begin{split}
  n\K^i_j &\to \frac{2}{d-1} W \delta^i_j -\frac12 \partial_r \mfe^i_j
  +\frac12 \left( \partial^i \mfd_j +\partial_j \mfd^i 
  + 2 \frac{\partial^i \partial_j}{\Box} \mfc \right) 
  +\frac12 \mfe^i_k \partial_r \mfe^k_j \\
  &\quad 
  - \frac12 \left[ 
  \mfe^i_k \left( \partial^k \mfd_j +\partial_j \mfd^k +
  2 \frac{\partial_j \partial^k}{\Box} \mfc \right) 
  + \left( \mfd^k +\frac{\partial^k}{\Box} \mfc \right) 
  \left( \partial^i \mfe_{jk} + \partial_j \mfe^i_k 
        -\partial_k \mfe^i_j \right)\right]~,
\end{split}\\
\intertext{and its trace is}
\label{nK_tr_exp}
  n\K^i_i &\to \frac{2d}{d-1} W 
  + \mfc +\frac12 \mfe^i_k \partial_r \mfe^k_i 
  - \mfe^i_k \left( \partial^k \mfd_i + 
  \frac{\partial_i \partial^k}{\Box} \mfc \right)~.
\end{align}
The intrinsic Ricci tensor is replaced by
\begin{align}
\label{R_ij_exp}
\begin{split}
  R^i_j &\to -\frac12 \e{-2A} \left[ \Box \mfe^i_j 
  + \mfe^k_l \left( \partial^i \partial_k \mfe^l_j 
                  + \partial_j \partial^l \mfe^i_k 
		  - \partial_k \partial^l \mfe^i_j 
                  - \partial^i \partial_j \mfe^l_k \right) 
  - \mfe^i_k \Box \mfe^k_j \phantom{\frac12} \right. \\
  &\quad \left. 
  -\frac12 (\partial^i \mfe^k_l)(\partial_j \mfe^l_k) 
         + (\partial_l \mfe^i_k)(\partial^k \mfe^l_j) 
         - (\partial_l \mfe^k_j)(\partial^l \mfe^i_k)\right]~,
\end{split}\\
\intertext{and the Ricci scalar becomes}
\label{R_exp}
  R &\to  \e{-2A} \left[ \mfe^i_j \Box \mfe^j_i
  + \frac34 \partial_i \mfe^j_k \partial^i \mfe^k_j 
  -\frac12 \partial_i \mfe^k_j \partial^j \mfe^i_k \right]~.
\end{align}

\section{Quadratic source terms}
\label{sources}
In this appendix, we provide the explicit expressions for the source
terms $J_a$, $J$, $J^i$ and $J^i_j$ to quadratic order, which appear
in the equations of motion \eqref{eqa1}, \eqref{eqc}, \eqref{eqbd} and
\eqref{eqe}, respectively. The field $\mfd^i$ has been dropped
everywhere, since its solution \eqref{d_sol} is of second order.
Moreover, we have used the linear equations of motion in order to
eliminate terms, in particular the relation
\begin{equation}
\label{bc_useful}
  \partial_r \mfc -\frac{2d}{d-1}W \mfc - \e{-2A} \Box \mfb =0~,
\end{equation}
which follows from \eqref{b_sol}, \eqref{c_sol} and \eqref{eqphi}. 

\begin{equation}
\label{Ja}
\begin{split}
  J^a &= \frac12 \left[ V^a_{\;\;|bc} -\R^a_{\;\;bcd} V^d -
  \left(\R^{a}_{\;\;bcd|e} -\R^a_{\;\;deb|c} \right) W^d W^e \right]
  \mfa^b\mfa^c -\\
  &\quad
  -2 \R^a_{\;\;bcd} W^d (D_r \mfa^b) \mfa^c+ 2V^a_{\;\;|b} \mfa^b \mfb +
  (D_r \mfa^a) (\mfc +\partial_r \mfb) +\\
  &\quad 
  + 2 V^a \mfb^2 
  + 2 \left( \frac{\partial^i}{\Box} \mfc \right)\partial_i D_r\mfa^a 
  - \e{-2A} \left( 2\mfb \Box \mfa^a 
    - \mfe^i_j \partial_i \partial^j \mfa^a \right) -
  \\ 
  &\quad
  -\underline{V^a \mfb^2} 
  +W^a \left[ -\underline{\mfb \partial_r \mfb}
    + \underline{\frac12 \mfe^i_j \partial_r \mfe^j_i} 
    - \underline{\mfe^i_j \frac{\partial_i \partial^j}{\Box} \mfc} 
    - \underline{(\partial_i \mfb) \frac{\partial^i}{\Box} \mfc} \right]~.
\end{split} 
\end{equation}

\begin{equation}
\label{J}
\begin{split}
  J &= 2 V_{a|b} \mfa^a \mfa^b -2 (D_r \mfa^a)(D_r \mfa_a)
  +2\R^a_{\;\;bcd} W_a W^c \mfa^b \mfa^d 
  +2 \e{-2A} (\partial^i \mfa^a)(\partial_i \mfa_a) +\\
  &\quad 
  + 8V_a \mfa^a \mfb 
  +\underline{4 W_a (\partial_i \mfa^a) \frac{\partial^i}{\Box} \mfc}  
  + 8 V \mfb^2 -\underline{4V\mfb^2} 
  + \left( \Pi^j_i \mfc \right) 
    \left(\frac{\partial^i \partial_j}{\Box} \mfc \right) + \\
  &\quad 
  + \left( \frac{\partial_i\partial^j}{\Box} \mfc \right) 
    \left( \partial_r \mfe^i_j - \underline{4W \mfe^i_j} \right) 
  - \frac14 (\partial_r \mfe^i_j)(\partial_r \mfe^j_i) 
  + \underline{2W \mfe^i_j \partial_r \mfe^j_i} -\\
  &\quad 
  - \e{-2A} \left[\mfe^i_j  \Box \mfe^j_i +\frac34
  (\partial_i \mfe^j_k)(\partial^i \mfe_j^k) -\frac12 (\partial_i
  \mfe^j_k)(\partial^k \mfe^i_j) \right]~.
\end{split}
\end{equation}

\begin{equation}
\label{Ji}
\begin{split}
  J_i &= 2 (\partial_i \mfa^a) D_r \mfa_a 
  - \underline{2 W \mfb \partial_i \mfb} 
  + (\partial_j \mfb) \Pi^j_i \mfc
  + \frac12 (\partial_j \mfb) \partial_r \mfe^j_i- \\
  &\quad
  - \frac12 \left( \frac{\partial_j}{\Box} \mfc \right) \Box \mfe^j_i 
  - \frac14 \partial_i \partial_r (\mfe^j_k \mfe^k_j) 
  + \frac12 \mfe^j_k \partial_r \partial_j \mfe^k_i
  + \frac14 (\partial_i \mfe^j_k)(\partial_r \mfe^k_j)~.
\end{split}
\end{equation}

The terms underlined in \eqref{Ja}, \eqref{J} and \eqref{Ji} can be
eliminated by the field redefinitions
\begin{align}
\label{redef1b}
  \mfb&\to \mfb +\frac12 \mfb^2~,\\
\label{redef1c}
  \mfc& \to \mfc -\frac12 \mfe^i_j \partial_r \mfe_i^j
                +(\partial_i \mfb) \frac{\partial^i}{\Box} \mfc 
		+ \mfe^i_j \frac{\partial_i \partial^j}{\Box} \mfc~.
\end{align}

\begin{equation}
\label{J_ij}
\begin{split}
  J^i_j &= \Pi^{ik}_{jl} \left\{ 
   2 \left( \frac{\partial^l\partial_m}{\Box} \mfc \right) 
     \left( \frac{\partial^m\partial_k}{\Box} \mfc \right) 
  -2 \left( \frac{\partial^l\partial_k}{\Box} \mfc \right) 
     (\mfc +\partial_r \mfb) + \right.\\
  &\quad 
  +2 (\partial_m\partial_r \mfe^l_k) \left(\frac{\partial^m}{\Box}
     \mfc \right) 
  + (\partial_r \mfe^l_k) (\mfc +\partial_r \mfb) 
  + \underline{(\partial_r \mfe^l_m) (\partial_r \mfe^m_k)} +\\
  &\quad 
  + \e{-2A} \left[ 2(\partial^l \mfb)(\partial_k \mfb) 
      -4 (\partial^l \mfa^a)(\partial_k \mfa_a) 
      -2 \mfb \Box \mfe^l_k 
      -2\mfe^m_n \partial^l \partial_m \mfe^n_k +
  \phantom{\frac12} \right.  \\ 
  &\quad \left. \left. 
     + \mfe^m_n \partial_m \partial^n \mfe^l_k
     -\frac12 (\partial^l \mfe^m_n)(\partial_k \mfe^n_m) 
     -(\partial_m \mfe^l_n)(\partial^n \mfe^m_k) 
     +\underline{(\partial_m \mfe^n_k)(\partial^m \mfe^l_n)} 
     \right] \right\}~.
\end{split}
\end{equation}
The underlined terms in \eqref{J_ij} can be eliminated by the field
redefinition 
\begin{equation}
\label{redef1e}
  \mfe^i_j \to \mfe^i_j +\frac12 \Pi^{ik}_{jl} (\mfe^l_m \mfe^m_k)~.
\end{equation}

%
%
\section{Matrices for the KS background}
\label{sec:KSmatrices}
 
In this appendix we give the explicit form of the matrices appearing
in \eqref{KS:MNdef} in the KS background. In order to keep the
formulas under (typographical) control, we introduce the following
notation
\begin{equation}
\label{abbrev}
\begin{split}
  H_1 &= \e{6 q+4 f}~, \\
  H_2 &= \e{10 q+6 f}~ , \\
  H_3 &= \e{-4 q+4 f}~, \\
  \Upsilon_1 &= \e{-\Phi} [2\cosh(2 y)P+\sinh(2 y)(2P- N_2+ N_1)]~,\\
  \Upsilon_2 &= \e{-\Phi} [2\sinh(2 y)P+\cosh(2 y)(2P- N_2+ N_1)]~,\\
  \Upsilon_3 &= Q - P (s- N_1 - N_2) + \frac12 (N_1^2 - N_2^2)~,\\
  \Upsilon_4 &= -4\e{6 q} -6\e{10 f+6 q} \cosh y +5\sqrt{27} \e{6 f} 
                   \Upsilon_3~,\\
  \Upsilon_5 &= \cosh y -\e{-10 f}~,\\
  \Upsilon_6 &= \e{-\Phi} (N_1+ N_2)~,\\
  \Upsilon_7 &= -\frac{2}{15}\left[8\e{-10 f}+12\cosh y 
                -\frac{25\sqrt{27}\Upsilon_3}{H_1}\right]~,
\end{split}
\end{equation}
where the fields denote the background values given in
Sec.~\ref{KSbackground}. With these abbreviations, the matrices are
given by 


\setlength{\arraycolsep}{1mm}
\begin{eqnarray}
\label{GWks}
  \hspace{-1cm} \G{a}{bc} W^c = H_3 \times && \nonumber \\
   && \hspace{-3cm} {\scriptstyle \begin{pmatrix} 
    0 & 0 & 0 & -\frac{\sqrt{27}}{10}\frac{P}{H_1} & 0 &
    \frac{\sqrt{27}}{10}\frac{N_1+P}{H_1}&
    -\frac{\sqrt{27}}{10}\frac{N_2-P}{H_1}\\
    0 & 0 & 0 & -\frac{\sqrt{27}}{10}\frac{P}{H_1}& 0 &
    \frac{\sqrt{27}}{10}\frac{N_1+P}{H_1}& 
    -\frac{\sqrt{27}}{10}\frac{N_2-P}{H_1} \\ 
    0 & 0 & 0 & -\sqrt{27}\frac{P}{H_1} & 0 &
    -\sqrt{27}\frac{N_2-P}{H_1} & \sqrt{27}\frac{N_1+P}{H_1} \\ 
    9\Upsilon_1 & 6\Upsilon_1 & \frac32 \Upsilon_1 & 
    \frac{\sqrt{27}\Upsilon_3-2\e{6(q-f)}}{H_1}& 
    3\Upsilon_2 & -3 \sinh y& -3 \sinh y \\ 
    0 & 0 & 0 & -\frac{\sqrt{27}}{2}\frac{N_1- N_2+2P}{H_1}& 0 &  
    \sqrt{27}\frac{P}{H_1} & \sqrt{27}\frac{P}{H_1}\\
    -\frac92(\Upsilon_2+\Upsilon_6) & -3(\Upsilon_2+\Upsilon_6)& 
    -\frac34(\Upsilon_2-\Upsilon_6)& -\frac32 \sinh y & 
    -\frac32 \Upsilon_1 & \frac{\sqrt{27}\Upsilon_3- 2\e{6(q-f)}}{H_1} 
    & 0 \\ 
    -\frac92(\Upsilon_2-\Upsilon_6) & -3(\Upsilon_2-\Upsilon_6)& 
    -\frac34(\Upsilon_2+\Upsilon_6) & -\frac32 \sinh y& 
    -\frac32 \Upsilon_1& 0 & 
    \frac{\sqrt{27}\Upsilon_3-2\e{6(q-f)}}{H_1} 
  \end{pmatrix}}~, \nonumber
\end{eqnarray}

\setlength{\arraycolsep}{.4mm}
\begin{eqnarray}
\label{W2ks}
   \hspace{-1.5cm} \partial_b W^a = H_3 \times && \nonumber\\ 
   && \hspace{-2.5cm} {\scriptstyle \begin{pmatrix} 
    \Upsilon_7 &
    \frac85\Upsilon_5 & 0 & \frac{\sqrt{3}P}{H_1} & \frac25 \sinh y & 
    -\sqrt{3} \frac{N_1 + P}{H_1} & \sqrt{3} \frac{N_2 - P}{H_1} \\
    \frac{12}{5}\Upsilon_5 & -\frac65 (2\cosh y+3 \e{-10 f}) & 0 & 0 & 
    -\frac35 \sinh y & 0 & 0\\
    0& 0& 0& 0& 0& 0& 0\\
    12\Upsilon_1 & -12\Upsilon_1 & -3\Upsilon_1 & 0 & -6\Upsilon_2 & -
    3\sinh(2 y) \e{\Phi} & 3\sinh(2 y) \e{\Phi} \\
    12 \sinh y & -12 \sinh y & 0 & 0 & -3\cosh y & 0 & 0\\
    -6(\Upsilon_2+\Upsilon_6) & 6(\Upsilon_2+\Upsilon_6) & 
    \frac32(\Upsilon_2-\Upsilon_6) & 0 & 3\Upsilon_1 & 
    \frac32 [\e{-\Phi}+\e{\Phi} \cosh(2 y)] & 
    \frac32 [\e{-\Phi}-\e{\Phi} \cosh(2 y)] \\
    -6(\Upsilon_2-\Upsilon_6) & 6(\Upsilon_2-\Upsilon_6) & 
    \frac32(\Upsilon_2+\Upsilon_6) & 0 & 3\Upsilon_1 & 
    -\frac32 [\e{-\Phi}-\e{\Phi}\cosh(2 y)] & 
    -\frac32 [\e{-\Phi}+\e{\Phi}\cosh(2 y)] 
  \end{pmatrix}}~, \nonumber
\end{eqnarray}

\setlength{\arraycolsep}{1mm}
\begin{eqnarray}
\label{WWks}
  \hspace{-1.5cm} W^aW_b = H_3^2 \times && \nonumber \\
  && \hspace{-2.9cm} {\scriptstyle \begin{pmatrix} 
    \frac{\Upsilon_4^2}{15H_2^2H_3^2} & 
    \frac{2 \Upsilon_4 \Upsilon_5}{5 H_2 H_3} & 0 & 
    \frac{\sqrt{3} P \Upsilon_4}{10 H_1 H_2 H_3} & 
    \frac{\Upsilon_4 \sinh y}{10 H_2 H_3} & 
    -\frac{\sqrt{3} (N_1+P) \Upsilon_4}{10 H_1 H_2 H_3} & 
    \frac{\sqrt{3} (N_2-P) \Upsilon_4}{10 H_1 H_2 H_3}  \\
    \frac{3 \Upsilon_4\Upsilon_5}{5 H_2 H_3} & 
    \frac{18}{5} \Upsilon_5^2 & 0 & 
    \frac{3 \sqrt{27} P \Upsilon_5}{10 H_1} & 
    \frac{9}{10} \sinh y \Upsilon_5 & 
    -\frac{3 \sqrt{27} (N_1+P) \Upsilon_5}{10 H_1} & 
    \frac{3 \sqrt{27} (N_2-P) \Upsilon_5}{10 H_1}\\
    0& 0& 0& 0& 0& 0& 0\\
    \frac{3\Upsilon_1 \Upsilon_4}{H_2 H_3} & 
    18 \Upsilon_1 \Upsilon_5& 0&  
    \frac{3 \sqrt{27} P \Upsilon_1}{2 H_1} & 
    \frac92 \sinh y \Upsilon_1 & 
    -\frac{3 \sqrt{27} (N_1+P) \Upsilon_1}{2 H_1} & 
    \frac{3 \sqrt{27} (N_2-P) \Upsilon_1}{2 H_1}\\
    \frac{3\Upsilon_4 \sinh y}{H_2 H_3} & 18\Upsilon_5\sinh y& 0& 
    \frac{3 \sqrt{27} P \sinh y}{2 H_1} & \frac92\sinh^2 y & 
    -\frac{3 \sqrt{27} (N_1+P) \sinh y}{2 H_1} & 
    \frac{3 \sqrt{27} (N_2-P) \sinh y}{2 H_1} \\
    -\frac{3 \Upsilon_4 (\Upsilon_2+\Upsilon_6)}{2H_2H_3} & 
    -9 \Upsilon_5(\Upsilon_2+\Upsilon_6)& 0& 
    -\frac{3 \sqrt{27} P (\Upsilon_2 +\Upsilon_6)}{4H_1} & 
    -\frac94 (\Upsilon_2+\Upsilon_6) \sinh y & 
    \frac{3 \sqrt{27} (N_1+P) (\Upsilon_2 +\Upsilon_6)}{4H_1} & 
    -\frac{3 \sqrt{27} (N_2-P) (\Upsilon_2 +\Upsilon_6)}{4H_1} \\
    -\frac{3\Upsilon_4(\Upsilon_2-\Upsilon_6)}{2 H_2H_3} & 
    -9 \Upsilon_5 (\Upsilon_2-\Upsilon_6) & 0 & 
    -\frac{3 \sqrt{27} P (\Upsilon_2 -\Upsilon_6)}{4H_1} & 
    -\frac94 (\Upsilon_2-\Upsilon_6) \sinh y & 
    \frac{3 \sqrt{27} (N_1+P) (\Upsilon_2 -\Upsilon_6)}{4H_1} & 
    -\frac{3 \sqrt{27} (N_2-P) (\Upsilon_2 -\Upsilon_6)}{4H_1}
  \end{pmatrix}}~. \nonumber
\end{eqnarray}
\newpage

\end{appendix}

\bibliographystyle{JHEP}
\bibliography{gaugeinv}

\end{document}